\documentclass[12pt]{article}
\usepackage{rotate,epsf}
\usepackage{hyperref}
\newcommand{\insertfig}[2]{\mbox{\epsfxsize=#1cm \epsfbox{#2.eps}}}
\newcommand{\ft}[2]{{\textstyle\frac{#1}{#2}}}

\newcommand{\Bx}{x_{\rm B}}
\newcommand{\bit}[1]{\mbox{\boldmath$#1$}}
\font\cmss=cmss12 
\def\1{\hbox{{1}\kern-.25em\hbox{l}}}
\def\bfZ{\relax{\hbox{\cmss Z\kern-.4em Z}}}

\begin{document}

\begin{titlepage}

\begin{flushright}
DOE/ER/40762-264 \\ [-2mm] UMD-PP\#03-055
\end{flushright}

\centerline{\large \bf Probing generalized parton distributions with}
\centerline{\large \bf electroproduction of lepton pairs off the nucleon}

\vspace{15mm}

\centerline{\bf A.V. Belitsky$^a$, D. M\"uller$^b$}

\vspace{15mm}

\centerline{\it $^a$Department of Physics}
\centerline{\it University of Maryland at College Park}
\centerline{\it College Park, MD 20742-4111, USA}

\vspace{5mm}

\centerline{\it $^b$Fachbereich Physik, Universit\"at  Wuppertal}
\centerline{\it D-42097 Wuppertal, Germany}

\vspace{15mm}

\centerline{\bf Abstract}

\vspace{0.5cm}

\noindent

We evaluate the differential cross section for the electroproduction of
lepton pairs off a polarized nucleon target in the generalized Bjorken
region to leading power accuracy in hard momentum. We discuss the importance
of this process for phenomenology of generalized parton distributions. A
special attention is given to the sensitivity of physical observables, i.e.,
diverse asymmetries, to their dependence on scaling variables that allows
to map directly the functional two-dimensional surface of generalized parton
distributions.

\vspace{5cm}


\vspace{0.3cm}


\end{titlepage}

\section{Introduction}

A quantum mechanical system is determined by its wave function. Acquiring
the latter from theoretical considerations, like solving the Schr\"odinger
equation, or experimental measurements allows one to predict any physical
observable of the system. The bulk of experimentally accessible quantities
is sensitive only to the absolute value of the wave function and the phase
of the latter is essentially unattainable. To circumvent the difficulty one
has to measure correlations of wave functions, --- or more generally the
density matrix, --- where the phase difference of wave functions can be
probed. The interference of a test system with a reference source,
possessing an a priori known characteristics, serves the purpose of
reconstructing the missing phase of the wave function and thus one acquires
complete information on the quantum mechanical system in question.

The nucleon represents a relativistic multi-particle quantum system in a
bound state whose dynamic is driven by strong interactions. It is the
subject of intensive studies for several decades. Recently, it was realized
that the most powerful theoretical tools in the analysis of the nucleon
structure are one-quark and one-gluon correlations, dubbed the generalized
parton distributions (GPDs) \cite{MueRobGeyDitHor94,Ji97,Rad97}. They are
analogous to a field-theoretical generalization of the phase-space Wigner
quasi-probability function of non-relativistic quantum mechanics
\cite{Bel03,Ji03}, --- a specific Fourier transform of a density matrix
alluded to above. A GPD depends on several
kinematical variables and it has to be mapped as a function of all of them
in order to predict such a fundamental quantity as the angular momentum of
nucleon's constituents, which is given by their second moment \cite{Ji96}
\begin{equation}
\label{SpinSumRule}
\int_{-1}^1 d \xi \, \xi \,
\Big(
H_{q,g} (\xi, \eta, \Delta^2) + E_{q,g} (\xi, \eta, \Delta^2)
\Big)
=
2 J_{q,g} \, .
\end{equation}
There are essentially three experimentally feasible processes that have
a clear theoretical understanding and which can be used for direct
measurements of GPDs: electroproduction of
the photon $e N \to e' N' \gamma$ which is sensitive to the deeply
virtual Compton scattering (DVCS) amplitude
\cite{Ji96,GouDiePirRal97,BelMulNieSch00,BelMulKir01}, photoproduction
of a lepton pair $\gamma N \to \ell \bar\ell N'$ \cite{BerDiePir01}, and
electroproduction of a lepton pair $e N \to e' N' \ell \bar\ell$
\cite{BelMul02,GuiVan02}. However, among these only the latter provides
a setup necessary for an independent measurement of a GPD
as a function of both scaling variables $\xi$ and $\eta$.
The former two reactions cannot entirely serve the purpose of testing
the sum rule due to the reality of the final- or initial-state photons,
respectively, which leads to the restriction $\xi = \mp \eta$. The last
process is the most challenging from the experimental point of view due
to small cross sections involved and requires high energy of the beam
and high luminosity, on the one hand, and full exclusivity of the final
state, on the other.

The process $e N \to e' N' \ell \bar\ell$ will be the subject of our
present study, --- a generalization to the full complexity of a short
preliminary note \cite{BelMul02}. The outline of the paper is as
follows. After presenting the general structure of the cross section,
kinematical variables and addressing the issue of reference frames in
the next section, we turn to the discussion of the factorization of the
process within perturbative QCD in terms of GPDs in section
\ref{FactorizationComptonAmplitude}. In section \ref{XsectionSection},
we perform the computation of the cross section. We derive first a
generating function for the squared of the virtual Compton scattering
(VCS) amplitude and its interference with Bethe-Heitler (BH) amplitudes.
It has a completely worked out leptonic part while the hadronic piece is
expressed in terms of Dirac bilinears involving Compton and
electromagnetic form factors. Next, we evaluate the products of
bilinears for an unpolarized nucleon target, leaving the longitudinally
and transversely polarized options for appendices. In section
\ref{Asymmetry}, we give a detailed discussion of diverse asymmetries to
be used for extraction of GPDs. Finally, we conclude.

\section{Kinematics}

\begin{figure}[t]
\begin{center}
\mbox{
\begin{picture}(0,87)(175,0)
\put(0,0){\insertfig{12}{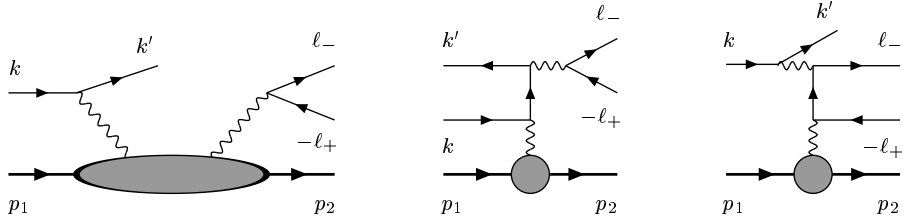}}
\end{picture}
}
\end{center}
\caption{\label{BHandDDVCS} Subprocesses contributing to electroproduction
of muon pairs.}
\end{figure}

The reaction we are dealing with consists of three interfering
processes, depicted in Fig.~\ref{BHandDDVCS} with implied crossed
contributions. However, only one of them is sensitive to the
one-particle correlations in the nucleon when at least one of the photon
virtualities is large compared to a typical hadronic scale. It arises
from the virtual Compton scattering amplitude, shown on the left hand
side in Fig.~\ref{BHandDDVCS}. The other two amplitudes represent the
Bethe-Heitler background. Presently, we discuss the production of a
lepton pair of a different flavor compared to the one of the beam, i.e.,
the muon pair. The consideration of the electroproduction of electron
pairs requires the addition of exchange contributions due to identity of
the electrons in the final state, i.e., $k' \to \ell_-$.

\subsection{Phase space}

\begin{figure}[t]
\begin{center}
\mbox{
\begin{picture}(0,310)(160,0)
\put(-30,-15){\insertfig{12}{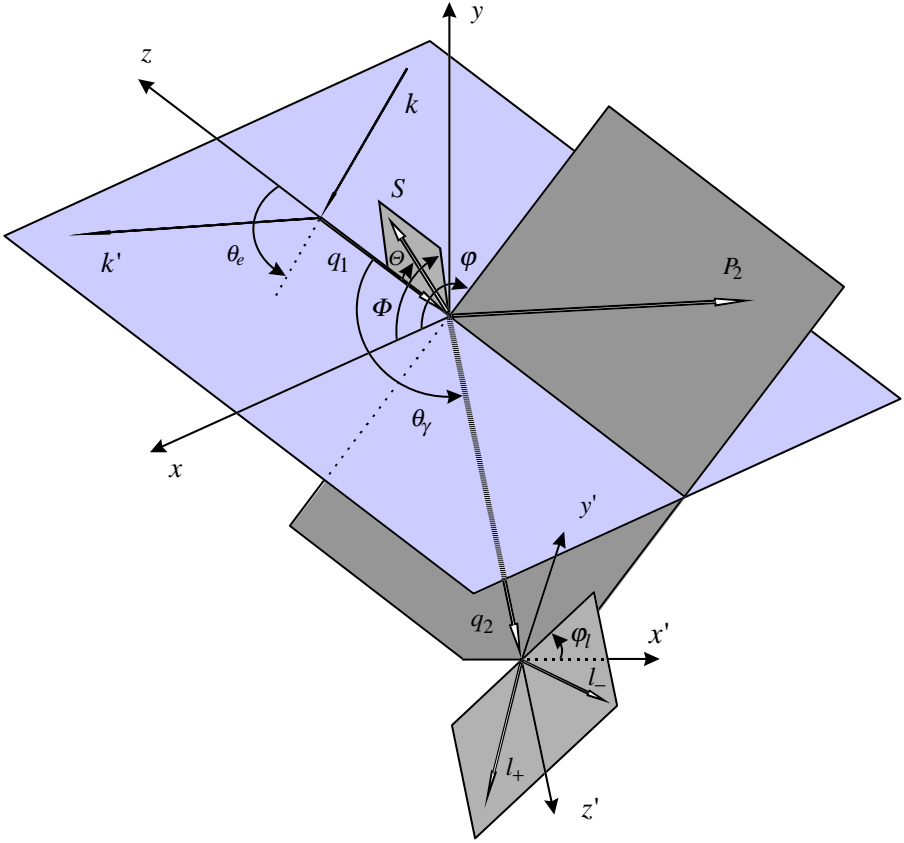}}
\end{picture}
}
\end{center}
\caption{\label{LeptonPairKinetic}
The kinematics of the lepton pair production in elastic electron-nucleon
scattering. The coordinate system with the $z$-axis being counter-aligned
to the spacelike virtual photon is termed TRF-I, while the one with $z'$
along the three-momentum of the timelike photon is named TRF-II.}
\end{figure}

The generic form of the cross section of the exclusive electroproduction of
lepton pairs off the nucleon, $e (k) N (p_1) \to e (k') N (p_2) \ell (\ell_-)
\bar\ell (\ell_+)$, is
\begin{equation}
d \sigma = \frac{1}{4 p_1 \cdot k} |{\cal T}|^2 \, d {\rm LIPS}_4
\, ,
\end{equation}
where ${\cal T}$ is a sum of the amplitude of the virtual Compton scattering
 and two Bethe-Heitler processes, ${\cal T} = {\cal T}_{\rm VCS} +
{\cal T}_{{\rm BH}_1} + {\cal T}_{{\rm BH}_2}$, displayed in Fig.\
\ref{BHandDDVCS}. The four-particle Lorentz invariant phase space
\begin{equation}
\label{LIPS4}
d {\rm LIPS}_4
=
d M_{\ell \bar\ell}^2 \, d {\rm LIPS}_3 \, d {\mit \Phi}_{\ell \bar\ell}
\, ,
\end{equation}
is factorized, by introducing the integration over the invariant mass of the
lepton pair $M_{\ell \bar\ell}^2$, into Lorentz invariant phase-space factors
for the production of a heavy timelike photon off a nucleon
\begin{equation}
d {\rm LIPS}_3
=
(2 \pi)^4 \delta^{(4)} \left( k + p_1 - k' - p_2 - q_2 \right)
\frac{d^4 p_2}{(2 \pi)^3} \delta_+ \left( p_2^2 - M_N^2 \right)
\frac{d^4 k'}{(2 \pi)^3} \delta_+ \left( {k'}^2 \right)
\frac{d^4 q_2}{(2 \pi)^3} \delta_+ \left( q_2^2 - M_{\ell \bar\ell}^2 \right)
\, ,
\end{equation}
and its subsequent decay into a lepton pair
\begin{equation}
d {\mit \Phi}_{\ell \bar\ell}
=
\frac{d^4 \ell_-}{(2 \pi)^3}
\delta_+ \left( \ell_-^2 - m_\ell^2 \right)
\delta_+ \left( (q_2 - \ell_-)^2 - m_\ell^2 \right)
\, .
\end{equation}
A simple calculation gives for them
\begin{equation}
\label{LIPS3-Phi}
d {\rm LIPS}_3
=
\frac{d \Bx d y d (- \Delta^2) d \phi}{16 (2 \pi)^4 \sqrt{1 + \varepsilon^2}}
\, , \qquad
d {\mit \Phi}_{\ell \bar\ell}
= \frac{\beta \, d {\mit \Omega}_\ell}{8 (2 \pi)^3}
\, ,
\end{equation}
respectively, where the solid angle of the final state lepton in the
$\ell \bar\ell$ center-of-mass frame is $d {\mit\Omega}_\ell = \sin
\theta_\ell d \theta_\ell d \varphi_\ell$. Here $\Delta^2 = (p_2 -
p_1)^2$ is the momentum transfer in the $t$-channel and $y = p \cdot
q_1/p \cdot k$ is the incoming lepton energy loss. The conventions for
angles are obvious from Fig.~\ref{LeptonPairKinetic}. Here we introduced
the Bjorken variable $\Bx$ and we will denote the virtualities of the
space- and timelike photons as
\begin{equation}
\Bx \equiv \frac{{\cal Q}^2}{ 2 p_1 \cdot q_1}
\, , \qquad
q_1^2
\equiv
- {\cal Q}^2
\, , \qquad
q_2^2
\equiv
M_{\ell \bar\ell}^2
\, ,
\end{equation}
respectively. The nucleon mass effects are encoded into the variable
$\varepsilon \equiv 2 \Bx M_N / {\cal Q}$ and the final state lepton
velocity in the $\ell \bar\ell$ center-of-mass frame reads
\begin{equation}
\beta = \sqrt{1 - 4 m_\ell^2/M_{\ell \bar\ell}^2}
\, .
\end{equation}

Extracting the lepton charge from the amplitudes, one gets for the cross
section, expressed in terms of experimentally measurable variables,
\begin{equation}
d \sigma
= \frac{\alpha_{\rm em}^4}{16 (2 \pi)^3}
\frac{\Bx y \beta}{{\cal Q}^2 \sqrt{1 + \varepsilon^2}}
\left| \frac{{\cal T}}{e^4} \right|^2
d \Bx d y d (- \Delta^2) d \phi d M_{\ell \bar\ell}^2 d {\mit \Omega}_\ell
\, .
\end{equation}

\subsection{Reference frames}
\label{ReferenceFrames}

Let us discuss appropriate reference frames to be used in analytical
computations. To have a finite Fourier series of the cross section in
terms of the azimuthal angles, we have just introduced above the target
rest frame in which the $z$-axis is directed in the counter-direction
of motion of the spacelike virtual photon, called the target rest
frame I (TRF-I), see Fig.\ \ref{LeptonPairKinetic}. In this frame we
obviously have
\begin{equation}
p_1 = (M_N, \, 0, \, 0, \, 0)
\, , \qquad
q_1 = (\omega_1, \, 0, \, 0, \, - q_1^z)
\, ,
\end{equation}
where
\begin{equation}
\omega_1 = \frac{{\cal Q}}{\varepsilon}
\, , \qquad
q_1^z = \frac{{\cal Q}}{\varepsilon} \sqrt{1 + \varepsilon^2}
\, .
\end{equation}
For the remaining four-vectors we find:

$\bullet$ The outgoing nucleon momentum $p_2 = (E_2, \vec{p}_2)$ has the components
\begin{equation}
E_2 = M_N - \frac{\Delta^2}{2 M_N}
\, , \qquad
|\vec{p}_2| = \sqrt{- \Delta^2 \left( 1 - \Delta^2/(4 M_N^2) \right)}
\, ,
\end{equation}
and the scattering angle of the recoiled nucleon is
\begin{equation}
\cos \theta_N =
-
\frac{
\varepsilon^2 \left( {\cal Q}^2 + M_{\ell \bar\ell}^2 - \Delta^2 \right)
- 2 \Bx \Delta^2
}{
4 \Bx M_N |{\vec p}_2| \sqrt{1 + \varepsilon^2}
}
\, .
\end{equation}

$\bullet$ The incoming electron four-momentum is
\begin{equation}
k = (E, \, k^x, \, 0, \, k^z) = E (1, \, \sin \theta_e, \, 0, \, \cos \theta_e) \, ,
\end{equation}
with
\begin{equation}
E = \frac{{\cal Q}}{y \varepsilon}
\, , \qquad
\cos \theta_e = - \frac{1 + y \varepsilon^2/2}{\sqrt{1 + \varepsilon^2}}
\, ,
\end{equation}
and as a consequence $\sin \theta_e = \varepsilon \sqrt{1 - y - y^2 \varepsilon^2/4}
/ \sqrt{1 + \varepsilon^2}$.

$\bullet$ The four-vector of the timelike virtual photon reads
\begin{equation}
q_2 = (\omega_2, \, \vec{v} \, \omega_2)
\, ,
\end{equation}
with
\begin{equation}
\label{Velocity}
\omega_2 = \frac{{\cal Q}}{\varepsilon} + \frac{\Delta^2}{2 M_N}
\, , \qquad
v \equiv |\vec{v}| = \sqrt{1 - M_{\ell \bar\ell}^2/\omega_2^2}
\, ,
\end{equation}

To evaluate scalar products which arise in contractions of the leptonic
and hadronic tensors, it will be convenient to transform all
four-vectors to a frame, termed TRF-II, where the $z^\prime$-axis is
directed along $\vec q_2$. The latter is achieved by rotating the TRF-I
$z$-axis along the three-velocity $\vec v$ of the timelike photon by the
scattering angle $\theta_\gamma$ (which lies in the hadron scattering
plane),
\begin{equation}
\cos \theta_\gamma
= - \frac{
\varepsilon
\left(
{\cal Q}^2 - M_{\ell \bar\ell}^2 + \Delta^2
\right)
+
2 {\cal Q} \omega_2
}{
2 {\cal Q} \omega_2 v \sqrt{1 + \varepsilon^2}
}
\, .
\end{equation}
In TRF-II, $q_2 = (\omega_2, \, 0, \, 0, \, \omega_2 v)$, and for the other
momenta we get
\begin{equation}
q_1
=
(\omega_1, \, q_1^z \sin \theta_\gamma, \, 0, \, - q_1^z \cos \theta_\gamma)
\, ,
\end{equation}
and
\begin{equation}
\label{Def-FouVec-k}
k =
E \left(
1
, \,
\sin \theta_e \cos \theta_\gamma \cos \varphi_\gamma
-
\cos \theta_e \sin \theta_\gamma
, \,
- \sin \theta_e \sin \varphi_\gamma
, \,
\sin \theta_e \sin \theta_\gamma \cos \varphi_\gamma
+
\cos \theta_e \cos \theta_\gamma
\right)
\, ,
\end{equation}
with $\varphi_\gamma = \pi + \phi$. In TRF-II the vector $p_1$ is unchanged.
In these formulas, the sinus of $\theta_\gamma$ is given by
\begin{equation}
\sin \theta_\gamma
=
\frac{
\sqrt{4 \Bx (1 - \Bx) + \varepsilon^2}
}{
2 {\cal Q} \omega_2 v \sqrt{1 + \varepsilon^2}
}
\sqrt{
- \left(
\Delta^2 - \Delta^2_{\rm min}
\right)
\left(
\Delta^2 - \Delta^2_{\rm max}
\right)
} \, ,
\end{equation}
in terms of the maximal and minimal momentum transfer in the $t$-channel,
\begin{eqnarray}
\label{Def-Delta-minmax}
\Delta^2_{\rm min, max}
\!\!\!&=&\!\!\!
- \frac{1}{ 4 \Bx (1 - \Bx) + \varepsilon^2}
\Bigg\{
2
\left(
(1 - \Bx) {\cal Q}^2 - \Bx M_{\ell \bar\ell}^2
\right)
+ \varepsilon^2
\left(
{\cal Q}^2 - M_{\ell \bar\ell}^2
\right)
\nonumber\\
&&
\mp 2 \sqrt{1 + \varepsilon^2}
\sqrt{
\left(
(1 - \Bx) {\cal Q}^2 - \Bx M_{\ell \bar\ell}^2
\right)^2
-
\varepsilon^2
{\cal Q}^2 M_{\ell \bar\ell}^2
}
\Bigg\}
\, ,
\end{eqnarray}
with $-$ ($+$) corresponding to $\Delta^2_{\rm min}$ ($\Delta^2_{\rm max}$).

A boost from the timelike photon rest frame to the TRF-II along the direction
of motion of the photon with velocity $\vec v$, see Eq.\ (\ref{Velocity}), yields
\begin{equation}
\label{Def-FouVec-lm}
\ell_-
=
\left(
\frac{1}{2} \omega_2 \left( 1 + v \beta \cos \theta_\ell \right)
, \,
\frac{1}{2} M_{\ell \bar\ell} \, \beta \sin \theta_\ell \cos \varphi_\ell
, \,
\frac{1}{2} M_{\ell \bar\ell} \, \beta \sin \theta_\ell \sin \varphi_\ell
, \,
\frac{1}{2} \omega_2 \left( v + \beta \cos \theta_\ell \right)
\right) \, ,
\end{equation}
where $\theta_\ell$ and $\varphi_\ell$ are the solid angles of $\ell_-$
in the $\ell \bar\ell$ center-of-mass frame alluded to above. The vector
$\ell_+$ is simply deduced by a reflection, $\varphi_\ell \to \varphi_\ell
+ \pi $ and $\theta_\ell \to \pi-\theta_\ell$, from $\ell_-$ (or
equivalently by the substitution $\beta \to - \beta$ ).

Using the explicit form of four-vectors in the TRF-II, we can readily compute
the invariant products. Namely, the most nontrivial scalar products are
\begin{eqnarray}
\label{DefkdotDelta}
k \cdot \Delta
\!\!\!&=&\!\!\!
- \frac{1}{2 y (1 + \varepsilon^2)}
\left\{
\left(
{\cal Q}^2 + M_{\ell \bar\ell}^2
\right)
\left(\!
1 - 2 K \cos \varphi_\gamma + \frac{y \varepsilon^2}{2}
\!\right)
- \Delta^2
\left(\!
1 - \Bx (2 - y) + \frac{y \varepsilon^2}{2}\!
\right)
\right\}
, \\
\ell_- \cdot \Delta
\!\!\!&=&\!\!\!
\label{DeflmdotDelta}
- \frac{\beta}{4 v}
\Bigg\{
\left( {\cal Q}^2 + M_{\ell \bar\ell}^2 \right)
\left(
\frac{v}{\beta}
+
\cos \theta_\ell
+
2 \frac{{\cal Q} M_{\ell \bar\ell}}{{\cal Q}^2 + \Bx \Delta^2}
\frac{K \sin \theta_\ell \cos \varphi_\ell}{\sqrt{1 - y - y^2 \varepsilon^2/4}}
\right)
\nonumber\\
&&\qquad\qquad\quad \ \, + \Delta^2
\left(
\frac{v}{\beta}
+
\frac{
{\cal Q}^2 - 2 \Bx M_{\ell \bar\ell}^2 + \Bx \Delta^2
}{
{\cal Q}^2 + \Bx \Delta^2
}
\cos \theta_\ell
\right)
\Bigg\}
\, .
\end{eqnarray}
where
\begin{equation}
\label{DefK}
K
\equiv
\frac{1}{2 \left( {\cal Q}^2 + M_{\ell \bar\ell}^2 \right)}
\sqrt{
- \left( 1 - y - y^2 \varepsilon^2/4 \right)
\left(
4 \Bx (1 - \Bx) + \varepsilon^2
\right)
\left(
\Delta^2 - \Delta^2_{\rm min}
\right)
\left(
\Delta^2 - \Delta^2_{\rm max}
\right)
}
\, .
\end{equation}
Eq.\ (\ref{DefkdotDelta}) reduces to the known expression from Ref.\
\cite{BelMulKir01} for the real final-state photon $M_{\ell \bar\ell} =
0$. Finally, since the expression for $\ell_- \cdot k$ is too complex to
be presented here, we will give below its expanded form which is used in
all practical calculations. To make the results look symmetric, we
introduce the variable
\begin{equation}
\frac{1}{\widetilde y}
\equiv
\frac{p_1\cdot \ell_-}{p_1\cdot q_2}
=
\frac{1 + v \beta\cos \theta_\ell}{2}
\simeq
\frac{1 + \cos \theta_\ell}{2}
\, ,
\end{equation}
which varies in the interval $1 \leq \widetilde y \leq \infty$.

\subsection{Symmetric variables}

{}For our subsequent use, we introduce the symmetric combinations of
momenta
\begin{equation}
q = \frac{1}{2} (q_1 + q_2)
\, , \qquad
p = p_1 + p_2
\, , \qquad
\Delta = p_2 - p_1 = q_1 - q_2 \, ,
\end{equation}
and the invariants built from them
\begin{equation}
q^2 = - Q^2
\, , \qquad
\xi = \frac{Q^2}{p \cdot q}
\, , \qquad
\eta = \frac{\Delta \cdot q}{p \cdot q} \, ,
\end{equation}
with the latter two being the generalized Bjorken and skewness variables,
respectively. These can be re-expressed in terms of experimental ones,
discussed in the previous section, via the equations:
\begin{equation}
\label{ResolutionScale}
Q^2 = \frac{1}{2}
\left(
{\cal Q}^2 - M_{\ell \bar\ell}^2 + \frac{\Delta^2}{2}
\right) \, ,
\end{equation}
for the inverse resolution scales, and
\begin{equation}
\label{XitoEta}
\xi = - \eta\,
\frac{
{\cal Q}^2 - M_{\ell \bar\ell}^2 + \Delta^2/2
}{
{\cal Q}^2 + M_{\ell \bar\ell}^2
}
\, , \qquad
\eta =-
\frac{
{\cal Q}^2 + M_{\ell \bar\ell}^2
}{
2{\cal Q}^2/ \Bx-  {\cal Q}^2 - M_{\ell \bar\ell}^2 + \Delta^2
}
\, ,
\end{equation}
for the scaling variables. Note that $Q^2$ and $\xi$ can take both positive
(${\cal Q}^2 > M_{\ell \bar\ell}^2$) and negative (${\cal Q}^2 <
M_{\ell \bar\ell}^2$) values depending on the relative magnitude of spacelike
and timelike photon virtualities, while $\eta < 0$ and $\pm \xi - \eta > 0$.
To complete the set of formulas, the inverse transformations read
\begin{equation}
M_{\ell \bar\ell}^2
= - \left( 1 + \frac{\eta}{\xi} \right) Q^2 + \frac{\Delta^2}{4}
\, , \qquad
{\cal Q}^2
=
\left( 1 - \frac{\eta}{\xi} \right) Q^2 - \frac{\Delta^2}{4}
\, ,
\end{equation}
and
\begin{equation}
\Bx =
\frac{(\xi - \eta) Q^2 - \xi \Delta^2/4}{(1 - \eta) Q^2 - \xi \Delta^2/2} \, .
\end{equation}

The bulk of results given in subsequent sections will be presented in
the symmetric variables. The following scalar products are needed in
course of evaluations. Neglecting the nucleon mass corrections $\sim
M_N^2/Q^2$, the above formulas (\ref{Def-Delta-minmax}) simplify
considerably
\begin{eqnarray*}
\Delta^2_{\rm min} \approx - 4 M_N^2 \frac{\eta^2}{1 - \eta^2}
\, , \qquad
\Delta^2_{\rm max} \approx - Q^2 \frac{1 - \eta^2}{\xi (1 - \xi)}
\, ,.
\end{eqnarray*}
and the transverse momentum transfer admits the form $\Delta_\perp^2
\approx (1 - \eta^2) \left( \Delta^2 - \Delta^2_{\rm min} \right)$.
The $1/(p \cdot q)$-expansion of the scalar products (\ref{DefkdotDelta}),
(\ref{DeflmdotDelta}), and $\ell_-\cdot k$, keeping the leading and
sub-leading terms only, results into expressions
\begin{eqnarray}
k \cdot \Delta
\!\!\!&\approx&\!\!\!
\frac{Q^2}{y} \frac{\eta}{\xi}
\left( 1 - 2 K \cos \varphi_\gamma \right)
\, , \\
\ell_- \cdot \Delta
\!\!\!&\approx&\!\!\!
\frac{Q^2}{\widetilde y} \frac{\eta}{\xi}
\left(
1 + 2 \widetilde K \cos \varphi_\ell
\right)
\, , \\
\label{Appldotk}
\ell_- \cdot k
\!\!\!&\approx&\!\!\!
\frac{Q^2}{y \widetilde y} \frac{1}{\xi}
\Bigg\{
\frac{1}{2} (\xi + \eta) ( 1 - \widetilde y )
+
\frac{1}{2} (\xi - \eta) (1 - y)
\nonumber\\
&&\qquad
+
\sigma
\sqrt{(1 - y) (1 - \widetilde y) (\xi^2 - \eta^2)}
\cos (\varphi_\gamma - \varphi_\ell)
+
2 \eta K \cos \varphi_\gamma + 2 \eta \widetilde K \cos \varphi_\ell
\Bigg\} \, .
\end{eqnarray}
with
\begin{equation}
\label{Def-K}
\left\{
\begin{array}{c}
K
\\
\widetilde K
\end{array}
\right\}
\approx - \frac{1}{2 \eta}
\sqrt{- \xi \frac{\Delta^2}{Q^2}}
\sqrt{ 1 - \frac{\Delta^2_{\rm min}}{\Delta^2} }
\sqrt{\frac{1 + \eta}{1 - \eta}}
\times
\left\{
\begin{array}{c}
\sqrt{(1 - y) (\xi - \eta)}
\\
\sqrt{(1 - \widetilde y) (\xi + \eta)}
\end{array}
\right\}
\, .
\end{equation}
We note that in Eq.\ (\ref{Appldotk}) we replaced the original square
root $\sqrt{(1 - y) (1 - \widetilde y) (1 - \eta^2/\xi^2) Q^4}$ in front
of $\cos(\varphi_\gamma - \varphi_\ell)$ by $\sigma Q^2/\xi \sqrt{(1 -
y) (1 - \widetilde y) (\xi^2 - \eta^2)}$, since $Q^2/\xi$ is positive
for the kinematics we are considering, $\sigma$ is $+1$. However, below
in discussion of relations between the amplitudes we will make use of
exchange symmetries that induce the interchange $\xi \to - \xi$ and $Q^2
\to Q^2$ in the underlying formulas. Under these substitutions, $\sigma$
changes the sign and takes the negative value $-1$. Making use of the
results, we have just derived, we are now in a position to proceed with
the computation of amplitudes. Before doing this, let us discuss in the
next section factorization properties and the structure of the virtual
Compton scattering amplitude.

\section{Factorization of Compton amplitude}
\label{FactorizationComptonAmplitude}

The amplitude ${\cal T}_{\rm VCS}$ is the object of interest since it
involves the off-forward Compton scattering amplitude on the nucleon
$T_{\mu\nu}$. This tensor
\begin{equation}
\label{ComptonAmplitude}
T_{\mu\nu}
=
i \int d^4 z \, {\rm e}^{i q \cdot z}
\langle p_2 |
T \left\{ j_\mu (z/2) j_\nu (- z/2) \right\}
| p_1 \rangle
\end{equation}
is expressed as a hadronic matrix element of the time-ordered product of
the two quark electromagnetic currents
\begin{equation}
j_\mu (z) = \sum_q Q_q \bar \psi_q (z) \gamma_\mu \psi_q (z)
\, .
\end{equation}
Here we extracted the absolute value of the electron charge so that $Q_q$ is
the quark's fractional charge.

\subsection{Structure of amplitudes in the generalized Bjorken limit}

Let us demonstrate the factorization of the VCS amplitude using the
example of a cubic scalar model%
\footnote{The absence of a stable vacuum state in this model is
irrelevant for the demonstration of generic perturbative properties
of scattering amplitudes.}. Although theorems of this sort already
exist in the literature \cite{Rad97,ColFre99}, it is rather instructive
to have a fresh inspection. We only need to analyze the singularity
structure stemming from the denominators of particle propagators.
Numerators present in QCD case will not be relevant for our present
needs. Complications due to the gauge invariance in the realistic case
will be treated in a straightforward manner. We will disregard
throughout the contribution of crossed diagrams since they do not
bring any new insights into the factorization property of the hadronic
tensor.

\begin{figure}[t]
\begin{center}
\mbox{
\begin{picture}(0,100)(150,0)
\put(0,0){\insertfig{10}{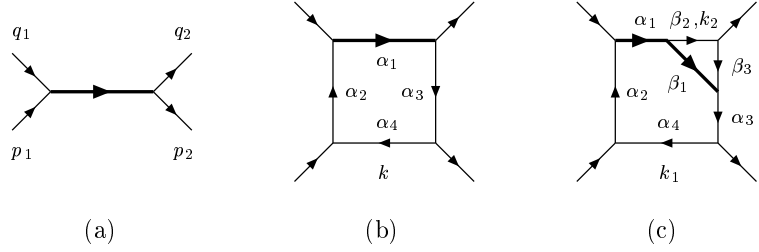}}
\end{picture}
}
\end{center}
\caption{\label{BoxDiagrams} Lowest order perturbative diagrams exhibiting
possible short-distance regimes contributing to the asymptotics of Compton
scattering amplitude. Thick lines correspond to the highly virtual
propagators.}
\end{figure}

The tree-level Compton amplitude, see Fig.\ \ref{BoxDiagrams} (a), reads
\begin{equation}
T_{(0)} = \frac{1}{(p_1 + q_1)^2 + i 0}
=
\frac{1}{(p \cdot q)}
\frac{1}{1 - \xi + \ft12 \epsilon + i 0}
\, ,
\end{equation}
where the small correction to the scaling contribution is $\epsilon \equiv
(2 M_N^2 - \Delta^2/2)/ (p \cdot q)$. From here it is obvious that the expansion
parameter is $1/(p \cdot q) = \xi/Q^2$, not $Q^2$ as it is suggested by the definition
(\ref{ComptonAmplitude}). Actually, this scale approaches zero when the
virtualities of incoming and outgoing photons are of the same magnitude. When
$p \cdot q \to \infty$ and $\xi$ is fixed, one recovers the scaling coefficient
function $1/(x - \xi + i 0)$ convoluted with the ``perturbative" GPD which,
to this order, is $H_{(0)} (x, \eta, \Delta^2) =
\delta (1 - x)$.

The one-loop expression for the amplitude displayed in Fig.\ \ref{BoxDiagrams}
(b) is
\begin{equation}
T_{(1)}
=
i g^2 \int \frac{d^4 k}{(2 \pi)^4}
\frac{1}{k^2 (k + p_1)^2 (k + p_2)^2 (k + p_1 + q_1)^2}
= \frac{g^2}{(4 \pi)^2}
\int_0^\infty \prod_{j = 1}^4 d \alpha_j \,
\frac{{\rm e}^{i \left( E_1 + i 0_\alpha \right)}}{\alpha^2}
\, ,
\end{equation}
($0_\alpha \equiv 0 \cdot \alpha$) where
\begin{eqnarray}
E_1
\!\!\!&=&\!\!\!
\alpha_1
\left\{ (1 - \xi + \ft12 \epsilon) \left(1 - \frac{\alpha_1}{\alpha}\right)
-
(1 - \eta + \epsilon) \frac{\alpha_2}{\alpha}
-
(1 + \eta + \epsilon) \frac{\alpha_3}{\alpha}
\right\} (p \cdot q)
+
\frac{\alpha_2 \alpha_3}{\alpha} \Delta^2
\nonumber\\
&&+
(\alpha_2 + \alpha_3) \left( 1 - \frac{\alpha_2 + \alpha_3}{\alpha} \right) M_N^2
\, ,
\end{eqnarray}
and $\alpha \equiv \sum_{j = 1}^4 \alpha_j$. Integrating in the vicinity of
$\alpha_1 \to 0$, which corresponds to the large virtuality of the
corresponding line in the Feynman diagram, --- the propagator between the
photon vertices, --- one gets, to leading order in $1/(p\cdot q)$, the
contribution
\begin{eqnarray}
\label{FactorizedForm}
T^{\mbox{\tiny SD}_1}_{(1)}
=
\frac{1}{(p \cdot q)}
\int_{-1}^1 d x \frac{H_{(1)} (x, \eta, \Delta^2)}{x - \xi + i 0}
\, ,
\end{eqnarray}
where we introduced the one-loop GPD, which absorbs mass singularities,
\begin{equation}
H_{(1)} (x, \eta, \Delta^2)
=
\frac{i g^2}{(4 \pi)^2}
\int_0^\infty \prod_{j = 2}^4 d \alpha_j \,
\delta
\left(
x - 1
+
(1 - \eta) \frac{\alpha_2}{\tilde\alpha}
+
(1 + \eta) \frac{\alpha_3}{\tilde\alpha}
\right)
\frac{{\rm e}^{i \left(\widetilde E_1 + i 0_{\tilde\alpha} \right)}}{\tilde\alpha^2}
\,
\end{equation}
Here $\tilde \alpha = \alpha_2 + \alpha_3 + \alpha_4$ and $\widetilde E_1 =
E_1 [\alpha_1 = 0]$ does not depend on large scales. Summing the tree and
one-loop contributions, one gets the factorized expression of the form
(\ref{FactorizedForm}) with $H_{(1)}$ being replaced by $H = H_{(0)} +
H_{(1)} + \dots$. One can easily convince oneself that the aforementioned
perturbative expansion of the GPD $H (x, \eta, \Delta^2)$ arises from the
light-cone operators matrix element:
\begin{equation}
\label{ScalarGPD}
H (x, \eta, \Delta^2)
= p_+ \int \frac{d \xi_-}{2 \pi}
{\rm e}^{i x \xi_- p_+}
\langle p_2 |
\phi \left( - \xi_-, 0_+, \bit{0} \right)
\phi \left( \xi_-, 0_+, \bit{0} \right)
| p_1 \rangle
\, .
\end{equation}
We refer to  appendix \ref{LCvectors} for the definition of the
light-cone vectors.

Other short-distance and infrared regimes lead to power-suppressed
contributions compared to the leading one (\ref{FactorizedForm}), see
Ref.\ \cite{Rad97}. More particle attachments from the hadronic line to
the hard propagator obviously lead to a stronger dumping of amplitudes,
except for the effect of longitudinally polarized gauge bosons in gauge
field theories. The latter restore the gauge invariance of the naive
GPD (\ref{ScalarGPD}), with the $\phi$-scalars begin replaced by
the quark fields $\psi$. The analysis of the most general hand-bag
diagram, i.e., with a single propagator between the photon vertices,
does not bring new complications compared to the already discussed
one-loop example and can be treated in an analogous manner. In QCD case,
due to the fact that quarks are fermions, the numerators of their
propagators cancel the power of $p \cdot q$ stemming from the
denominator, so that $T^{\mbox{\tiny SD}_1} \sim {\cal O} (p \cdot
q^0)$.

\begin{figure}[t]
\begin{center}
\mbox{
\begin{picture}(0,140)(100,0)
\put(0,0){\insertfig{7}{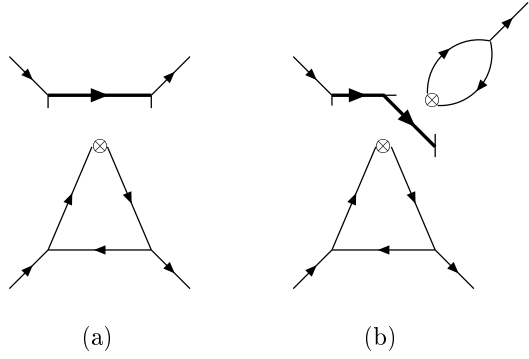}}
\end{picture}
}
\end{center}
\caption{\label{ShortDistanceRegimes} Short-distance regimes corresponding
to loop diagrams in Fig.\ \protect\ref{BoxDiagrams}.}
\end{figure}

Due to the timelike nature of the final-state photon virtuality, the
quark-antiquark pair can form an on-shell intermediate hadronic state
before annihilating into the heavy photon. This happens when the hard
momentum is re-routed around the photon vertex as demonstrated in the
diagram \ref{BoxDiagrams} (c). This configuration can potentially generate
leading asymptotic behavior when the photon virtuality is low. Let us
demonstrate that it is actually suppressed. The two-loop diagram
\ref{BoxDiagrams} (c) has the form
\begin{eqnarray}
T_{(3)}
\!\!\!&=&\!\!\!
- g^4
\int \frac{d^4 k_1}{(2 \pi)^4}
\frac{1}{k_1^2 (k_1 + p_1)^2 (k_1 + p_2)^2 (k_1 + p_1 + p_2)^2}
\\
&\times&\!\!\!
\int \frac{d^4 k_2}{(2 \pi)^4}
\frac{1}{k_2^2 (k_2 - q_2)^2 (k_2 - k_1 - q_1 - p_1)^2}
=
i \frac{g^4}{(4 \pi)^4}
\int_0^\infty \prod_{i = 1}^4 d \alpha_i \prod_{j = 1}^3 d \beta_j
\frac{
{\rm e}^{i (E_2 + i 0_{\alpha + \beta})}
}{
[\alpha \beta + \beta_1 (\beta - \beta_1)]^2
}
\nonumber
\end{eqnarray}
with $\alpha \equiv \sum_{i = 1}^4 \alpha_i$, $\beta \equiv \sum_{i = 1}^3
\beta_i$. The exponential reads
\begin{eqnarray}
E_2 \!\!\!&=&\!\!\!
\alpha
\frac{
\alpha_1 + \beta_1 \left( 1 - \frac{\beta_1 + \beta_3}{\beta} \right)
}{
\alpha + \beta_1 \left( 1 - \frac{\beta_1}{\beta} \right)
}
\nonumber\\
&&\times
\Bigg\{
(1 - \xi + \ft12 \epsilon)
\left(
1 - \frac{\alpha_1}{\alpha} + \frac{\beta_1 \beta_3}{\alpha \beta}
\right)
-
(1 - \eta + \epsilon) \frac{\alpha_2}{\alpha}
-
(1 + \eta + \epsilon)
\left(
\frac{\alpha_3}{\alpha} + \frac{\beta_1 \beta_3}{\alpha \beta}
\right)
\Bigg\} (p \cdot q)
\nonumber\\
&&+
\frac{
\alpha_2
\left(
\alpha_3 + \frac{\beta_1 \beta_3}{\beta}
\right)
}{
\alpha + \beta_1 \left(1 - \frac{\beta_1}{\beta} \right)
} \Delta^2
+
\left(
\alpha_2 + \alpha_3 + \frac{\beta_1 \beta_3}{\beta}
\right)
\left(
1
-
\frac{
\alpha_2 + \alpha_3 + \frac{\beta_1 \beta_3}{\beta}
}{
\alpha + \beta_1 \left(1 - \frac{\beta_1}{\beta} \right)
}
\right)
M_N^2
\nonumber\\
&&+
\beta_3 \left( 1 - \frac{\beta_1 + \beta_3}{\beta} \right) M_{\ell\bar\ell}^2
\, .
\end{eqnarray}
In the short-distance regime, i.e., $\alpha_1 \to 0$, $\beta_1 \to 0$, where
we define $\tilde \alpha = \alpha_2 + \alpha_3 + \alpha_4$ and $\tilde \beta
= \beta_2 + \beta_3$, we get, see Fig.\ \ref{ShortDistanceRegimes} (b),
\begin{equation}
\label{SD2}
T^{\mbox{\tiny SD}_2}_{(3)}
=
- \frac{g^2}{(p \cdot q)^2}
\int_{-1}^1 d x \int_0^1 d u
\frac{
H_{(1)} (x, \eta, \Delta^2) {\mit\Pi}_{(1)} (u, M_{\ell\bar\ell}^2)
}{
(1 - u)(x - \xi + i 0)^2
}
\, ,
\end{equation}
where $H_{(1)}$ was given above, since $\widetilde E_1 = E_2 [\alpha_1 =
\beta_i = 0]$, while
\begin{equation}
{\mit\Pi}_{(1)} (u, M_{\ell\bar\ell}^2)
=
\frac{1}{(4 \pi)^2}
\int_0^\infty
\prod_{j = 2}^3 d \beta_j \, \delta \left( u - \frac{\beta_3}{\tilde\beta} \right)
\exp
\left\{
i \beta_3 \left( 1 - \frac{\beta_3}{\tilde\beta} \right) M_{\ell\bar\ell}^2
\right\}
\end{equation}
is the first term in the perturbative expansion of the correlation function
\begin{equation}
{\mit\Pi} (u, M_{\ell\bar\ell}^2)
=
i q_{2 -} \int d^4 z \, {\rm e}^{i q_2 \cdot z}
\int \frac{d \xi_+}{2 \pi} {\rm e}^{- i u \xi_+ q_{2 -}}
\langle 0 |
T \left\{
\phi (0_-, 0_+, \bit{0}) \phi (0_-, \xi_+, \bit{0}) ,
\, j (z)
\right\}
| 0 \rangle
\, ,
\end{equation}
where the ``electromagnetic" current is $j (z) = \frac{1}{2} \phi^2 (z)$.
Therefore, we observe that this short-distance regime is suppressed
compared to the leading one, Eq.\ (\ref{FactorizedForm}).

In the QCD case, the suppression of contributions due to the hadronic
component of the photon is much milder than in the scalar example and it
is only $(p \cdot q)^{-1/2}$ compared to the handbag diagram. The structure
of the reduced amplitude is the same as in Eq.\ (\ref{SD2}), however with
only one power of the hard-scattering coefficient $1/(x - \xi + i 0)$
being involved and the vacuum correlator is of the form
\begin{eqnarray}
\label{Def-VacCor}
{\mit\Pi}_\mu (u, q_2)
\!\!\!&\equiv&\!\!\!
\left(
q_{2\mu} q_{2\nu}
-
M_{\ell\bar\ell}^2 \ g_{\mu\nu}
\right) n^\ast_\nu
{\mit\Pi} (u, M_{\ell\bar\ell}^2)
\nonumber\\
&=&\!\!\!
i \int d^4 z {\rm e}^{i z \cdot q_2}
\int \frac{d \xi_+}{2 \pi} {\rm e}^{- i u \xi_+ q_{2 -}}
\langle 0 |
T
\left\{
\bar\psi (0_-, 0_+, \bit{0}) \gamma_- \psi (0_-, \xi_+, \bit{0}) ,
j_\mu (z)
\right\}
|0 \rangle
\, .
\end{eqnarray}
The correlation function can be saturated by the $\rho$-meson and reads
\begin{equation}
\label{Mod-VacCor}
{\mit\Pi} (u, M_{\ell\bar\ell}^2)
=
- \frac{m_\rho^2}{g_\rho^2}
\frac{\varphi_\rho (u)}
{
M_{\ell\bar\ell}^2 - m_\rho^2 + i m_\rho {\mit\Gamma}_\rho
}
+
\frac{3}{4 \pi^2} u (1 - u)
\int_{s_0}^\infty \frac{d s}{s - M_{\ell\bar\ell}^2 - i 0}
\, ,
\end{equation}
where $\varphi_\rho (u)$ is the $\rho$-meson distribution amplitude normalized
according to $\int_0^1 d u \varphi_\rho (u) = 1$, while $g_\rho^2/(4 \pi) =
2.36 \pm 0.18$, $m_\rho = 770 \, {\rm MeV}$ and ${\mit\Gamma}_\rho = 150 \,
{\rm MeV}$ are the $\rho$-meson decay constant, mass and width, respectively.
The second term on the right-hand side comes from the perturbative contribution
to the correlator, known to two-loop order \cite{ShiVanZak79}, a part of which
is dual to the $\rho$-meson in the interval $s \in [0, s_0]$ and is absorbed
there. The parameter $s_0 \approx 0.8 \, {\rm GeV}^2$ is the continuum threshold%
\footnote{The approximation of the continuum contribution by a step-function
threshold in the spectral density causes a divergence in the real part at
$M_{\ell\bar\ell}^2 = s_0$, which is  spurious.}. Due to divergence in the correlation
function, one has to use a renormalized expression, stemming from the subtracted
dispersion relation, ${\mit\Pi}_R (u, M_{\ell\bar\ell}^2) = {\mit\Pi}
(u, M_{\ell\bar\ell}^2) - {\mit\Pi} (u, 0)$. The gauge invariance of the
light-ray operator involved in the correlation function is restored by means
of Wilson lines as discussed below.

{}From this result it is apparent that there is an extra imaginary part,
besides the conventional $s$-channel discontinuity, in the VCS
amplitude associated with the production of an on-shell intermediate hadronic
state by the quark-antiquark pair. Due to the current conservation, which
implies $\bar u(\ell_-) {\not\!q_2} u (- \ell_+) = 0$, only one Lorentz
structure contributes to the leptoproduction amplitude so that
\begin{equation}
\label{Con-Lep-VacCor}
{\mit\Pi}_\mu (u, q_2)
\frac{g_{\mu\nu}}{M_{\ell\bar\ell}^2}
\bar u(\ell_-) \gamma_\nu u (- \ell_+)
=
- {\mit\Pi} (u, M_{\ell\bar\ell}^2)
\bar u(\ell_-) \gamma_- u (- \ell_+)
\, .
\end{equation}

We should note, that due to the inequality, in general, of
the skewness and the generalized Bjorken variable $\eta \neq \mp \xi$, possible
complications due to the singular structure of the hard coefficient function
do not arise since the pole in the former is away from the ``turning" point
in GPDs when one of the struck partons has zero momentum fraction. Moreover,
it was demonstrated that GPDs are continuous at this point, i.e., have no
jumps, so that the region does not present a problem in the case $\eta = \mp \xi$
either \cite{Rad97,Rad99,ColFre99}.

\begin{figure}[t]
\begin{center}
\mbox{
\begin{picture}(0,95)(170,0)
\put(0,0){\insertfig{12}{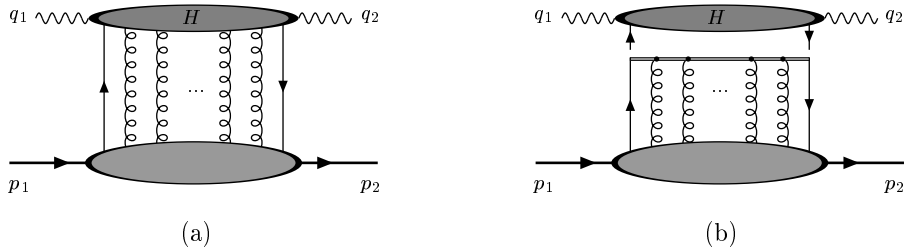}}
\end{picture}
}
\end{center}
\caption{\label{LeadingRegion} The leading asymptotic region in the Compton
scattering amplitude (a) and the factorization of longitudinal gluons into
the path-ordered exponential (b).}
\end{figure}
We finally conclude that the leading region in the amplitude is depicted in
Fig.\ \ref{LeadingRegion} (a), where in addition to considerations of the
scalar theory, one is allowed to attach an infinite number of zero-twist
longitudinally polarized gluons to the hard part. They do not induce power
suppressed contributions and factorize into the Wilson line along the
trajectory of motion of the struck quark in the hard subprocess, see Refs.\
\cite{EfrRad80,LabSte85,ColSopSte88} and \cite{BelJiYua02} for the most
recent and complete treatment of this issue. As a result the quark field
in the naive definition of the GPD gets replaced by
\begin{equation}
\psi (y_-, 0_+, \bit{0})
\to
P \exp \left( i g \int_{y_-}^\infty d z_- A_+ (z_-, 0_-, \bit{0}) \right)
\psi (y_-, 0, \bit{0})
\, .
\end{equation}
Due to the unitary cancellation of the eikonal lines beyond the photon
absorption and emission points, the path-ordered exponential extends only
between the quark fields, see Fig.\ \ref{LeadingRegion} (b).

\subsection{Lorentz decomposition of Compton amplitude}

The gauge invariant decomposition of the hadronic tensor
(\ref{ComptonAmplitude}) was found in Ref.\ \cite{BelMul00} by an
explicit twist-three analysis at leading order of perturbation theory
[see Refs.\ \cite{AniPirTer00,PenPolShuStr00,RadWei00} for independent
developments along this line]. Presently, we concentrate on the
twist-two sector at leading order in QCD coupling constant. As a
consequence, we will not discuss gluonic contributions, in particular
helicity-flip gluon effects which introduce new Lorentz structures into
the hadronic tensor. The Compton scattering amplitude $T_{\mu\nu}$ admits
the following Lorentz invariant decomposition in terms of Compton form
factors (CFFs)
\begin{eqnarray}
\label{VCSamplitude}
T_{\mu\nu}
\!\!\!&=&\!\!\!
- \frac{1}{2} \left(
g_{\mu\nu} - \frac{q_{1 \mu} \, q_{2 \nu}}{q_1 \cdot q_2}
\right)
{\cal V}_1 (\xi, \eta, \Delta^2)
+
\frac{1}{2 p \cdot q}
\left(
p_\mu - \frac{p \cdot q_2}{q_1 \cdot q_2} \, q_{1 \mu}
\right)
\left(
p_\nu - \frac{p \cdot q_1}{q_1 \cdot q_2} \, q_{2 \nu}
\right)
{\cal V}_2 (\xi, \eta, \Delta^2)
\nonumber\\
&&\!\!\!+
\frac{i}{2 p \cdot q} \varepsilon_{\theta \lambda \rho \sigma} p_\rho \, q_\sigma
\left(
g_{\mu \theta} - \frac{p_\mu \, q_{2 \theta}}{p \cdot q_2}
\right)
\left(
g_{\nu \lambda} - \frac{p_\nu \, q_{1 \lambda}}{p \cdot q_1}
\right)
{\cal A} (\xi, \eta, \Delta^2)
\, .
\end{eqnarray}
Here $\varepsilon^{0123} = 1$. To simplify notations we will set in what
follows ${\cal V}_1 \equiv {\cal V}$ and use the longitudinal-longitudinal
helicity%
\footnote{Referring to incoming-outgoing virtual photons.}
amplitude
\begin{eqnarray}
{\cal V}_{L}
\equiv
\frac{1}{\xi} {\cal V}_2 - {\cal V}_1
\end{eqnarray}
analogous to the longitudinal structure function $F_L$ of deeply inelastic
scattering.

The Compton form factors are decomposed into Dirac structures via
\begin{equation}
\label{Def-VA-Amp}
{\cal V} = h_+ {\cal H} + e_+ {\cal E}
\, , \qquad
{\cal V}_L = h_+ {\cal H}_L + e_+ {\cal E}_L
\, , \qquad
{\cal A} = \widetilde h_+ \widetilde {\cal H}
+
\widetilde e_+ \widetilde {\cal E}
\, .
\end{equation}
with Dirac bilinears which read
\begin{eqnarray}
&& h_\mu
= \bar u (p_2) \gamma_\mu u (p_1)
\, , \qquad\qquad
e_\mu
= \bar u (p_2) i \sigma_{\mu\nu} \frac{\Delta_\nu}{2 M_N} u (p_1)
\, , \nonumber\\
&&\widetilde h_\mu
= \bar u (p_2) \gamma_\mu \gamma_5  u (p_1)
\, , \qquad\quad\,
\widetilde e_\mu
= \frac{\Delta_\mu}{2 M_N} \bar u (p_2) \gamma_5 u (p_1)
\, ,
\label{DiracBilinears}
\end{eqnarray}
where $u$ is the nucleon bispinor normalized as $\bar u (p) u (p) = 2 M_N$
and $\gamma_5 = \left( {0 \, 1 \atop 1 \, 0}\right)$.

As we demonstrated in the previous section, the inverse $s$-channel energy
$p \cdot q$ sets the distance between the quark fields in the Compton
scattering amplitude. Therefore, for the QCD factorization to be applicable
one has to impose the condition
\begin{equation}
\label{ApplicabilityPQCD}
(p \cdot q)
\gg
{\rm max} \{ M_N^2, |\Delta^2| \}
\, .
\end{equation}
In this domain, the Compton form factors are factorized, as we have
shown above, into calculable coefficient functions and GPDs via, cf.\
(\ref{FactorizedForm}),
\begin{eqnarray}
\label{Def-ComForFac}
\Big( {\cal H}, {\cal E} \Big) (\xi, \eta, \Delta^2)
\!\!\!&=&\!\!\!
\sum_{i=u,d,s,G}
\int_{-1}^1 d x \, C_i^{(-)} (x, \xi) \,
\Big( H_i, E_i \Big) (x, \eta, \Delta^2)
\, , \nonumber\\
\Big( \widetilde {\cal H}, \widetilde {\cal E} \Big) (\xi, \eta, \Delta^2)
\!\!\!&=&\!\!\!
\sum_{i=u,d,s,G}
\int_{-1}^1 d x \, C_i^{(+)} (x, \xi) \,
\Big( \widetilde H_i, \widetilde E_i \Big) (x, \eta, \Delta^2)
\, ,
\end{eqnarray}
where the sum runs over all parton species. Analogous factorized formula
holds for ${\cal H}_L$ and ${\cal E}_L$ with the coefficient function $C^{(-)}
(x, \xi)$ being replaced by $C_L (x, \xi)$. To zeroth-order in the QCD
coupling constant only the quark coefficient functions $i = q$
\begin{equation}
\label{HarScaAmp}
C_q^{(\mp)} (x, \xi)
=
\frac{Q_q^2}{\xi -x  - i 0}
\mp
\frac{Q_q^2}{\xi + x  - i 0}
+ {\cal O} (\alpha_s) \, ,
\end{equation}
give a nonvanishing contribution. The Compton form factor ${\cal V}_L$
does not enter in leading order. This is a consequence of a (generalized)
Callan-Gross relation, ${\cal V}_2 = \xi {\cal V}_1 + {\cal O} (\alpha_s)$.
Thus, $C_L (x, \xi)$ starts at next-to-leading order, i.e., ${\cal O}
(\alpha_s)$, as we emphasized above. Perturbative corrections to $C^{(\pm)}_q$
and gluon contributions are available at one-loop order
\cite{JiOsb98,BelMul97,ManPilSteVanWei97,JiHoo98,BelMul00d}.

\section{Cross section}
\label{XsectionSection}

This section is devoted to the calculation of the electroproduction
cross section with the leading-twist Compton scattering amplitude
analyzed above. In the next section we compute a generic form of the
squared amplitude, where the hadronic part is left untouched and the
leptonic sector is fully worked out. Then this intermediate result is
used in section \ref{SubSec-AngDep} for the evaluation of the cross
section for an unpolarized nucleon and is further used for deriving
the results in appendix \ref{App-PolTar}, when the polarization of
the target is available as an option.

\subsection{Generating function}
\label{Sec-GenFun}

The square of the total amplitude, ${\cal T} = {\cal T}_{\rm VCS} +
{\cal T}_{{\rm BH}_1} + {\cal T}_{{\rm BH}_2}$, involves three
essentially different contributions
\begin{eqnarray}
{\cal T}^2 \!\!\!&=&\!\!\! | {\cal T}_{\rm VCS} |^2
+
{\cal I}
+
| {\cal T}_{{\rm BH}_1} + {\cal T}_{{\rm BH}_2} |^2
\, ,
\end{eqnarray}
the square of the virtual Compton scattering amplitude, --- bilinear in
Compton form factors, --- the square of the Bethe-Heitler processes, ---
independent on GPDs and expressed solely in terms of elastic form factors,
--- and, at last, the interference term
\begin{equation}
{\cal I}
=
{\cal T}_{\rm VCS} {\cal T}_{{\rm BH}_1}^\dagger
+
{\cal T}_{\rm VCS} {\cal T}_{{\rm BH}_2}^\dagger
+
{\cal T}_{\rm VCS}^\dagger {\cal T}_{{\rm BH}_1}
+
{\cal T}_{\rm VCS}^\dagger {\cal T}_{{\rm BH}_2}
\, ,
\end{equation}
which is linear in Compton form factors.

{}For the electron beam, the separate contributions to the total amplitude
read
\begin{eqnarray}
\label{Def-AmpVCS}
{\cal T}_{\rm VCS}
\!\!\!&=&\!\!\!
\frac{e^4}{q_1^2 q_2^2} \
\bar u (\ell_-) \gamma_\mu u (- \ell_+) \
\bar u (k') \gamma_\nu u (k) \
T_{\mu\nu}
\, , \\
\label{Def-AmpBH-1}
{\cal T}_{{\rm BH}_1}
\!\!\!&=&\!\!\!
\frac{e^4}{q_2^2 \Delta^2} \
\bar u (\ell_-) \gamma_\mu u (- \ell_+) \
\bar u (k')
\left(
\gamma_\mu \frac{1}{ {\not\!k} - {\not\!\!\Delta} } \gamma_\nu
+
\gamma_\nu \frac{1}{ {\not\!k'} + {\not\!\!\Delta} } \gamma_\mu
\right)
u (k) \ J_\nu
\, , \\
\label{Def-AmpBH-2}
{\cal T}_{{\rm BH}_2}
\!\!\!&=&\!\!\!
\frac{e^4}{q_1^2 \Delta^2} \
\bar u (k') \gamma_\mu u (k) \
\bar u (\ell_-)
\left(
\gamma_\mu \frac{1}{ -{\not\!\ell_+} - {\not\!\!\Delta} } \gamma_\nu
+
\gamma_\nu \frac{1}{ {\not\!\ell_-} + {\not\!\!\Delta} } \gamma_\mu
\right)
u (- \ell_+) \
J_\nu
\, ,
\end{eqnarray}
corresponding to diagrams (a), (b), and (c) in Fig.~\ref{BHandDDVCS},
respectively, including the crossed contributions in the latter two
cases. The VCS tensor $T_{\mu\nu}$ was given previously in Eq.\
(\ref{VCSamplitude}), while the nucleon electromagnetic current is
parametrized in terms of the Dirac and Pauli form factors
\begin{equation}
J_\mu
\equiv
\langle p_2 | j_\mu (0) | p_1 \rangle
=
F_1 (\Delta^2) h_\mu
+
F_2 (\Delta^2) e_\mu
\, ,
\end{equation}
using the bilinears from (\ref{DiracBilinears}). The two amplitudes
${\cal T}_{\rm VCS}$ and ${\cal T}_{{\rm BH}_2}$ change the overall sign
when one switches from the electron to the positron beam, and so do the
interference terms involving them, while the ${\cal T}_{{\rm BH}_1}$ does
not. Obviously, both BH amplitudes are related by the interchange of the
momenta $k^\prime \leftrightarrow \ell_-$ and $k \leftrightarrow -\ell_+$.
Moreover, we find that the VCS and the first BH amplitude are even under
the interchange of the produced leptons in the pair, while the second BH
amplitude is odd. This symmetry property in the timelike DVCS plays an
analogous role as the charge asymmetry in the spacelike case \cite{BerDiePir01}.

The evaluation of separate terms yields expressions which are represented
as a Fourier sum of a few harmonics in the difference of the azimuthal
angles $\varphi_l - \varphi_\gamma = \varphi_l - \phi - \pi$. Namely, we
get for particular contributions to the total amplitude squared.

$\bullet$ The VCS amplitude squared
\begin{equation}
\label{Def-VCS2}
| {\cal T}_{\rm VCS} |^2
=
\frac{2 \xi^2 e^8}{Q^4 y^2 \widetilde y^2 (\eta^2 - \xi^2)}
\sum_{n = 0}^{2}
\left(
a^{\rm VCS}_{n} + \lambda b^{\rm VCS}_{n}
\right)
\cos \Big(n ( \varphi_l - \phi ) \Big) \, ,
\end{equation}
has the following expansion coefficients
\begin{eqnarray}
\label{a0VCS}
a^{\rm VCS}_0
\!\!\!&=&\!\!\!
\frac12 (2 - 2y + y^2) (2 - 2 \widetilde y + \widetilde y^2)
\left(
{\cal V} {\cal V}^\dagger
+
{\cal A} {\cal A}^\dagger
\right)
+
4(1 - y)(1 - \widetilde{y}) \frac{\xi^2 - \eta^2}{\xi^2}
{\cal V}_{L} {\cal V}_{L}^\dagger
\, , \\
a^{\rm VCS}_1
\!\!\!&=&\!\!\!
- \frac{\sigma}{\xi} \sqrt{(1 - y)(1 - \widetilde{y})(\xi^2 - \eta^2)}
(2 - y) (2 - \widetilde{y})
\left(
{\cal V} {\cal V}_{L}^\dagger
+
{\cal V}_{L} {\cal V}^\dagger
\right)
\, , \\
a^{\rm VCS}_2
\!\!\!&=&\!\!\!
2 (1 - y) (1 - \widetilde y)
\left(
{\cal V} {\cal V}^\dagger
-
{\cal A} {\cal A}^\dagger
\right)
\, , \\
b^{\rm VCS}_0
\!\!\!&=&\!\!\!
\frac12 y (2 - y) (2 - 2 \widetilde y + \widetilde y^2)
\left(
{\cal V} {\cal A}^\dagger
+
{\cal A} {\cal V}^\dagger
\right)
\, , \\
b^{\rm VCS}_1
\!\!\!&=&\!\!\!
- \frac{\sigma}{\xi} \sqrt{(1 - y)(1 - \widetilde{y})(\xi^2 - \eta^2) }
y (2 - \widetilde{y})
\left(
{\cal V}_{L} {\cal A}^\dagger
+
{\cal A} {\cal V}_{L}^\dagger
\right)
\, , \\
b^{\rm VCS}_2
\!\!\!&=&\!\!\!
0 \, .
\label{b2VCS}
\end{eqnarray}
Note that contrary to the DVCS case, due to the virtuality of both the
incoming and outgoing photons, the Lorentz structure accompanying
${\cal V}_L$ does indeed contribute to the cross section and generates,
e.g., the coefficient $a_1^{\rm VCS}$.

The interference of the VCS amplitude with the BH ones will
involve lepton propagators from the latter which will bring [conveniently
rescaled] factors in the denominator
\begin{eqnarray}
\label{Def-P1P2}
&&(k^\prime + \Delta)^2 \equiv - 2 \eta \, p \cdot q \, {\cal P}_1 (k)
\, , \qquad
(k - \Delta)^2 \equiv - 2 \eta \, p \cdot q \, {\cal P}_2 (k)
\, , \\
\label{Def-P3P4}
&&(\ell_+ + \Delta )^2 \equiv 2 \eta \, p \cdot q \, {\cal P}_3 (\ell_-)
\, , \qquad
(\ell_- + \Delta)^2 \equiv 2 \eta \, p \cdot q \, {\cal P}_4 (\ell_-)
\, .
\end{eqnarray}
The expressions  are rather lengthy  and are obtained by means of
substitution of Eqs.\ (\ref{DefkdotDelta}) and (\ref{DefK}) into the
left-hand side of the above definitions. We also introduce the following
shorthand notations for the structures involving the nucleon matrix
element of the quark electromagnetic current to make the formulas look
as concise as possible,
\begin{eqnarray}
\label{Def-S1}
{\cal S}_1
\!\!\!&\equiv&\!\!\!
\eta \left( k - \frac{1}{y} q_1 \right) \cdot J^\dagger
-
\frac{1}{p \cdot q} \left( k - \frac{1}{y} q_1 \right) \cdot \Delta
\ q_1 \cdot J^\dagger \, , \\
\label{Def-S2}
{\cal S}_2
\!\!\!&\equiv&\!\!\!
\eta \left( \ell_- - \frac{1}{\widetilde y} q_2 \right) \cdot J^\dagger
-
\frac{1}{p \cdot q} \left( \ell_- - \frac{1}{\widetilde y} q_2 \right) \cdot \Delta
\ q_2 \cdot J^\dagger
\, , \\
\label{Def-R12}
{\cal R}_1
\!\!\!&\equiv&\!\!\!
\frac{i}{p \cdot q}
\epsilon_{\mu\nu\rho\sigma} q_\mu k_\nu \Delta_\rho J^\dagger_\sigma
\, , \quad
{\cal R}_2
\equiv
\frac{i}{p \cdot q}
\epsilon_{\mu\nu\rho\sigma} q_\mu \ell_{- \nu} \Delta_\rho J^\dagger_\sigma
\, .
\end{eqnarray}
With these results at hand, we find, similarly to the previous analysis
of $|{\cal T}_{\rm VCS}|^2$, the interference contributions from the VCS
and BH amplitudes.

$\bullet$ The interference ${\cal T}_{\rm VCS} {\cal T}_{{\rm BH}_1}^\dagger$:
\begin{equation}
\label{Def-BHINT1}
{\cal T}_{\rm VCS} {\cal T}_{{\rm BH}_1}^\dagger
=
\frac{2 \xi^2 e^8}
{y^2 \widetilde y^2 \eta^2 (\eta^2 - \xi^2) Q^4 \Delta^2
{\cal P}_1 (k) {\cal P}_2 (k)
}
\sum_{n = 0}^{2}
\left( a^1_n + \lambda b^1_n \right)
\cos \Big( n ( \varphi_l - \phi ) \Big) \, ,
\end{equation}
where
\begin{eqnarray}
a^1_0
\!\!\!&=&\!\!\!
4 (1 - y) (1 - \widetilde y)
\left(
\eta {\cal S}_1 {\cal V}
-
\xi {\cal R}_1 {\cal A}
+
2 \frac{\xi^2 - \eta^2}{\xi} {\cal S}_1 {\cal V}_L
\right)
\nonumber\\
&&\hspace{0cm}
-
(2 - 2 y + y^2) (2 - 2 \widetilde y + \widetilde y^2)
\Big(
\xi {\cal S}_1 {\cal V}
+
\eta {\cal R}_1 {\cal A}
\Big)
\\
&&
-
2 \frac{\widetilde y}{y} (1 - y)(2 - y)(2 - \widetilde y) (\xi - \eta)
\left(
{\cal R}_2 {\cal A}
+
\frac{\eta}{\xi} {\cal S}_2 {\cal V}_L
\right)
\, ,
\nonumber\\
a^1_1
\!\!\!&=&\!\!\!
2 \sigma \sqrt{(1 - y)(1 - \widetilde y)(\xi^2 - \eta^2)}
\\
&&\times
\Bigg\{
(2 - y) (2 - \widetilde y)
\left(
{\cal S}_1 {\cal V}
+
{\cal R}_1 {\cal A}
-
\frac{\xi - \eta}{\xi}{\cal S}_1 {\cal V}_L
\right)
-
4 \frac{\widetilde y}{y} \frac{1 - y}{\xi + \eta}
\Big(
\eta {\cal S}_2 {\cal V}
-
\xi {\cal R}_2 {\cal A}
\Big)
\Bigg\}
\, , \nonumber\\
a^1_2
\!\!\!&=&\!\!\!
- 4 (\xi - \eta) (1 - y) (1 - \widetilde y)
\Big( {\cal S}_1 {\cal V} + {\cal R}_1 {\cal A} \Big)
\, , \\
b^1_0
\!\!\!&=&\!\!\!
- y (2 - y) (2 - 2 \widetilde y + \widetilde y^2)
\Big( \xi {\cal S}_1 {\cal A} + \eta {\cal R}_1 {\cal V} \Big)
- 2 (\xi - \eta) (1 - y) \widetilde y (2 - \widetilde y)
{\cal R}_2 \left({\cal V}+{\cal V}_L \right)
\, , \\
b^1_1
\!\!\!&=&\!\!\!
2 \sigma \sqrt{(1 - y) (1 - \widetilde y) (\xi^2 - \eta^2)}
y (2 - \widetilde y)
\left(
{\cal S}_1 {\cal A}
+
{\cal R}_1 {\cal V}
+
\frac{\xi - \eta}{\xi} {\cal R}_1 {\cal V}_L
\right)
\, , \\
\label{Def-BHINT1-b2}
b^1_2
\!\!\!&=&\!\!\! 0
\, .
\end{eqnarray}

$\bullet$ The interference ${\cal T}_{\rm VCS} {\cal T}_{{\rm BH}_2}^\dagger$:
\begin{equation}
\label{Def-BHINT2}
{\cal T}_{\rm VCS} {\cal T}_{{\rm BH}_2}^\dagger
=
\frac{2 \xi^2 e^8}
{y^2 \widetilde y^2 \eta^2 (\eta^2 - \xi^2) Q^4 \Delta^2
{\cal P}_3 (\ell_-) {\cal P}_4 (\ell_-)
}
\sum_{n = 0}^{2}
\left( a^2_n + \lambda b^2_n \right)
\cos \Big( n ( \varphi_l - \phi ) \Big) \, ,
\end{equation}
with
\begin{eqnarray}
\label{Def-BHINT2-a0}
a^2_0
\!\!\!&=&\!\!\!
4 (1 - y) (1 - \widetilde y)
\left(
\eta {\cal S}_2 {\cal V}
-
\xi {\cal R}_2 {\cal A}
-
2 \frac{\xi^2 - \eta^2}{\xi} {\cal S}_2 {\cal V}_L
\right)
\nonumber\\
&& +
(2 - 2 y + y^2) (2 - 2 \widetilde y + \widetilde y^2)
\Big(
\xi {\cal S}_2 {\cal V}
+
\eta {\cal R}_2 {\cal A}
\Big)
\\
&&
- 2 \frac{y}{\widetilde y} (1 - \widetilde y) (2 - y)(2 - \widetilde y) (\xi + \eta)
\left(
{\cal R}_1 {\cal A}
+
\frac{\eta}{\xi} {\cal S}_1 {\cal V}_L
\right)
\, ,
\nonumber \\
a^2_1
\!\!\!&=&\!\!\!
-2 \sigma \sqrt{(1 - y)(1 - \widetilde y)(\xi^2 - \eta^2)}
\\
&&\times
\Bigg\{
(2 - y) (2 - \widetilde y)
\left(
{\cal S}_2 {\cal V}
-
{\cal R}_2 {\cal A}
-
\frac{\xi+\eta}{\xi} {\cal S}_{2} {\cal V}_L
\right)
+ 4 \frac{y}{\widetilde y} \frac{1 - \widetilde y}{\xi - \eta}
\Big(
\eta {\cal S}_1 {\cal V}
-
\xi {\cal R}_1 {\cal A}
\Big)
\Bigg\}
\, , \nonumber\\
a^2_2
\!\!\!&=&\!\!\!
4 (\xi + \eta) (1 - y) (1 - \widetilde y)
\Big(
{\cal S}_2 {\cal V}
-
{\cal R}_2 {\cal A}
\Big)
\, , \\
b^2_0
\!\!\!&=&\!\!\!
y (2 - y) (2 - 2 \widetilde y + \widetilde y^2)
\Big(
\xi {\cal S}_2 {\cal A} + \eta {\cal R}_2 {\cal V}
\Big)
-
2 (\xi + \eta)
\frac{y^2}{\widetilde y} (2 - \widetilde y) (1 - \widetilde y)
{\cal R}_1 \left( {\cal V} + {\cal V}_L \right)
\, , \\
b^2_1
\!\!\!&=&\!\!\!
- 2 \sigma \sqrt{(1 - y) (1 - \widetilde y) (\xi^2 - \eta^2)}
y (2 - \widetilde y)
\Big(
{\cal S}_2 {\cal A}
-
{\cal R}_2 {\cal V}
-
\frac{\xi + \eta}{\xi} {\cal R}_2 {\cal V}_L
\Big)
\, , \\
\label{Def-BHINT2-b2}
b^2_2
\!\!\!&=&\!\!\! 0
\, .
\end{eqnarray}

The unpolarized parts of the two interference terms must obey a symmetry
relation, since the BH amplitudes (\ref{Def-AmpBH-1}) and (\ref{Def-AmpBH-2})
are related by the exchange $k \leftrightarrow - \ell_+$ and $k^\prime
\leftrightarrow \ell_-$. Obviously, $q_1 \leftrightarrow - q_2$ under it,
while the Bose symmetry ensures the invariance of the Compton amplitude
(\ref{Def-AmpVCS}) with respect to this replacement. As we mentioned above,
all the amplitudes have definite symmetry properties under the permutation
of leptons in the pair $\ell_- \leftrightarrow \ell_+$ and, thus, we take
the advantages of combining both transformations together, the above with
$k \leftrightarrow -\ell_-$ and $k^\prime \leftrightarrow \ell_+$. From the
definitions of the four-vectors (\ref{Def-FouVec-k}) and (\ref{Def-FouVec-lm})
one can read off, after some algebra, the complete set of substitution rules
\begin{eqnarray}
\label{SymRel}
Q^2 \rightarrow Q^2
\, , \quad
\xi \rightarrow - \xi
\, , \quad
\sigma \rightarrow - \sigma
\, , \quad
\Delta \rightarrow \Delta
\, , \quad
\eta \rightarrow \eta
\, , \quad
y \leftrightarrow \widetilde y
\, , \quad
\varphi_{\ell} \leftrightarrow \phi
\, .
\end{eqnarray}
Next, we remark that the product of the BH propagators (\ref{Def-P1P2})
and (\ref{Def-P3P4}) obeys the symmetry relation
\begin{eqnarray}
{\cal P}_3 {\cal P}_4
(Q^2, \Delta^2, \xi, \eta, y, \widetilde y, \varphi_{\ell})
=
{\cal P}_1 {\cal P}_2
(Q^2, \Delta^2, - \xi, \eta, \widetilde y, y, \phi = \varphi_{\ell})
\, .
\end{eqnarray}
The prefactors in the interference terms (\ref{Def-BHINT1}) and
(\ref{Def-BHINT2}) are even under the transformation (\ref{SymRel}). Moreover,
from the definitions (\ref{Def-S1}), (\ref{Def-S2}), and (\ref{Def-R12}) we
conclude also that ${\cal S}_1 \leftrightarrow - {\cal S}_2; {\cal R}_1
\leftrightarrow - {\cal R}_2$ with (\ref{SymRel}). Taking all of our results
together, we deduce that the Fourier coefficients satisfy the equalities
\begin{equation}
a_n^2
=
-
\left.
a_n^1
\right|{\textstyle{
{\cal S}_1 \leftrightarrow -{\cal S}_2;
{\cal R}_1 \leftrightarrow - {\cal R}_2
\atop
y \leftrightarrow \widetilde y;
\xi \leftrightarrow - \xi \hfill
}}
\, ,
\end{equation}
where ${\cal V}$ and ${\cal A}$ are even and odd functions in $\xi$,
respectively. This property is a consequence of the definitions
(\ref{Def-VA-Amp}) and (\ref{Def-ComForFac}), where in the hard-scattering
amplitude (\ref{HarScaAmp}) one has to replace $\xi-i0$ by $-\xi+i0$.

\subsection{Angular dependence of the cross section}
\label{SubSec-AngDep}

After giving the generic expression for the total amplitude squared, we
will elaborate the hadronic part and present the result as a Fourier
expansion in terms of the azimuthal angles, $\phi$ of the recoiled
nucleon and $\varphi_\ell$ of a lepton in the lepton pair. Before doing
so, we write down the general angular decomposition of the squared
amplitudes, which results from the Lorentz structure of the leptonic
tensors contracted with the hadronic ones. The harmonics, appearing
here, can be classified with respect to the underlying twist expansion
of the hadronic tensor (\ref{ComptonAmplitude}) and the Fourier
coefficients are in one-to-one correspondence with helicity amplitudes,
defined in the target rest frame. Extracting certain kinematical factors
in order to match the normalization adopted for the leptoproduction
cross section of a real photon in Ref.\ \cite{BelMulKir01}, the square
of the VCS amplitude and its interference with the BH amplitudes as
well as the squared BH amplitudes admit the following expansion in
these azimuthal angles
\begin{eqnarray}
\label{Dec-VCS2}
&&| {\cal T}_{\rm VCS} |^2
=
\frac{2 \xi^2 e^8}{Q^4 y^2 \widetilde y^2 (\eta^2 - \xi^2)}
\sum_{n = 0}^2
\left\{
{\rm c}^{\rm VCS}_n (\varphi_\ell) \cos (n \phi)
+
{\rm s}^{\rm VCS}_n (\varphi_\ell) \sin (n \phi)
\right\}
\, , \\
\label{Dec-BHVCS}
&&{\cal I}
=
\frac{2 \xi (1 - \eta) e^8}{y^3 \widetilde y^3 (\eta^2 - \xi^2) Q^2 \Delta^2}
\sum_{n = 0}^3
\Bigg\{
\pm
\frac{\widetilde y}{{\cal P}_1 {\cal P}_2 (\phi)}
\left(
{\rm c}^1_n (\varphi_\ell) \cos (n \phi)
+
{\rm s}^1_n (\varphi_\ell) \sin (n \phi)
\right)
\\
&&\qquad\qquad\qquad\qquad\qquad\qquad\ \ \
+
\frac{y}{{\cal P}_3 {\cal P}_4(\varphi_\ell)}
\left(
{\rm c}^2_n (\phi) \cos (n \varphi_\ell)
+
{\rm s}^2_n (\phi) \sin (n \varphi_\ell)
\right)
\Bigg\}
\, ,
\nonumber\\
\label{Dec-BHsq}
&&| {\cal T}_{\rm BH} |^2
=
-
\frac{\xi (1 - \eta )^2}{y^4 {\widetilde y}^4 \Delta^2 Q^2 \eta (\eta^2 - \xi^2)}
\left\{ \sum_{n=0}^4
\Bigg\{ \frac{{\widetilde y}^2}{{\cal P}_1^2 {\cal P}_2^2 (\phi)}
\left(
{\rm c}^{11}_n (\varphi_\ell) \cos (n \phi)
+
{\rm s}^{11}_n (\varphi_\ell) \sin (n \phi)
\right)\right.
\\
&&\qquad\qquad\qquad\qquad\qquad\qquad\qquad\quad\,\,\,
+ \frac{y^2}{ {\cal P}_3^2 {\cal P}_4^2 (\varphi_\ell)}
\left(
{\rm c}^{22}_n (\phi) \cos (n \varphi_\ell)
+
{\rm s}^{\rm 22}_n (\phi ) \sin (n \varphi_\ell)
\right)\Bigg\}
\nonumber\\
&&\qquad\qquad\qquad\qquad\qquad\qquad\quad\,\,\,
\left.
\left.
\pm \sum_{n = 0}^3
\frac{y \widetilde y}{ {\cal P}_1 {\cal P}_2 {\cal P}_3 {\cal P}_4}
\left(
{\rm c}^{\rm 12}_n (\varphi_\ell) \cos (n \phi)
+
{\rm s}^{\rm 12}_n (\varphi_\ell) \sin (n \phi)
\right)
\right\}
\right\}
\, .
\nonumber
\end{eqnarray}
Here the $+$ ($-$) sign stands for the electron (positron) beam and in
the expansion we have used the relation between the azimuthal angles
$\varphi_\gamma = \pi + \phi$. In turn, the Fourier coefficients for the
squared VCS term ($i = {\rm VCS}$), the interference term with the first
BH amplitude (\ref{Def-AmpBH-1}) ($i=1$), and the squared  BH amplitude
(\ref{Def-AmpBH-1}) ($i = 11$) are expanded up to the second order
harmonics in the azimuthal angle $\varphi_\ell$ of the lepton pair
\begin{eqnarray}
\label{Dec-FC-gen}
{\rm c}^i_n (\varphi_\ell)
\!\!\!&=&\!\!\!
\sum_{m = 0}^2
\left\{
{\rm cc}^i_{n m} \cos (m \varphi_\ell)
+
{\rm cs}^i_{n m} \sin (m \varphi_\ell)
\right\}
\, ,
\nonumber\\
{\rm s}^i_n (\varphi_\ell)
\!\!\!&=&\!\!\!
\sum_{m = 0}^2
\left\{
{\rm sc}^i_{n m} \cos (m \varphi_\ell)
+
{\rm ss}^i_{n m} \sin (m \varphi_\ell)
\right\}
\, .
\end{eqnarray}
While for the interference of the second BH amplitude (\ref{Def-AmpBH-2})
with the VCS one ($i = 2$) and its square ($i = 22$), an analogous expansion
is performed in terms of the azimuthal angle $\phi$.  The interference of
both BH amplitudes ($i = 12$) is  analogous to Eq.\ (\ref{Dec-FC-gen}),
however, it contains now  a third order harmonics in the azimuthal angle
$\varphi_\ell$. The Fourier coefficients lineary depend on the polarization
vector of the nucleon, see Eq.\ (\ref{Def-DecFC}). At the edge of the phase
space the overall coefficient in the BH amplitude gets corrected according to
\begin{equation}
\frac{
\left( 1 - \eta \right)^2
}{
\eta^2 - \xi^2
}
\to
\frac{
\left(1 - \eta + \frac{\xi \Delta^2}{2 Q^2}\right)^2
}{
\eta^2
-
\left( 1 - \frac{\Delta^2}{4 Q^2} \right)^2 \xi^2
}
\, .
\end{equation}
We emphasize that $1/(\eta^2-\xi^2)$ expressions in the squared VCS
(\ref{Dec-VCS2}) and interference (\ref{Dec-BHVCS}) terms are corrected
in analogous manner to ensure their correct behavior in the limits
${\cal Q}^2 \to 0$ and $M_{\ell\bar\ell}^2 \to 0$.

{}Finally, we remark that all BH propagators, defined in Eqs.\
(\ref{Def-P1P2}) and (\ref{Def-P3P4}), are even functions in the
azimuthal angle $\varphi$:
\begin{eqnarray}
{\cal P}_i (\varphi)
=
{\cal P}_i (2 \pi - \varphi)
\quad \mbox{for} \quad i = \{1, \dots, 4\}
\, ,
\end{eqnarray}
and, thus, even and odd harmonics can be clearly separated from each other.
It is also worth to mention that ${\cal P}_3 {\cal P}_4$ as a function of
the lepton-pair solid angles $\varphi_{\ell}$ and $\theta_{\ell}$ satisfy
the symmetry relation
\begin{eqnarray}
{\cal P}_3 {\cal P}_4 (\theta_{\ell},\varphi_{\ell})
=
{\cal P}_3 {\cal P}_4 (\pi-\theta_{\ell},\varphi_{\ell}+\pi)
\, .
\end{eqnarray}
For later use, we mention that as a consequence of this symmetry the
integration over $d\theta_\ell$ in a symmetric interval around the point
$\theta_{\ell} = \pi/2$ gives for any definite symmetric moment in
$\theta_{\ell}$ the following characteristic $\cos$-Fourier expansions
(for any number $r$)
\begin{eqnarray}
\int_{\pi/2 - \vartheta}^{\pi/2 + \vartheta} d \cos\theta_{\ell} \;
\frac{\tau(\theta_{\ell}) }{[ {\cal P}_3 {\cal P}_4]^r}
=
\sum_{n = 0,1,2,\cdots} \tau_n (\vartheta)
\left\{
\cos([2 n + 1] \varphi_{\ell})
\atop
\cos(2 n\varphi_{\ell})
\right\}
\quad \mbox{for} \quad
\left\{
\tau(\theta_{\ell}) = - \tau(\pi - \theta_{\ell})
\atop
\tau(\theta_{\ell}) = \tau(\pi - \theta_{\ell})
\right\} \, ,
\end{eqnarray}
where $\vartheta \le \frac{\pi}{2}$.

\subsubsection{Squared virtual Compton amplitude}

At leading twist, it turns out that $| {\cal T}_{\rm VCS} |^2$ depends only
on the harmonics $\cos \Big( n (\varphi_\ell - \phi) \Big)$ with $n = 0, 1,
2$. Consequently, we find in this approximation, with the help of the addition
theorem $\cos \Big( n (\varphi_\ell - \phi) \Big) = \cos (n \varphi_\ell)
\cos(n\phi) + \sin (n \varphi_\ell) \sin(n\phi)$, the following relations
between the Fourier coefficients:
\begin{eqnarray}
\label{Rel-FC-VCS2-Tw2}
{\rm ss}^{\rm VCS}_{nn} \!\!\!&\simeq&\!\!\! {\rm cc}^{\rm VCS}_{nn}
\, , \nonumber\\
{\rm sc}^{\rm VCS}_{nm} \!\!\!&\simeq&\!\!\! {\rm cs}^{\rm VCS}_{nm} \simeq 0
\, , \\
{\rm ss}^{\rm VCS}_{nm} \!\!\!&\simeq&\!\!\! {\rm cc}^{\rm VCS}_{nm} \simeq 0
\, , \qquad n \neq m
\, . \nonumber
\end{eqnarray}
The nonvanishing Fourier coefficients ${\rm cc}^{\rm VCS}_{nn}$ can be
easily evaluated from the generic Eqs.\ (\ref{a0VCS})-(\ref{b2VCS}) and
products of Compton form factors. For the unpolarized target one defines
\begin{equation}
\label{Def-VV}
\frac{1}{4} {\cal V} {\cal V}^\dagger
\equiv
{\cal C}^{\rm VCS}_{{\cal VV}, {\rm unp}}
\, , \qquad
\frac{1}{4} {\cal V} {\cal A}^\dagger
\equiv
{\cal C}^{\rm VCS}_{{\cal VA}, {\rm unp}}
\, , \qquad
\frac{1}{4} {\cal A} {\cal A}^\dagger
\equiv
{\cal C}^{\rm VCS}_{{\cal AA}, {\rm unp}}
\, ,
\end{equation}
where the functions ${\cal C} ({\cal F},{\cal F}^\ast)$ depend on the set of
Compton form factors. For this case we find at leading order in $1/(p \cdot q)$
\begin{eqnarray}
\label{Res-C-unpVV}
{\cal C}^{\rm VCS}_{{\cal V}{\cal V}, {\rm unp}} ({\cal F}, {\cal F}^\ast)
\!\!\!&=&\!\!\!
(1 - \eta^2) {\cal H} {\cal H}^\ast
-
\eta^2 ({\cal H} {\cal E}^\ast
+
{\cal E} {\cal H}^\ast)
-
\left( \frac{\Delta^2}{4 M^2} + \eta^2 \right) {\cal E} {\cal E}^\ast
\, , \\
\label{Res-C-unpAA}
{\cal C}^{\rm VCS}_{{\cal A}{\cal A}, {\rm unp}} ({\cal F}, {\cal F}^\ast)
\!\!\!&=&\!\!\!
(1 - \eta^2) \widetilde {\cal H} \widetilde {\cal H}^\ast
-
\eta^2
(
\widetilde  {\cal H} \widetilde {\cal E}^\ast
+
\widetilde  {\cal E} \widetilde {\cal H}^\ast
)
- \eta^2 \frac{\Delta^2}{4 M^2} \widetilde {\cal E} \widetilde {\cal E}^\ast
\, , \\
{\cal C}^{\rm VCS}_{{\cal V}{\cal A}, {\rm unp}} ({\cal F}, {\cal F}^\ast)
\!\!\!&=&\!\!\!
0
\, .
\phantom{\left(\frac{\Delta^2}{4 M^2}\right)}
\end{eqnarray}
The spin-dependent results, including both longitudinally and transversely
polarized target options, are presented in the appendix \ref{C-coef}.

Note that in the (spacelike) DVCS limit, i.e., when one sets $\eta \simeq - \xi$, we
retrieve our previous result from Ref.\ \cite{BelMulKir01}
\begin{eqnarray*}
{\cal C}^{\rm DVCS}_{\rm unp}
\stackrel{\mbox{\tiny DVCS}}{=}
{\cal C}^{\rm VCS}_{{\cal V}{\cal V},\rm unp}
+
{\cal C}^{\rm VCS}_{{\cal A}{\cal A},\rm unp}
\, ,
\end{eqnarray*}
and analogous relations for the polarized case, see Eq.\ (\ref{Cor-DVCS2}).
In this way, we find for the unpolarized target the following nonvanishing
Fourier coefficients in the twist-two sector
\begin{eqnarray}
\label{Def-FC-unp}
{\rm cc}^{\rm VCS}_{00, {\rm unp}}
\!\!\!&=&\!\!\!
2 (2 - 2y + y^2) (2 - 2 \widetilde y + \widetilde y^2)
\left\{
{\cal C}^{\rm VCS}_{{\cal VV}, {\rm unp}} ({\cal F}, {\cal F}^\ast)
+
{\cal C}^{\rm VCS}_{{\cal AA}, {\rm unp}} ({\cal F}, {\cal F}^\ast)
\right\}   \nonumber\\
&&
+
\frac{16}{\xi^2} (1 - y) (1 - \widetilde y) ( \xi^2 - \eta^2 )
{\cal C}^{\rm VCS}_{{\cal VV}, {\rm unp}} ({\cal F}_L, {\cal F}^\ast_L)
\, , \\
{\rm cc}^{\rm VCS}_{11,{\rm unp}}
\!\!\!&=&\!\!\!
\frac{4 \sigma}{\xi}
\sqrt{(1 - y)(1 - \widetilde y)(\xi^2 - \eta^2)} (2 - y)(2 - \widetilde y)\!
\left\{
{\cal C}^{\rm VCS}_{{\cal VV}, {\rm unp}} ({\cal F}, {\cal F}^\ast_L)
+
{\cal C}^{\rm VCS}_{{\cal VV}, {\rm unp}} ({\cal F}_L, {\cal F}^\ast) \!
\right\}
, \\
{\rm cc}^{\rm VCS}_{22,{\rm unp}}
\!\!\!&=&\!\!\!
8 (1 - y) (1 - \widetilde y)
\left\{
{\cal C}^{\rm VCS}_{{\cal VV}, {\rm unp}} ({\cal F}, {\cal F}^\ast)
-
{\cal C}^{\rm VCS}_{{\cal AA}, {\rm unp}} ({\cal F}, {\cal F}^\ast)
\right\}
\, .
\end{eqnarray}
All other coefficients are expressed making use of Eq.\ (\ref{Rel-FC-VCS2-Tw2}).
Note, however, that the tensor-gluon contribution induces further second order
harmonics, which are not displayed here since they are suppressed by a power
of $\alpha_s$. The Fourier coefficients for the polarized target are collected
in the appendix \ref{C-coef}.

\subsubsection{Interference of virtual Compton and Bethe-Heitler amplitudes}
\label{SubSec-Int}

In the leading-twist approximation the following general relations between
the Fourier coefficients of the interference term are established
\begin{eqnarray}
\label{Rel-FC-INT-Tw2}
{\rm ss}^{\rm INT}_{nm}
\!\!\! &\simeq\!\!\!&
{\rm cc}^{\rm INT}_{nm}
\quad\mbox{and}\quad
{\rm cs}^{\rm INT}_{nm}
\simeq
- {\rm sc}^{\rm INT}_{nm}
\quad\mbox{for} \quad  \{ nm \} = \{ 12, 21, 32 \}
\, , \nonumber\\
{\rm ss}^{\rm INT}_{nm}
\!\!\!&\simeq\!\!\!&
{\rm cc}^{\rm INT}_{nm}
\simeq
{\rm cs}^{\rm INT}_{nm}
\simeq
{\rm sc}^{\rm INT}_{nm}
\simeq
0
\quad\mbox{for} \quad n \neq m \pm 1
\, ,
\end{eqnarray}
where $n,m+1 \le 3$ for INT$=\{1,2\}$. For the unpolarized target, five
nontrivial entries in the case of the unpolarized lepton beam appear,
namely, ${\rm cc}^{\rm INT}_{01,{\rm unp}}$, ${\rm cc}^{\rm INT}_{10,{\rm unp}}$,
${\rm cc}^{\rm INT}_{12,{\rm unp}}$, ${\rm cc}^{\rm INT}_{21,{\rm unp}}$,
and ${\rm cc}^{\rm INT}_{32,{\rm unp}}$, which will be supplemented by
three further Fourier coefficients in the polarized-beam case:
${\rm cs}^{\rm INT}_{01,{\rm unp}}$, ${\rm sc}^{\rm INT}_{10,{\rm unp}}$,
and ${\rm sc}^{\rm INT}_{21,{\rm unp}}$, while ${\rm sc}^{\rm INT}_{12,{\rm unp}}
\simeq {\rm sc}^{\rm INT}_{32,{\rm unp}} \simeq 0$. To find their explicit
form we have evaluated the products of Dirac bilinears. Again, for the
unpolarized nucleon target we get
\begin{eqnarray}
\label{TraSV}
\left\{
\!\!\!
\begin{array}{c}
{\cal S}_1
\\
{\cal S}_2
\end{array}
\!\!\!
\right\}\!
{\cal V}
\!\!\!&=&\!\!\!
-
4 Q^2 \frac{(1 - \eta) \eta}{y \widetilde y \xi}
\left\{
\!\!\!
\begin{array}{c}
\widetilde y K \cos \phi
\\
y \widetilde K \cos \varphi_\ell
\end{array}
\!\!\!
\right\}
{\cal C}_{{\cal V},{\rm unp}} ({\cal F})
\, , \\
\nonumber\\
\left\{
\!\!\!
\begin{array}{c}
{\cal R}_1
\\
{\cal R}_2
\end{array}
\!\!\!
\right\}\!
{\cal V}
\!\!\!&=&\!\!\!
4 i Q^2 \frac{(1 - \eta) \eta}{y \widetilde y \xi}
\left\{
\!\!\!
\begin{array}{c}
\widetilde y K \sin \phi
\\
y \widetilde K \sin \varphi_\ell
\end{array}
\!\!\!
\right\}
{\cal C}_{{\cal V}, {\rm unp}} (\cal F)
\, , \\
\nonumber\\
\left\{
\!\!\!
\begin{array}{c}
{\cal S}_1
\\
{\cal S}_2
\end{array}
\!\!\!
\right\}\!
{\cal A} \!\!\!&=&\!\!\!
- 4 i Q^2 \frac{(1 - \eta) \eta}{y \widetilde y \xi}
\left\{
\!\!\!
\begin{array}{c}
\widetilde y K \sin \phi
\\
y \widetilde K \sin \varphi_\ell
\end{array}
\!\!\!
\right\}
{\cal C}_{{\cal A},{\rm unp}} ({\cal F})
\, , \\
\nonumber\\
\label{TraRA}
\left\{
\!\!\!
\begin{array}{c}
{\cal R}_1
\\
{\cal R}_2
\end{array}
\!\!\!
\right\}\!
{\cal A} \!\!\!&=&\!\!\!
4  Q^2 \frac{(1 - \eta) \eta}{y \widetilde y \xi}
\left\{
\!\!\!
\begin{array}{c}
\widetilde y K \cos \phi
\\
y \widetilde K \cos \varphi_\ell
\end{array}
\!\!\!
\right\}
{\cal C}_{{\cal A},{\rm unp}} ({\cal F})
\, .
\end{eqnarray}
We introduced here universal electric- and magnetic-like combinations
of the form factors intertwined with CFFs
\begin{equation}
\label{Int-dif-Com}
{\cal C}_{{\cal V}, {\rm unp}}
=
F_1 {\cal H} - \frac{\Delta^2}{4 M_N^2} F_2 {\cal E}
\, , \qquad
{\cal C}_{{\cal A}, {\rm unp}}
= - \eta (F_1 + F_2) \widetilde {\cal H}
\, .
\end{equation}
Then from Eqs.\ (\ref{Def-BHINT1})--(\ref{Def-BHINT1-b2}) and
(\ref{Dec-BHVCS}) the following nonzero Fourier coefficients for the
first interference term are evaluated in a straightforward manner:
\begin{eqnarray}
\label{cc-INT1-01}
{\rm cc}^1_{01,{\rm unp}}
\!\!\!&=&\!\!\!
- 8 \widetilde K (1 - y)(2 - y)  (2 - \widetilde y)
\frac{\xi - \eta}{\eta}
\Re{\rm e}\!
\left\{\!
{\cal C}_{{\cal V},{\rm unp}}({\cal F})
+
{\cal C}_{{\cal A},{\rm unp}}({\cal F})
-
\frac{\xi + \eta}{\xi} {\cal C}_{{\cal V},{\rm unp}}({\cal F}_L)
\!\right\}\!
, \\
{\rm cs}^1_{01,{\rm unp}}
\!\!\!&=&\!\!\!
- 8 \lambda \widetilde K y (1 - y)   (2 - \widetilde y)
\frac{\xi - \eta}{\eta}
\Im{\rm m}
\left\{
{\cal C}_{{\cal V},{\rm unp}}({\cal F})
+
{\cal C}_{{\cal A},{\rm unp}}({\cal F})
+
\frac{\xi + \eta}{\xi} {\cal C}_{{\cal V},{\rm unp}}({\cal F}_L)
\right\}
, \\
{\rm cc}^1_{10,{\rm unp}}
\!\!\!&=&\!\!\!
8 K \;
\Re{\rm e}
\left\{
(2 - 2 y + y^2) (2 - 2 \widetilde y + \widetilde y^2)
\left(
\frac{\xi}{\eta} {\cal C}_{{\cal V},{\rm unp}} ({\cal F})
-
{\cal C}_{{\cal A},{\rm unp}}({\cal F})
\right)
\right.
\nonumber \\
&&\qquad\qquad
\left.
- 8 (1 - y) (1 - \widetilde y)\frac{\xi^2 - \eta^2}{\eta \xi}
{\cal C}_{{\cal V},{\rm unp}}({\cal F}_L)
\right\}
, \\
{\rm cc}^1_{12,{\rm unp}}
\!\!\!&=&\!\!\!
16 K (1 - y)  (1 - \widetilde y)
\frac{\xi + \eta}{\xi}
\Re{\rm e}
\left\{
{\cal C}_{{\cal V},{\rm unp}}({\cal F})
+
{\cal C}_{{\cal A},{\rm unp}}({\cal F})
\right\}
, \\
{\rm sc}^1_{10,{\rm unp}}
\!\!\!&=&\!\!\!
- 8 \lambda K y   (2-y)(2 - 2\widetilde y + \widetilde y^2)
\Im{\rm m}
\left\{
{\cal C}_{{\cal V},{\rm unp}}({\cal F})
-
\frac{\xi}{\eta} {\cal C}_{{\cal A}, {\rm unp}}({\cal F})
\right\}
, \\
{\rm cc}^1_{21,{\rm unp}}
\!\!\!&=&\!\!\!
- 8 \widetilde K (1 - y) (2 - y)  (2 - \widetilde y) \frac{\xi - \eta}{\eta}
\Re{\rm e}\!
\left\{\!
{\cal C}_{{\cal V},{\rm unp}}({\cal F})
-
{\cal C}_{{\cal A},{\rm unp}}({\cal F})
-
\frac{\xi - \eta}{\xi} {\cal C}_{{\cal V},{\rm unp}}({\cal F}_L)
\right\}\!
, \\
{\rm sc}^1_{21,{\rm unp}}
\!\!\!&=&\!\!\!
8 \lambda\widetilde K y (1 - y)   (2 - \widetilde y) \frac{\xi - \eta}{\eta}
\Im{\rm m}
\left\{
{\cal C}_{{\cal V},{\rm unp}}({\cal F})
-
{\cal C}_{{\cal A},{\rm unp}}({\cal F})
+
\frac{\xi - \eta}{\xi} {\cal C}_{{\cal V},{\rm unp}}({\cal F}_L)
\right\}
, \\
\label{cc-INT1-32}
{\rm cc}^1_{32,{\rm unp}}
\!\!\!&=&\!\!\!
16  K (1 - y)    (1 - \widetilde y) \frac{\xi - \eta}{\eta}
\Re{\rm e}
\left\{
{\cal C}_{{\cal V},{\rm unp}}({\cal F})
-
{\cal C}_{{\cal A},{\rm unp}}({\cal F})
\right\}
,
\end{eqnarray}
suplemented by
${\rm ss}^1_{12,{\rm unp}}$,
${\rm ss}^1_{21,{\rm unp}}$,
 ${\rm cs}^1_{21,{\rm unp}}$,  and
${\rm ss}^1_{32,{\rm unp}}$, which arise from Eq.\ (\ref{Rel-FC-INT-Tw2}).
The corresponding expressions for the second interference term follow from
Eqs.\ (\ref{Def-BHINT2})--(\ref{Def-BHINT2-b2}) and (\ref{Dec-BHVCS}). For
the unpolarized lepton beam  they can be obtained from the symmetry under the
exchange $k\leftrightarrow -\ell_-$ and $k^\prime \leftrightarrow \ell_+$ as
discussed above in section \ref{Sec-GenFun},
\begin{eqnarray}
\left\{
{\rm cc}^2_{01},
{\rm cc}^2_{10},
{\rm cc}^2_{12},
{\rm cc}^2_{21},
{\rm cc}^2_{32}
\right\}_{\rm unp}
=
\left\{
{\rm cc}^1_{01},
{\rm cc}^1_{10},
{\rm cc}^1_{12},
{\rm cc}^1_{21},
{\rm cc}^1_{32}
\right\}_{\rm unp}
\Big|
{\textstyle {y \leftrightarrow \widetilde y\hfill \atop \xi \to - \xi}}
\, .
\end{eqnarray}
Here one has to keep in mind that ${\cal C}_{\cal V}$ and ${\cal C}_{\cal A}$
are even and odd functions in $\xi-i 0$, respectively, and  $\widetilde K
(\xi, \widetilde y) = K(- \xi, y = \widetilde y)$. It turns out that the
remaining coefficients for the polarized lepton beam satisfy the following
symmetry relations
\begin{eqnarray}
\label{FC-Int-Sym}
\left\{  {\rm cs}^2_{01},{\rm sc}^2_{21} \right\}_{\rm unp}
=
\sqrt{\frac{\widetilde y-1}{1 - y}}
\left\{  {\rm cs}^1_{01},{\rm sc}^1_{21} \right\}_{\rm unp}
\Big|
{\textstyle { \atop \xi \to - \xi}}
\, , \qquad
{\rm sc}^2_{10,{\rm unp}}
=
- \sqrt{\frac{\widetilde y-1}{1 - y}}  {\rm sc}^1_{10,{\rm unp}}
\Big|
{\textstyle { \atop \xi \to - \xi}}
\, .
\end{eqnarray}
We note also that $\widetilde K(\xi,\widetilde y) \approx
\sqrt{(\widetilde y - 1)/(1 - y)} K(- \xi,y)$. The explicit expressions
of the Fourier coefficients for the second interference term are given
in appendix \ref{App-IntPolTar} together with the general structure
and results for the polarized nucleon.

\subsubsection{Squared Bethe-Heitler amplitude}

The expressions for the Fourier coefficients of the pure BH term
(\ref{Dec-BHsq}) are extremely lengthy and, therefore, will not be
displayed here in an analytical form. In the following, we merely
limit ourselves to leading terms in the asymptotic expansion as
$Q^2/\xi \to \infty$. Namely
\begin{eqnarray}
{\rm cc}_{00,{\rm unp}}^{11}
\!\!\!&\approx&\!\!\!
- 2 \frac{1 + \eta}{1 - \eta} \left(1 - \frac{\xi}{\eta}\right)
(1 - y)
\Bigg\{
(2 - 2 y + y^2) (2 - 2 \widetilde y + \widetilde{y}^2)
\left(1 + \frac{\xi^2}{\eta^2}\right)
\\
\!\!\!&-&\!\!\!
8 (1 - y)(1 - \widetilde y)
\left(1 - \frac{\xi^2}{\eta^2}\right)
\Bigg\}
\left\{
\left( 1 - \frac{\Delta^2_{\rm min}}{\Delta^2} \right)
\left( F_1^2 - \frac{\Delta^2}{4M^2}F_2^2 \right)
+
\frac{2\eta^2}{1-\eta^2} (F_1 + F_2)^2
\right\}
\, , \nonumber\\
{\rm cc}_{02,{\rm unp}}^{11}
\!\!\!&\approx&\!\!\!
2 \frac{1 + \eta}{1 - \eta} \left( 1 - \frac{\xi}{\eta} \right)
\left( 1 - \frac{\xi^2}{\eta^2} \right)
\left(1 - \frac{\Delta^2_{\rm min}}{\Delta^2}\right)  (1 - y)
\Bigg\{(2 - 2 y + y^2) (2 - 2 \widetilde y + \widetilde{y}^2)
\\
&+&\!\!\!
8 (1 - y) (1 - \widetilde y)
\Bigg\}
\left(F_1^2 - \frac{\Delta^2}{4M^2}F_2^2 \right)
\, , \nonumber\\
{\rm cc}_{11,{\rm unp}}^{11}
\!\!\!&\approx&\!\!\!
4 \frac{1 + \eta}{1 - \eta} \left(1 - \frac{\xi}{\eta}\right)
\frac{\sigma}{\eta}
\sqrt{(1 - y)(1 - \widetilde y)\left(\xi^2 - \eta^2\right)}
\\
&\times&\!\!\!
(1 - y) (2 - y) (2 - \widetilde y)
\left\{ \left(1 + 3 \frac{\xi}{\eta} \right)
\left( 1 - \frac{\Delta^2_{\rm min}}{\Delta^2} \right)
\left( F_1^2 - \frac{\Delta^2}{4M^2} F_2^2 \right)
+
\frac{4\xi\eta}{1 - \eta^2} (F_1 + F_2)^2
\right\}
\, , \nonumber\\
{\rm cc}_{13,{\rm unp}}^{11}
\!\!\!&\approx&\!\!\!
- 4 \frac{1 + \eta}{1 - \eta} \left( 1 - \frac{\xi}{\eta} \right)^2
\left( 1 - \frac{\Delta^2_{\rm min}}{\Delta^2} \right)
\frac{\sigma}{\eta}
\sqrt{(1 - y)(1 - \widetilde y) \left( \xi^2 - \eta^2 \right)}
\, , \\
&\times&\!\!\!
(1 - y) (2-y) (2-\widetilde y)
\left( F_1^2 - \frac{\Delta^2}{4M^2}F_2^2 \right)
\, , \nonumber\\
{\rm cc}_{20,{\rm unp}}^{11}
\!\!\!&\approx&\!\!\!
- 4 \frac{1 + \eta}{1 - \eta}
\left(1 + \frac{\xi}{\eta}\right) \left(1 - \frac{\xi^2}{\eta^2}\right)
\left( 1 - \frac{\Delta^2_{\rm min}}{\Delta^2} \right)
(1 - y)^2 (1-\widetilde y)
\left(F_1^2 - \frac{\Delta^2}{4M^2}F_2^2 \right)
\, , \\
{\rm cc}_{22,{\rm unp}}^{11}
\!\!\!&\approx&\!\!\!
8 \frac{1 + \eta}{1 - \eta}
\left(1 - \frac{\xi}{\eta}\right) \left(1 - \frac{\xi^2}{\eta^2}\right)
(1 - y)^2 (1 - \widetilde y)
\\
&\times&\!\!\!
\Bigg\{
\left(1 - \frac{\Delta^2_{\rm min}}{\Delta^2}\right)
\left(F_1^2 - \frac{\Delta^2}{4M^2}F_2^2 \right)
+
\frac{2\eta^2}{1 - \eta^2} (F_1 + F_2)^2
\Bigg\}
\, , \nonumber
\end{eqnarray}
and the rest are expressed via the already known coefficients
\begin{eqnarray}
{\rm cc}_{24,{\rm unp}}^{11}
\!\!\!&\approx&\!\!\!
\frac{(\eta - \xi)^2}{(\eta + \xi)^2} {\rm cc}_{20,{\rm unp}}^{11}
\, , \qquad
{\rm ss}_{11,{\rm unp}}^{11}
\approx
{\rm cc}_{11,{\rm unp}}^{11}
+
2 \frac{\eta + \xi}{\eta - \xi} {\rm cc}_{13,{\rm unp}}^{11}
\, , \\
{\rm ss}_{13,{\rm unp}}^{11}
\!\!\!&\approx&\!\!\!
{\rm cc}_{13,{\rm unp}}^{11}
\, , \qquad\qquad\quad\;\;
{\rm ss}_{22,{\rm unp}}^{11}
\approx
{\rm cc}_{22,{\rm unp}}^{11}
\, , \qquad\qquad\qquad\quad
{\rm ss}_{24,{\rm unp}}^{11}
\approx
{\rm cc}_{24,{\rm unp}}^{11}
\, . \nonumber
\end{eqnarray}
One has to realize that this expansion is only valid if one stays away
from kinematical boundaries, e.g., $y \ll 1$ is required. The reason
for this is that the leading terms vanish with $(1 - y)$ and subleading
corrections become important as $y \to 1$. Contrary to the DVCS case, it
appears that no partial cancellation occurs between the numerator and the
denominator in the BH amplitude squared, so that in general the Fourier
decomposition goes as high as up to the forth-order harmonics.

We note that the Fourier coefficients for the second BH-amplitude
squared simply follow from the symmetry under $k \leftrightarrow - \ell_- $
and $k^\prime \leftrightarrow \ell_+ $.
Since we extracted one power of $\xi$ in front of the squared BH
amplitude (\ref{Dec-BHsq}), we obtain the substitution rule
\begin{eqnarray}
{\rm cc}_{nm}^{22} = - {\rm cc}_{nm}^{11}\Big|
{\textstyle {y \leftrightarrow \widetilde y \hfill \atop \xi \to - \xi}}
\quad\mbox{and}\quad
{\rm ss}_{nm}^{22} = - {\rm ss}_{nm}^{11}\Big|
{\textstyle {y \leftrightarrow \widetilde y \hfill \atop \xi \to - \xi}}
\, ,
\end{eqnarray}
while the remaining variables $\{\eta,\Delta^2,Q^2\}$ are kept unchanged.
For the interference term of the first and second BH amplitudes the
Fourier coefficients are
\begin{eqnarray}
{\rm cc}_{00,{\rm unp}}^{12}
\!\!\!&\approx&\!\!\!
- 8 \frac{1 + \eta}{1 - \eta} \frac{\xi}{\eta}
\left( 1 - \frac{\xi^2}{\eta^2} \right)
(1 - y) (2 - y) (1 - \widetilde y) (2 - \widetilde y)
\\
&\times&\!\!\!
\left\{
\left( 1 - \frac{\Delta^2_{\rm min}}{\Delta^2} \right)
\left( F_1^2 - \frac{\Delta^2}{4M^2}F_2^2 \right)
+
\frac{2\eta^2}{1-\eta^2} (F_1+F_2)^2
\right\}
\, , \nonumber
\\
{\rm cc}_{02,{\rm unp}}^{12}
\!\!\!&\approx&\!\!\!
8 \frac{1 + \eta}{1 - \eta}
\left(1 - \frac{\xi}{\eta}\right) \left(1 - \frac{\xi^2}{\eta^2}\right)
(1 - y) (2 - y) (1 - \widetilde y) (2 - \widetilde y)
\left(1 - \frac{\Delta^2_{\rm min}}{\Delta^2}\right)
\left(F_1^2 - \frac{\Delta^2}{4M^2}F_2^2 \right)
\, ,
\nonumber\\
\\
{\rm cc}_{11,{\rm unp}}^{12}
\!\!\!&\approx&\!\!\!
- 8 \frac{1 + \eta}{1 - \eta}
\frac{\sigma}{\eta}
\sqrt{(1-y)(1-\widetilde y)\left(\xi^2-\eta^2\right)}
\\
&\times&\!\!\!
\Bigg\{
(2 - 2 y + y^2) (2 - 2 \widetilde y + \widetilde y^2)
\Bigg[
\frac{\xi^2}{\eta^2}
\left( 1 - \frac{\Delta^2_{\rm min}}{\Delta^2} \right)
\left( F_1^2 - \frac{\Delta^2}{4M^2}F_2^2 \right)
+
\frac{\xi^2 + \eta^2}{1 - \eta^2} \left( F_1 + F_2 \right)^2
\Bigg]
\nonumber\\
&-&\!\!\!
(1 - y)(1 - \widetilde y) \left( 1 - \frac{\xi^2}{\eta^2} \right)
\Bigg[
9 \left( 1 - \frac{\Delta^2_{\rm min}}{\Delta^2} \right)
\left( F_1^2 - \frac{\Delta^2}{4M^2}F_2^2 \right)
+ 10 \frac{\eta^2}{1 - \eta^2} \left( F_1 + F_2 \right)^2
\Bigg]
\Bigg\}
\, , \nonumber\\
{\rm cc}_{20,{\rm unp}}^{12}
\!\!\!&\approx&\!\!\!\!
- 8 \frac{1 + \eta}{1 - \eta}
\left(1 + \frac{\xi}{\eta}\right) \left(1 - \frac{\xi^2}{\eta^2}\right)
\left(1 - \frac{\Delta^2_{\rm min}}{\Delta^2}\right)
(1 - y)(2 - y) (1-\widetilde y) (2-\widetilde y)
\left( F_1^2 - \frac{\Delta^2}{4M^2}F_2^2 \right) \! ,
\nonumber\\
\\
{\rm cc}_{22,{\rm unp}}^{12}
\!\!\!&\approx&\!\!\!
- 8 \frac{1 + \eta}{1 - \eta} \frac{\xi}{\eta}
\left(1 - \frac{\xi^2}{\eta^2}\right)
\\
&\times&\!\!\!
(1 - y)(2 - y) (1 - \widetilde y) (2 - \widetilde y)
\Bigg\{
\left(1 - \frac{\Delta^2_{\rm min}}{\Delta^2}\right)
\left(F_1^2 - \frac{\Delta^2}{4M^2}F_2^2 \right)
+
\frac{2\eta^2}{1 - \eta^2} (F_1 + F_2)^2
\Bigg\}
\, , \nonumber\\
{\rm ss}_{11,{\rm unp}}^{12}
\!\!\!&\approx&\!\!\!
{\rm cc}_{11,{\rm unp}}^{12}
-
8 \frac{1 + \eta}{1 - \eta}
\frac{\sigma}{\eta}
\sqrt{(1 - y)(1 - \widetilde y) \left(\xi^2 - \eta^2\right)}
\left( 1 - \frac{\xi^2}{\eta^2} \right)
\\
&\times&\!\!\!
\left\{
(2 - 2 y + y^2) (2 - 2 \widetilde y + \widetilde y^2) + 8 (1 - y)(1 - \widetilde y)
\right\}
\left( 1 - \frac{\Delta^2_{\rm min}}{\Delta^2} \right)
\left( F_1^2 - \frac{\Delta^2}{4M^2}F_2^2 \right)
\, ,
\nonumber\\
{\rm ss}_{22,{\rm unp}}^{12}
\!\!\!&\approx&\!\!\!
{\rm cc}_{22,{\rm unp}}^{12}
\, .
\end{eqnarray}
Again, under the exchanges $k \leftrightarrow - \ell_-$ and $k^\prime
\leftrightarrow \ell_+$ both amplitudes are odd, i.e., ${\cal T}_{\rm BH_1}
\leftrightarrow - {\cal T}_{\rm BH_2}$, and so their interference term is
invariant. Thus, the Fourier coefficients must satisfy the relation
\begin{eqnarray}
{\rm cc}_{nm}^{12} = - {\rm cc}_{mn}^{12}\Big|
{\textstyle {y \leftrightarrow \widetilde y \hfill \atop \xi \to - \xi}}
\quad\mbox{and}\quad
{\rm ss}_{nm}^{12} = - {\rm ss}_{mn}^{12}\Big|
{\textstyle {y \leftrightarrow \widetilde y \hfill \atop \xi \to - \xi}}
\, ,
\end{eqnarray}
which is the case for our result.

\section{Measurements of GPDs}
\label{Asymmetry}

Having computed the cross section, we will discuss now the observables
that can be used for the experimental exploration of GPDs.

\subsection{Extraction of GPDs from electroproduction of electron pairs?}

The major complication in the experimental measurement of the process is
a rather small magnitude of the cross section which is suppressed by two
powers of the electromagnetic coupling constant $\alpha_{\rm em}$
compared to a typical deeply inelastic event. The other obstacle is the
contamination of the heavy-photon events by the background of meson
production. The latter can be circumvented in a relatively
straightforward manner by avoiding the regions of $M_{\ell\bar\ell}^2$
close to meson-resonance thresholds. However, this certainly also
restrict the phase space in the measurements of GPDs. For a general
discussion of this issue we refer to Ref. \cite{GouDiePirRal97}. We also
note that a numerical estimate of this contamination can be done by
means of Eqs.\ (\ref{Def-VacCor})-(\ref{Mod-VacCor}). Indeed, the
contribution of the $\rho$ meson resonance to the beam spin asymmetry
turns out to be small in a perturbative QCD estimate \cite{GuiVan02}.

Before turning to the discussion of possible measurements of GPDs
in appropriate observables and presenting more quantitative estimates,
we have to emphasize that a clear study of GPDs can be done in
experiments in which the tagged flavor of the lepton pairs differs
from the one of the beam. In case they are the same, the results
deduced so far have to be supplemented by contributions in which the
final electrons are interchanged. Under this exchange, the outgoing
electron momenta jump places, i.e., $k^\prime \leftrightarrow \ell_-$.
This obviously yields an essentially different dependence of the VCS
amplitude on the external variables. To employ the process with
identical leptons for the extraction of information on GPDs, one has
to ensure that the momentum flow in the quark propagator in the
handbag diagram remains large. Indeed, the scalar product
\begin{eqnarray}
p \cdot q^\prime = - p \cdot q \frac{ 2 -  y - y \cos\theta_\ell}{2y}
\left\{
1
+
{\cal O}
\left(
\frac{\Delta_\perp}{\sqrt{p\cdot q}}
\right)
\right\}
\, ,
\end{eqnarray}
that sets the scale in the exchanged VCS amplitude remains large,
however, it is now timelike. Note that here and in the following we
denote the kinematical variables that enter the exchanged amplitudes
with a prime. The power-suppressed contributions depend on all
kinematical variables, especially, on $y$, $\widetilde y$ and both
azimuthal angels $\phi$ and $\varphi_{\ell}$. Besides $(2-y)/y >
\cos\theta_\ell$, which is fulfilled by the usual kinematical
restriction $y< 1$, no other {\sl kinematical cuts} are required to
ensure the applicability of perturbative QCD. Moreover, we find that
$\eta^\prime$ is given by $\eta$ in leading order
\begin{eqnarray}
\eta^\prime =
\eta + {\cal O}
\left(
\frac{\Delta_\perp}{\sqrt{p\cdot q}}
\right)
\, ,
\end{eqnarray}
while $\xi^\prime$ receives a strong dependence on the leptonic variables:
\begin{eqnarray}
\label{XiToXiprime}
\xi^\prime
 =
\xi
\frac{
2 \cos \theta_\ell - y(1 + \cos \theta_\ell)
}{
2 - y (1 + \cos \theta_\ell)
}
-
\frac{
2 \sqrt{1 - y} \sqrt{\eta^2 - \xi^2} \sin\theta_\ell
}{
2 - y (1 + \cos\theta_\ell)
}
\cos(\phi - \varphi_\ell)
+
{\cal O}
\left(
\frac{\Delta_\perp}{\sqrt{p\cdot q}}
\right)
\, .
\end{eqnarray}
We point out that if $\theta_\ell$  approaches the edge of phase
space, i.e., $\theta_\ell \to \{0, \pi\}$, the absolute
values of the scaling variables in Eq.\ (\ref{XiToXiprime}) become
identical $|\xi^\prime| \simeq |\xi|$. The conclusions we draw from our
kinematical considerations are as follows. There are no crucial
difficulties in the application of perturbative QCD as long as $p \cdot
q$ is large, however, the analytical evaluation of observables and
further studies are required to find an ``optimal'' method to deduce the
$(\xi, \eta)$ shape of GPDs from measurements of the reaction $e^\mp p
\to e^\mp p \ e^+ e^-$.

\subsection{Mapping the surface of GPDs}

The most valuable information on GPDs can be accessed in observables that
arise from the interference of the VCS and BH amplitudes, since they are
proportional to linear combinations of the former. Such observables can be
measured in single lepton- and hadron-spin asymmetries, in which the whole
BH term squared drops out in the considered leading order in $\alpha_{\rm em}$,
as well as in charge and angular asymmetries.

In the former case one accesses the imaginary part of the VCS amplitude,
where the contamination of the squared VCS amplitude is expected to be small.
Thus, to leading order accuracy in the QCD coupling $\alpha_s$, see Eqs.\
(\ref{HarScaAmp}) and (\ref{Def-ComForFac}), one measures  the GPDs in the
exclusive region. Schematically, we write
\begin{eqnarray}
\label{SSA-gen}
\frac{d\sigma^\uparrow - d\sigma^\downarrow}
{d\sigma^\uparrow + d\sigma^\downarrow}
\propto
\sum_{{\cal F} = {\cal H} \dots {\widetilde {\cal E}}_L}
{\cal K}_{\cal F} (\phi, \varphi)
\Im{\rm m} {\cal F} (\xi, \eta, \Delta^2)
\, ,
\end{eqnarray}
where the imaginary part $\Im{\rm m} {\cal F}$ for $\widetilde {\cal H}$
(${\cal H}$) and $\widetilde {\cal E}$ (${\cal E}$) is a (anti)symmetric
function of $\xi$ (however, always symmetric in $\eta$)
\begin{eqnarray}
\Im{\rm m} {\cal F} = \pi \sum_{q = u, d, s}
Q_q^2
\left\{
F_q (\xi, \eta, \Delta^2) \mp F_q ( - \xi, \eta, \Delta^2)
\right\}
+ {\cal O} (\alpha_s)
\, .
\end{eqnarray}
The kinematical factors ${\cal K}_F (\phi, \varphi)$ can be read off
from the Fourier coefficients, presented in section \ref{SubSec-AngDep}
and appendix \ref{App-PolTar}, where the squared BH amplitude, the
scaled propagators ${\cal P}_i$ and, at the edge of phase space, also the
$K$ and $\widetilde K$ factors should be exactly taken into account. In
principle, a separation of different types of GPDs is partially feasible
by a Fourier analysis or fully possible in case when single target-spin
asymmetries are also available.

For the other two asymmetries, mentioned above, one uses the fact
that the first (second) BH amplitude is even (odd) with respect to
the interchange of both the lepton charge of the beam, i.e., $e^-
\leftrightarrow e^+$, and the lepton's momenta in the pair $\ell^-
\leftrightarrow \ell^+$, i.e., with respect to the simultaneous
replacement $\theta_\ell \to \pi- \theta_\ell$ and $\varphi_\ell \to
\varphi_\ell + \pi$. On the other hand, the VCS amplitude is odd
under charge and even under angular exchanges. A combination of both
asymmetries offers the possibility to access the real part of the
VCS amplitude without the contamination from the BH background. We
point out that a certain separation of different types of Compton
form factors can be done again by means of a Fourier analysis of
unpolarized measurements, while the complete separation requires
double spin-flip experiments. Generically, these observables are
analogous to the one in Eq.\ (\ref{SSA-gen}), where $\Im{\rm m}
{\cal F}$ is replaced by the real part. To leading order in $\alpha_s$,
the real part of the Compton form factors is given by a ``dispersion''
relation:
\begin{eqnarray}
\Re{\rm e} {\cal F} (\xi, \eta, \Delta^2)
= \frac{1}{\pi} {\rm PV}
\int_{-1}^{1} dx\,
\frac{1}{\xi - x}\Im{\rm m} {\cal F}(x, \eta, \Delta^2)
+
{\cal O}(\alpha_s)
\, .
\end{eqnarray}
The value of this integral is also sensitive to the momentum fraction
$|\eta| \le |x|$, which is not probed in single spin asymmetries.

The unequal masses of the incoming and outgoing photons allow to probe
GPDs away from the diagonal $|\xi| = |\eta|$, the only kinematics which
is accessible in DVCS. The skewness variable $\eta$, given in Eq.
(\ref{XitoEta}), depends besides the photon energy $\omega_1 = {\cal
Q}^2/2\Bx M_N = y E$ (with respect to the target rest frame) mainly on
the sum of both squared photon virtualities, i.e., ${\cal Q}^2 +
M_{\ell\bar\ell}^2$, while $\xi$ essentially depends on their differences
${\cal Q}^2-M_{\ell\bar\ell}^2$. The boundaries of the ($\xi$, $\eta$)
region, probed in the process, are set by the following kinematical
constraints:
\begin{itemize}
\item The skewness parameter lies in the region $\eta_{\rm min} < \eta < 0$,
where the lower bound comes from the kinematical condition
$|\Delta^2| \geq |\Delta_{\rm min}^2|$:
\begin{equation}
-\eta_{\rm min} \leq  \sqrt{- \Delta^2/(4 M_N^2 - \Delta^2)}
\, .
\end{equation}
\item The upper and lower value of $\xi$ is a consequence of the quasi-real
limit of the space- or timelike photon
\begin{equation}
-|\eta| < \xi < |\eta|
\, .
\end{equation}
\item The minimal attained value of $|\xi| = 0$ stems from the condition
${\cal Q}^2 \simeq M_{\ell\bar\ell}^2$.
\end{itemize}

\begin{figure}[t]
\begin{center}
\mbox{
\begin{picture}(0,214)(90,0)
\put(0,0){\insertfig{7}{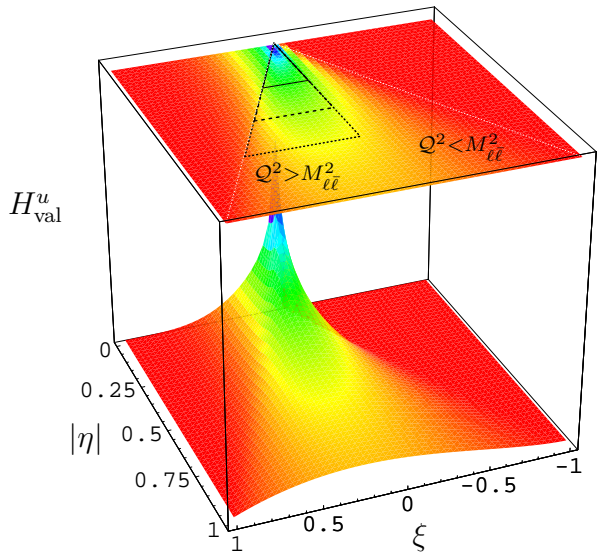}}
\put(128,5){$\xi$}
\put(-2,42){$|\eta|$}
\put(-25,130){$H^u_{\rm val}$}
\put(68,143){{$\scriptstyle {\cal Q}^2 > M_{\ell\bar\ell}^2$}}
\put(130,154){$\scriptstyle {\cal Q}^2 < M_{\ell\bar\ell}^2$}
\end{picture}
}
\end{center}
\caption{\label{XiEtaPlane} The coverage of the GPD surface (the valence
component of the $u$-quark distribution, as an example) with electron
beams of different energies: (solid contour) $E = 11 \, {\rm GeV}$,
(dashed contour) $E = 25 \, {\rm GeV}$, (dotted contour) $E = 40 \,
{\rm GeV}$, as described in the text. [For simplicity, we discarded in
this plot the change of the GPD with $\Delta^2$ for different kinematical
settings.]}
\end{figure}
In Fig.\ \ref{XiEtaPlane} it can be seen that the area of the surface
probed in the electroproduction of the lepton pair is quite extensive.
The three contours displayed in this figure embrace the areas to be
probed for different values of the electron beam energy $E$, the lepton
energy loss $y$, and the $t$-channel momentum transfer $\Delta^2$: (i)
solid contour corresponds to $E = 11 \, {\rm GeV}$, $y = 0.5$, and
$\Delta^2 = - 0.3 \, {\rm GeV}^2$ ($M_N^2 \leq {\cal Q}^2 \leq 10 \,
{\rm GeV}^2$), (ii) dashed contour corresponds to $E = 25 \, {\rm GeV}$,
$y = 0.75$, and $\Delta^2 = - 1 \, {\rm GeV}^2$ ($- 4 \Delta^2 \leq
{\cal Q}^2 \leq 20 \, {\rm GeV}^2$), (iii) dotted contour corresponds to
$E = 40 \, {\rm GeV}$, $y = 0.9$, and $\Delta^2 = - 3 \, {\rm GeV}^2$
($- 4 \Delta^2 \leq {\cal Q}^2 \leq 35 \, {\rm GeV}^2$). The
corresponding values of the Bjorken variable are computed with the
formula $\Bx = {\cal Q}^2 /(2 M_N E y)$. It is obvious that the higher
the energy of the lepton beam, the higher $\Delta^2$ are allowed with
observed applicability of the perturbative analysis of the Compton
amplitude, and thus the higher values of $|\eta|$ are achieved.

We have addressed above only the case ${\cal Q}^2 > M_{\ell\bar\ell}^2$
which probes $\xi > 0$ component of GPDs. For the reversed inequality,
one gets information on the region $\xi < 0$ and probes patches of the
two-dimensional surface analogous to the previous case. The positive
mass of the final-state photon allows to directly access only the
exclusive, or distribution amplitude-like, component $|\eta| > |\xi|$
of the function. The inclusive, or parton distribution-like, component
with $|\eta| < |\xi|$ requires spacelike virtuality for the outgoing
photon which arises in two-photon exchange events in elastic
electron-nucleon scattering. However, since the hadronic tensor
(\ref{ComptonAmplitude}) enters now via a loop integral, the single
spin asymmetry measurements cannot be used for a direct extraction
of GPDs.

\subsection{Compton form factors}

The magnitude of asymmetries depends on the relative strength of the BH
amplitudes with respect to the VCS one. To get a rough idea of what happens
we consider two limiting cases of the space- and timelike DVCS. In case
of the production of a quasi-real final-state photon off the unpolarized
proton target, the approximation of the amplitudes to leading power in
${\cal Q}^2$ for $- \Delta^2_{\rm min} \ll -\Delta^2 \ll 4 M_N^2 \ll
{\cal Q}^2$ and $\xi \ll 1$ gives
\begin{eqnarray}
\label{ratioDVCS/BH}
\frac{|{\cal T}^{\rm DVCS}|^2}{|{\cal T}^{\rm BH}|^2}
\sim
\frac{1-y}{y^2}
\frac{|\Delta^2|}{{\cal Q}^2} \;
\frac{\xi^2 |{\cal H} (\xi, \xi, \Delta^2)|^2}{F_1 (\Delta^2)^2}
\, ,
\end{eqnarray}
for $M_{\ell\bar\ell}^2 \to 0$ and large ${\cal Q}^2$. One expects that
$|{\cal H}(\xi,\xi,\Delta^2)|$ behaves like $\xi^{-1}$ and, thus, the
ratio (\ref{ratioDVCS/BH}) is essentially $(1 - y) |\Delta^2|/ y^2 {\cal
Q}^2$. Obviously, the kinematical suppression by $|\Delta^2|/ {\cal
Q}^2$ can be removed in the small-$y$ region and, therefore, the DVCS
cross section can be extracted in collider experiments. On the other
hand, for moderate values of $y$, one finds a rather sizeable (beam-spin)
asymmetries in fixed target experiments. In contrast, in case of the
timelike DVCS, replacing the variables $y \to \widetilde y$ and
${\cal Q}^2 \to - M_{\ell\bar\ell}^2$ in the above equation, we find
the ratio
\begin{eqnarray}
\label{ratioDVCS/BH-tim}
\frac{|{\cal T}^{\rm DVCS}|^2}{|{\cal T}^{\rm BH}|^2}
\sim
\frac{\sin^2\theta_\ell}{4}\, \frac{|\Delta^2|}{M_{\ell\bar\ell}^2}
\frac{\xi^2 |{\cal H} (\xi, \xi, \Delta^2)|^2}{F_1 (\Delta^2)^2}
\, ,
\end{eqnarray}
for ${\cal Q}^2 \to 0$ and large $M_{\ell\bar\ell}^2$. It is always
suppressed by $1/M_{\ell\bar\ell}^2$ and so the BH cross section
overwhelms the VCS ones \cite{BerDiePir01}. Thus, one anticipates that
the second BH amplitude dominates the cross section except when one
approaches the limit $M_{\ell\bar\ell} \to 0$. On this basis on would
naively argue that the asymmetries will be suppressed as well. To get
rid partly of this problem, instead of integrating over the full range
of the lepton-pair scattering angles, one rather has to sum over a
restricted domain and exclude the endpoint regions $\theta_\ell = \{
0,\pi \}$.

\begin{figure}[t]
\begin{center}
\mbox{
\begin{picture}(0,140)(270,0)
\put(42,2){\insertfig{7.6}{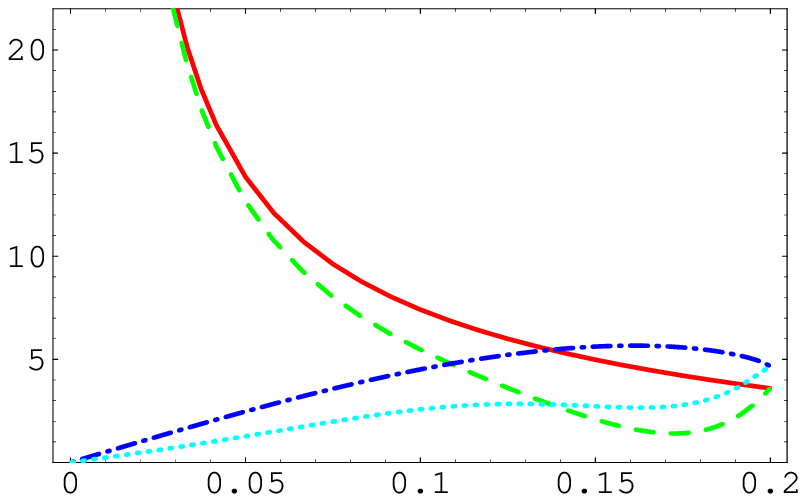}}
\put(25,40){\rotate{$\Im{\rm m}\ {\cal H}(\xi,\eta,\Delta^2)$}}
\put(290,-3){\insertfig{8}{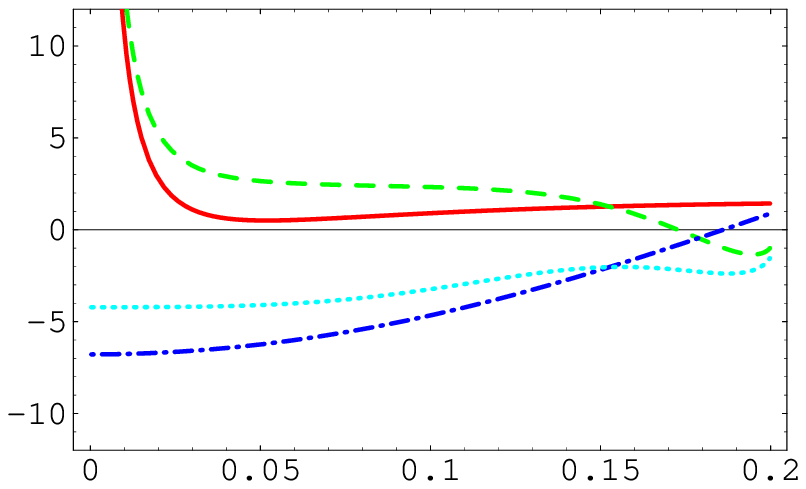}}
\put(280,40){\rotate{$\Re{\rm e}\ {\cal H}(\xi,\eta,\Delta^2)$}}
\put(230,-10){$\xi$}
\put(485,-10){$\xi$}
\put(188,115){${\Delta^2 = - 0.2\ {\rm GeV}^2 \atop \eta = - 0.2 \qquad \ \ }$}
\put(442,115){${\Delta^2 = - 0.2\ {\rm GeV}^2 \atop \eta = - 0.2 \qquad \ \ }$}
\end{picture}
}
\end{center}
\caption{\label{Fig-CFFs}
The imaginary and real part of ${\cal H} (\xi, \eta = - 0.2, \Delta^2 =
- 0.2 \, {\rm GeV}^{2})$ plotted versus $\xi$ in the left and right
panel, respectively, for different GPD models: $b^{\rm val} = b^{\rm
sea} = \infty$ and $B^{\rm sea} = 5 \ {\rm GeV}^{-2}$ (solid, dashed) as
well as for $b^{\rm val} = b^{\rm sea} = 1$ and $B^{\rm sea}= 9\ {\rm
GeV}^{-2}$ (dash-dotted, dotted) with (dashed, dotted) and without
(solid, dotted) $D$-term.}
\end{figure}

To give quantitative estimates and provide some insights into the
procedure of extracting GPDs from such experiments, we use several GPD
models with quite distinct behavior in the central region. For the GPDs
$H_q$, where $q$ labels the three light (anti-)quarks, we take the
factorized version of the $(\xi, \eta)$ and $\Delta^2$ dependence, i.e.,
$H_q (\xi, \eta, \Delta^2) = F_q (\Delta^2) H_q (\xi, \eta, \Delta^2 =
0)$, where the reduced GPD at zero momentum $\Delta^2 = 0$ is modeled
via a factorized double distribution (DD) ansatz \cite{Rad99}. For
specific details we refer the reader to Ref.\ \cite{BelMulKir01}. The
essential freedom left within this specification concerns the parameter
$b^q$ of the profile function, which controls the strength of the
$\eta$-dependence of the reduced GPDs $H_q (\xi, \eta, \Delta^2 = 0)$,
the slope of the partonic sea form factor,
\begin{eqnarray}
F^{\rm sea}(\Delta^2)
=
\left(1 - \frac{B^{\rm sea}}{3}\Delta^2\right)^{-3}
\, ,
\end{eqnarray}
and the $D$-term%
\footnote{Note that this $D$-term, taken from the quark soliton model
\cite{PolWei98,SchBofRad02} at a low scale $\mu \sim 0.6$ GeV, will mix under scale
evolution with a gluonic $D$-term. For the present considerations it is
suffice to neglect the scale dependence completely.},
given as an expansion in terms of Gegenbauer polynomials \cite{GeoPolVan01}
\begin{eqnarray}
D (\xi,\eta,\Delta^2)
=
\frac{
\theta(|\xi| \le |\eta|){\rm sign}(\eta)
}{
\left(1-\Delta^2/0.77 {\rm GeV}^2\right)^{3}
}
(1 - x^2)
\left(
- 4 C_1^{3/2} (x) - 1.2 C_3^{3/2} (x) - 0.4 C_5^{3/2} (x)
\right)\Big|_{x = \frac{\xi}{\eta}}
\, .
\end{eqnarray}
The latter is entirely concentrated in the central region $|\xi| \le |\eta|$.
The imaginary and real parts of the Compton form factors for these models are
shown in Fig.\ \ref{Fig-CFFs}. In case the skewness effect is eliminated
(as $b \to \infty$), we model the reduced GPDs by the usual forward parton
densities taken in MRS A$^\prime$ parameterization at the input scale $\mu^2
= 4 \ {\rm GeV}^2$. Consequently, for this so-called FPD-model the imaginary
and real parts strongly increase as $\xi$ gets smaller. This is displayed by
the solid and dashed lines in Fig.\ \ref{Fig-CFFs}. On the contrary, for
$b = 1$, --- for brevity we call this choice as the DD-model, --- (dash-dotted
and dotted lines), the contribution in the central region is suppressed and
goes to zero for the imaginary part when $\xi \to 0$. At the same time, the
real part approaches a constant, with the value determined by the inverse
moment of GPDs
\begin{eqnarray}
\label{ReaParxi0}
\Re{\rm e}{\cal H} (\xi = 0, \eta, \Delta^2)
=
-\int_{-1}^{1}\frac{d x}{x}
\sum_{q = u,d,s} Q_q^2
\left\{
H_q (x, \eta, \Delta^2) - H_q (- x, \eta, \Delta^2)
\right\}
\, .
\end{eqnarray}
The amplitudes supplemented by the $D$-term contributions are presented by
the dashed and dotted lines in Fig.\ \ref{Fig-CFFs}. It is clearly demonstrated
that their effect is especially prominent (within the parameter range chosen
for the estimates) in the real part of the Compton form factors where it
changes their sign in the vicinity of $\xi \to \pm \eta$.

\subsection{Single-spin asymmetries}

Now, having discussed the properties of the Compton form factors, we will
have a closer look on single-spin asymmetries, in particularly, the beam-spin
asymmetry for the proton
\begin{eqnarray}
d\sigma^{\uparrow} - d\sigma^{\downarrow}
\propto
\left(\pm  {\cal T}_{{\rm BH}_1}^\ast + {\cal T}^\ast_{{\rm BH}_2} \right)
\Im{\rm m}{\cal T}_{\rm VCS}
+
\cdots
\, .
\end{eqnarray}
Potentially, the interference term could be contaminated by the
imaginary part $\Im{\rm m} {\cal T}_{\rm VCS}{\cal T}_{\rm
VCS}^\dagger$, arising from the interference of twist-two and -three
Compton form factors. As for the DVCS process \cite{BelMulKir01} we
expect, assuming the smallness of three-particle correlations, that this
contribution can be safely neglected. As mentioned above, in leading
order of perturbation theory, the single spin asymmetries are directly
proportional to the linear combination of GPDs $\sum_{q = u, d, s} Q_q^2
\left\{ F_q (\xi, \eta, \Delta^2) \mp F_q (\xi, - \eta, \Delta^2)
\right\}$, where $(+)$ $-$ applies for (axial-) vector-type GPDs.
For instance, eight leading-twist observables are measurable in the
beam-spin asymmetry, which are coming in pairs from the interference
of the VCS with the first and second BH amplitudes%
\footnote{For brevity we neglect throughout this section in the Fourier
coefficients the subscript ``unp'', which refers to the unpolarized target.}:
${\rm cs}^1_{01}$,
${\rm sc}^1_{10}$, ${\rm sc}^1_{21}$, ${\rm cs}^1_{12}$ as well as
${\rm cs}^2_{01}$, ${\rm sc}^2_{10}$, ${\rm sc}^2_{21}$, ${\rm cs}^2_{12}$,
see Eq.\ (\ref{FC-Int-Sym}). However, they depend only on two different
linear combinations (\ref{Int-dif-Com}) of GPDs:
\begin{eqnarray}
\label{LinComBSA}
F_1 ({\cal H} + {\cal H}_L)
-
\frac{\Delta^2}{4 M_N^2} F_2 ({\cal E} + {\cal E}_L)
\, , \qquad
- \eta (F_1 + F_2) \widetilde {\cal H}  + \frac{\eta}{\xi}
\left(
F_1 {\cal H}_L - \frac{\Delta^2}{4 M_N^2} F_2  {\cal E}_L
\right)
\, .
\end{eqnarray}
Consequently, there there exist six constraints among the whole set of
coefficients, which can be expressed as:
\begin{eqnarray}
\label{FouCoeRel}
&&{\rm sc}^1_{10}
\simeq
-
\frac{
(2 - y)(2 - 2 \widetilde  y + \widetilde  y^2)
}{
2(1 - y)(2 - \widetilde y)
}
\frac{
\sqrt{(1 - y)(\xi - \eta)}
}{
\sqrt{(1-\widetilde y) (\xi + \eta) }
}
\left\{
{\rm cs}^1_{01} + \frac{ \xi+\eta}{\xi-\eta} {\rm sc}^1_{21}
\right\}
\, , \nonumber\\
&&\frac{{\rm cs}^2_{01}}{{\rm sc}^1_{21}}
\simeq
\frac{{\rm sc}^2_{10}}{{\rm sc}^1_{10}}
\simeq
\frac{{\rm sc}^2_{21}}{{\rm cs}^1_{01}}
\simeq
-
\frac{\sqrt{(1 - \widetilde y) (\xi + \eta) }}{\sqrt{(1 - y)(\xi - \eta)}}
\, ,
\end{eqnarray}
supplemented by the relation (\ref{Rel-FC-INT-Tw2}), i.e., ${\rm
cs}^{1}_{21} \simeq - {\rm sc}^{1}_{21}$ and ${\rm cs}^{2}_{21} \simeq -
{\rm sc}^{2}_{21}$. Another consequence of Eq.\ (\ref{LinComBSA}) is
that the beam-spin asymmetry gives us no handle on the Callan-Gross
relation, i.e., the longitudinal Compton form factors can not be
separated from the leading ones. The Fourier coefficients are projected
out by taking the following moments, when integrated over the scattering
and azimuthal angles:
\begin{eqnarray}
\label{Def-BSA}
&&\!\!\!\!\!\!\!\!\!
\sin\phi \to {\rm sc}_{10}^1
\, , \
\cos\theta_\ell \sin{\varphi_\ell} \rightarrow {\rm cs}_{01}^1
\, , \
\cos\theta_\ell \cos{\varphi_\ell} \sin(2\phi) \rightarrow {\rm sc}^1_{21}
\, , \
\cos\theta_\ell \sin{\varphi_\ell}  \cos(2\phi) \rightarrow {\rm cs}^1_{21}
\, , \nonumber\\
&&\!\!\!\!\!\!\!\!\!
\sin\phi_\ell \rightarrow {\rm sc}_{10}^2
\, , \
\cos\theta_\ell \sin{\phi} \rightarrow {\rm cs}_{01}^2
\, , \
\cos\theta_\ell \sin(2\varphi_\ell) \cos{\phi}  \rightarrow {\rm sc}^2_{21}
\, , \
\cos\theta_\ell \cos(2\varphi_\ell) \sin{\phi}  \rightarrow {\rm cs}^2_{21}
\, , \nonumber
\end{eqnarray}
where the weight in the first (second) row is even (odd) under the
reflection $\theta_{\ell} \to \pi - \theta_{\ell}$ and $\varphi_{\ell}
\to \pi + \varphi_{\ell}$. In the same line of thinking, one can study
the Fourier coefficients for the single target-spin asymmetries. Because
of the substitution rules (\ref{Def-FC-INT-LP}) and (\ref{Def-FC-INT-TP}),
the number of the Compton form factors will be the same as for the case of
charge and angular asymmetries, discussed below. Of course, single
target-spin asymmetries are given by the imaginary part of new linear
combinations in GPDs with a characteristic angular dependence.

As we conclude from Eq.\ (\ref{FouCoeRel}), the same information on
GPDs is obtained by taking the appropriate moments in $\phi$ or
$\varphi_\ell$. However, the size of the complementary beam-spin
asymmetries can vary. Moreover, if one takes the asymmetry from
the interference with the second BH amplitude, the weight must be odd
and, thus, we have no contamination from the squared VCS amplitude. To
suppress the squared BH amplitude, we integrate over the region $\pi/4
\le \theta_\ell \le 3\pi/4$ and form alternatively the $\sin\phi$ or
$\sin \varphi_\ell$ moments. This picks up the coefficients ${\rm
sc}^1_{10}$ and ${\rm sc}^2_{10}$, respectively, cf.\ Eqs.\
(\ref{Dec-BHVCS}), (\ref{Dec-FC-gen}), and (\ref{Rel-FC-INT-Tw2}). Thus,
the beam-spin asymmetry
\begin{eqnarray}
\left\{
A_{\rm LU}^{\sin\phi} \atop A_{\rm LU}^{\sin\varphi_{{\ell}}}
\right\}
&\!\!\!=\!\!\!&
\frac{1}{{\cal N}}
\int_{\pi/4}^{3\pi/4}\! d \theta_{\ell}
\int_{0}^{2\pi}\! d \varphi_\ell
\int_{0}^{2\pi}\! d \phi
\left\{ 2 \sin\phi \atop 2 \sin\varphi_\ell \right\}
\frac{d \sigma^\uparrow - d\sigma^\downarrow}{d {\mit\Omega}_{\ell}d\phi}
\nonumber\\
\!\!\!&\propto&\!\!\!
\Im{\rm m}
\left\{
F_1 {\cal H}
-
\frac{\Delta^2}{4 M_N^2} F_2 {\cal E}
+
\xi (F_1 + F_2) \widetilde {\cal H}
\right\} \, ,
\end{eqnarray}
with the normalization factor being
\begin{eqnarray*}
{\cal N}
=
\int_{\pi/4}^{3\pi/4}\! d \theta_{\ell}
\int_{0}^{2\pi}\! d \varphi_\ell
\int_{0}^{2\pi}\! d \phi
\frac{
d \sigma^\uparrow + d \sigma^\downarrow
}{d{\mit\Omega}_{\ell}d\phi}
\, ,
\end{eqnarray*}
is analogous to the one defined in the case of space- and timelike DVCS.
For the proton target it is mainly sensitive to the contribution $F_1
\Im{\rm m} {\cal H} $ and, therefore, we might neglect in our estimate
the other two Compton form factors.

\begin{figure}[t]
\begin{center}
\mbox{
\begin{picture}(0,140)(270,0)
\put(42,2){\insertfig{8}{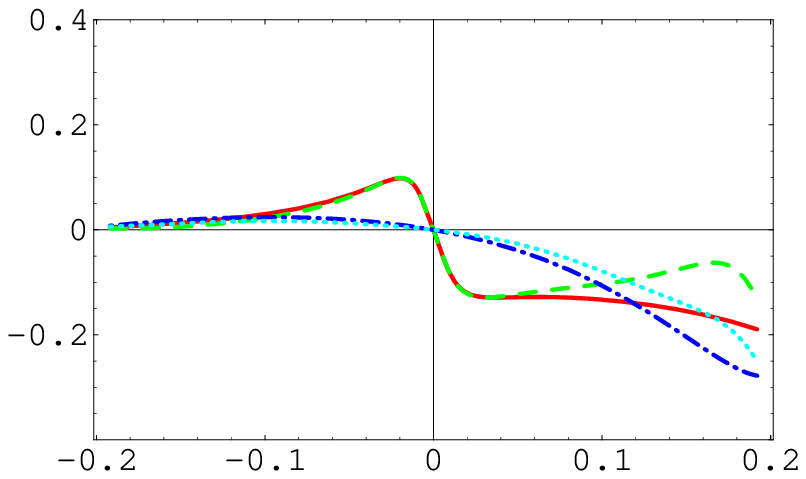}}
\put(25,40){\rotate{$A_{\rm LU}^{\sin\phi}(\xi,\eta,\Delta^2)$}}
\put(290,2){\insertfig{8}{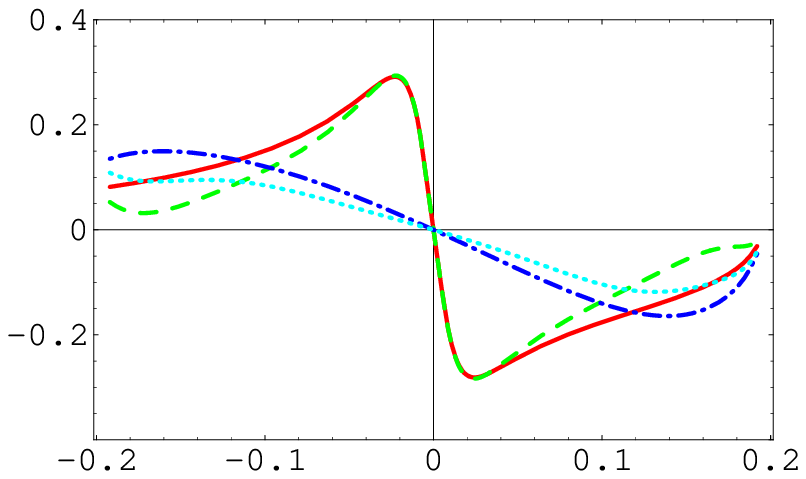}}
\put(275,40){\rotate{$A_{\rm LU}^{\sin\varphi_\ell}(\xi,\eta,\Delta^2)$}}
\put(230,-10){$\xi$}
\put(485,-10){$\xi$}
\put(188,115){${\Delta^2 = - 0.2\ {\rm GeV}^2 \atop \eta = - 0.2, \ y = 0.5 }$}
\put(442,115){${\Delta^2 = - 0.2\ {\rm GeV}^2 \atop \eta = - 0.2, \ y = 0.5 }$}
\end{picture}
}
\end{center}
\caption{\label{Fig-BSA}
Beam spin asymmetries of $A_{\rm LU}^{\sin \phi}$ and $A_{\rm LU}^{\sin
\varphi_\ell}$ as functions of $\xi$, respectively, for $\eta = - 0.2$,
$\Delta^2 = - 0.2 \, {\rm GeV}^{2}$, $y = 0.5$ and different GPD models,
specified in Fig.\ \ref{Fig-CFFs}.}
\end{figure}

In Fig.\ \ref{Fig-BSA}, we show the beam-spin asymmetries
(\ref{Def-BSA}) for an $11$ GeV electron beam and $\eta = - 0.2,\
\Delta^2 = - 0.2 \, {\rm GeV}^{2}\ (\Delta^2_{\rm min} \approx - 0.15 \,
{\rm GeV}^{2})$, and $y = 0.5$. We fix the value of ${\cal Q}^2+ {\cal
M}^2_{\ell\bar\ell} \approx 3.4 \, {\rm GeV}^2$, i.e., $Q^2/\xi \approx
8.5\, {\rm GeV^2}$, and scan the $-0.2 < \xi < 0.2$ region by varying
the virtuality ${\cal Q}^2$ in the range $0 \le {\cal Q}^2 \le 3.4 \,
{\rm GeV}^{2}$. In the spacelike DVCS limit (the left panel with $\xi
\to -\eta$, $\Bx \approx 0.33$), we uncover a typical beam-spin asymmetry
of order 20\% measured in fixed target experiments. This asymmetry is
getting smaller in the DD-model (dashed and dash-dotted line) with
increasing $\xi$, reflecting the fact that the imaginary part of the
Compton form factor goes to zero (see Fig. \ref{Fig-CFFs}). We stress
that this feature truly arises from the GPD model, since $\xi/Q^2$ is
fixed and so no essential kinematical suppression of the interference
term arises as $\xi \to 0$. The negative $D$-term slightly changes the
normalization away from the edges of phase spaces. In case of the
FPD-model, the interference term increases at $\xi \to 0$ with $\xi^{- 1
- \lambda}$ (for MRS A$^\prime$ parametrization $\lambda = 0.17$), as a
consequence of the small-$x$ behavior of the sea-quark densities. On the
other hand, the VCS cross section in the denominator increases like
$\xi^{-2- 2\lambda}$, overwhelming the BH contributions, and forces the
asymmetry to vanish at $\xi = 0$. This explains why with the FPD-model
the asymmetry remains sizeable over a large interval of $\xi$. The
asymmetry, formed with $\sin\phi$ is considerably smaller in the
timelike region, compared with the spacelike one, since the first BH
amplitude, responsible for this asymmetry, is getting smaller, while the
second BH amplitude, entering the denominator, increases for $\xi \to -
|\eta|$. Forming the moments with respect to the $\sin \varphi_\ell$
(the right panel of Fig.\ \ref{Fig-BSA}), the asymmetry is caused by the
interference with the second BH amplitude. As explained above, we expect
that this amplitude is in general larger than the first one and results
into a rather sizeable asymmetry not only in the time- but also in the
spacelike region.

\subsection{Charge and angular asymmetries}

Let us now comment on charge and angular asymmetries, in which the
Fourier coefficients ${\rm cc}^1_{01}$, ${\rm cc}^1_{10}$,
${\rm cc}^1_{12} \simeq {\rm ss}^1_{12}$, ${\rm cc}^1_{21} \simeq
{\rm ss}^1_{21}$, and ${\rm cc}^1_{32} \simeq {\rm ss}^1_{32}$ as
well as the complementary set of the second interference term are
attainable. As above, the Fourier coefficients of both interference
terms are related by
\begin{eqnarray}
\label{FouCoeRel-cc}
\frac{{\rm cc}^2_{01}}{{\rm cc}^1_{21}} \simeq
\frac{{\rm cc}^2_{10}}{{\rm cc}^1_{10}} \simeq
\frac{{\rm cc}^2_{12}}{{\rm cc}^1_{32}} \simeq
\frac{{\rm cc}^2_{21}}{{\rm cc}^1_{01}} \simeq
\frac{{\rm cc}^2_{32}}{{\rm cc}^1_{12}} \simeq
- \frac{\sqrt{(1-\widetilde y) (\xi+\eta) }}{\sqrt{(1 - y)(\xi-\eta)}}\,.
\end{eqnarray}
There are now three independent Compton form factors ${\cal C}_{\cal
V,{\rm unp}}({\cal F})$, ${\cal C}_{\cal V,{\rm unp}}({\cal F}_L)$, and
${\cal C}_{\cal A,{\rm unp}}({\cal F})$, given in Eq.\ (\ref{Int-dif-Com}),
which can be accessed there. Consequently, there exist two constraints
among five nontrivial Fourier coefficients. Provided the Callan-Gross
relation is assumed to be fulfilled, this number increases to three:
\begin{eqnarray}
\label{FouCoeRelCG}
{\rm cc}^1_{10}
\!\!\!&\simeq&\!\!\!
-
\frac{(2 - 2 y +  y^2)(2 - 2 \widetilde y
+
\widetilde  y^2)}{2(2 - y)(1 - \widetilde y)(2 - \widetilde y)}
\frac{\sqrt{(1-\widetilde y) (\xi + \eta) }}{\sqrt{(1 - y)(\xi - \eta)}}
\left\{
\frac{\xi - \eta}{\xi + \eta} {\rm cc}^1_{01} + {\rm cc}^1_{21}
\right\}
\, , \\
{\rm cc}^1_{12}
\!\!\!&\simeq&\!\!\!
\frac{- 2 (1 - y)}{(2 - y)(2 - \widetilde y)}
\frac{\sqrt{(1-\widetilde y) (\xi + \eta) }}{\sqrt{(1 - y)(\xi - \eta)}}
\, {\rm cc}^1_{01}
\, , \nonumber\\
{\rm cc}^1_{32}
\!\!\!&\simeq&\!\!\!
\frac{-2(1-\widetilde y)}{(2 - y)(2 - \widetilde y)}
\frac{\sqrt{(1 - y)(\xi-\eta)}}{\sqrt{(1-\widetilde y)(\xi+\eta)}}\,
{\rm cc}^1_{21}
\, . \nonumber
\end{eqnarray}
We add that charge and angular asymmetries can be combined with double
spin-flip experiments, which offer information on a new combination of
GPDs. As in the case of the beam-spin asymmetry, the number of independent
Compton form factors is, however, reduced to two, as a consequence of
our results (\ref{Def-FC-INT-LP}) and (\ref{Def-FC-INT-TP}).

The charge odd part is given by the interference of the first BH
amplitude with the VCS  as well as with the second BH ones. For
unpolarized settings the charge asymmetry reads
\begin{eqnarray}
d\sigma^{+} - d\sigma^{-}
\propto
{\cal T}^\ast_{{\rm BH}_1} {\cal T}_{{\rm BH}_2}
+
\Re{\rm e} \left( {\cal T}^\ast_{{\rm BH}_1} {\cal T}_{\rm VCS} \right)
\, .
\end{eqnarray}
Taking now  moments with respect to the solid angle of the final state
that are even under reflection, e.g., by means of the weight function
\begin{eqnarray}
w^{\rm even}(\phi_\ell,\theta_\ell)
=
\{
1 , \
\cos\phi_\ell \cos\theta_\ell , \
\cos(2\phi_\ell) , \
\sin\phi_\ell \cos\theta_\ell , \
\sin(2\phi_\ell) , \ \dots \,
\}
\, ,
\end{eqnarray}
the contamination of the BH interferences drops out:
\begin{eqnarray}
\int d {\mit\Omega}_{\ell}\, w^{\rm even} (\phi_\ell,\theta_\ell)
\frac{d \sigma^{+} - d \sigma^{-}}{d {\mit\Omega}_\ell}
\propto
\int d {\mit\Omega}_{\ell}\, w^{\rm even} (\phi_\ell,\theta_\ell)\,
\Re{\rm e} \left(  {\cal T}^\ast_{{\rm BH}_1} {\cal T}_{\rm VCS} \right)
\, .
\end{eqnarray}
Corresponding to the choice of the weight function, this average will
provide Fourier series in $\phi$, where the zeroth, first, second and
third harmonics lead to access to all leading-twist coefficients of
the first interference term. In case when only the lepton beam of a
specified single charge is available, on can form asymmetries with an
odd weight
\begin{eqnarray}
&&w^{\rm odd}(\phi_\ell,\theta_\ell)
\\
&&\quad=
\{
\cos\theta_\ell , \
\cos\varphi_\ell , \
\cos(2\varphi_\ell) \cos\theta_\ell , \
\cos(3\varphi_\ell) , \
\sin\varphi_\ell , \
\sin(2\varphi_\ell) \cos\theta_\ell , \
\sin(3\varphi_\ell) , \
\dots \,
\}
\, , \nonumber
\end{eqnarray}
so that the squared amplitudes exactly drop out
\begin{eqnarray}
\int d {\mit\Omega}_{\ell}\, w^{\rm odd} (\phi_\ell,\theta_\ell)
\frac{d\sigma}{d {\mit\Omega}_{\ell}}
\propto
\int d {\mit\Omega}_{\ell} \, w^{\rm odd}(\phi_\ell,\theta_\ell) \,
\left\{
\pm {\cal T}^\ast_{{\rm BH}_1} {\cal T}_{{\rm BH}_2}
+
\Re{\rm e} \left( {\cal T}^\ast_{{\rm BH}_2} {\cal T}_{\rm VCS} \right)
\right\}
\, .
\end{eqnarray}
After the subtraction of the remaining BH interference is done, one measures
the leading twist-two Fourier coefficients. Still, this procedure may allow
a handle on the real part of the Compton form factors. If both kinds of the
lepton-beam charges are available, the BH contribution drops in the charge
even combination
\begin{eqnarray}
\int d {\mit\Omega}_{\ell}\, w^{\rm odd}(\phi_\ell,\theta_\ell)
\frac{d\sigma^{+} + d\sigma^{-}}{d {\mit\Omega}_\ell}
\propto
\int d {\mit\Omega}_{\ell}\, w^{\rm odd}(\phi_\ell,\theta_\ell)\,
\Re{\rm e} \left( {\cal T}^\ast_{{\rm BH}_2} {\cal T}_{\rm VCS} \right)
\, .
\end{eqnarray}

\begin{figure}[t]
\begin{center}
\mbox{
\begin{picture}(0,140)(270,0)
\put(42,2){\insertfig{8}{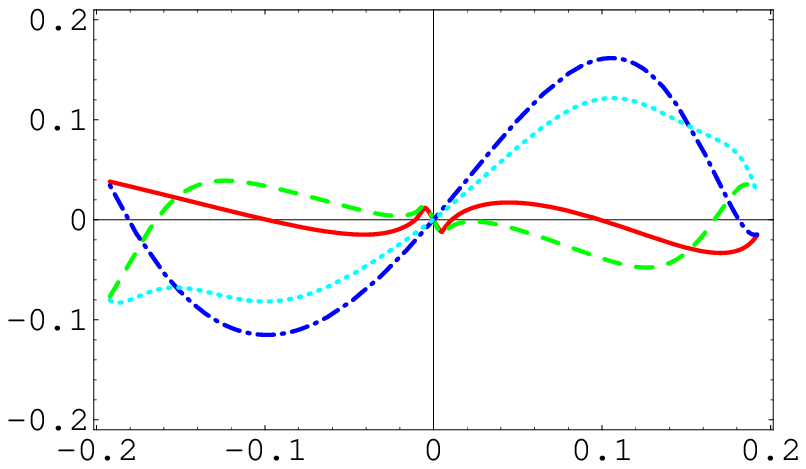}}
\put(25,40){\rotate{$A_{\rm CA}^{\cos\varphi_\ell}(\xi,\eta,\Delta^2)$}}
\put(290,2){\insertfig{8}{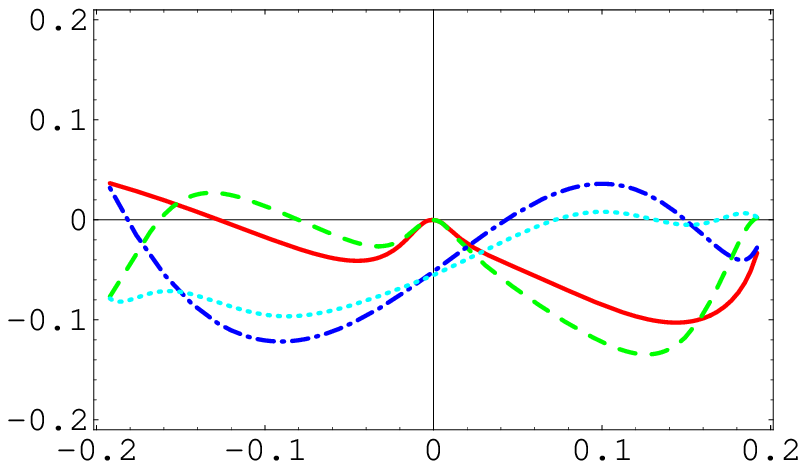}}
\put(275,40){\rotate{$A^{\cos\varphi_\ell}(\xi,\eta,\Delta^2)$}}
\put(230,-10){$\xi$}
\put(485,-10){$\xi$}
\put(78,115){${\Delta^2 = - 0.2\ {\rm GeV}^2 \atop \eta = - 0.2, \ y = 0.5 }$}
\put(442,115){${\Delta^2 = - 0.2\ {\rm GeV}^2 \atop \eta = - 0.2, \ y = 0.5 }$}
\end{picture}
}
\end{center}
\caption{\label{Fig-CA}
The charge asymmetry $A_{\rm CA}^{\cos \varphi_\ell}$ and angular asymmetry
$A^{\cos \varphi_\ell}$ are displayed versus $\xi$ in the left and right panel,
respectively, for the same kinematics as in Fig.\ \ref{Fig-BSA} and different GPD
models, specified in Fig.\ \ref{Fig-CFFs}.}
\end{figure}

To illustrate the feasibility of the subtraction procedure, we consider the
charge and the angular asymmetries
\begin{eqnarray}
\left\{
A_{\rm CA}^{\cos\varphi_{{\ell}}}
\atop
A^{\cos\varphi_{{\ell}}}
\right\}
&\!\!\!=\!\!\!&
\frac{1}{{\cal N}}
\int_{\pi/4}^{3\pi/4}\! d \theta_{\ell}
\int_{0}^{2\pi}\! d \phi
\int_{0}^{2\pi}\! d \varphi_\ell \;
2 \cos\varphi_\ell
\left\{
\left( d \sigma^+ + d \sigma^- \right)/2 d {\mit\Omega}_{\ell}d\phi
\atop
d \sigma^-/d {\mit\Omega}_{\ell}d\phi
\right\}
\, ,
\end{eqnarray}
performed with respect to $2 \cos\varphi_\ell$, where in both cases we
choose the normalization to be
\begin{eqnarray*}
{\cal N}
=
\int_{\pi/4}^{3\pi/4}\! d\theta_{\ell}
\int_{0}^{2\pi}\! d \phi
\int_{0}^{2\pi}\! d \varphi_\ell\,
\frac{d \sigma^-}{d {\mit\Omega}_{\ell} d \phi}
\, .
\end{eqnarray*}
These asymmetries project the Fourier coefficient ${\rm cc}_{10}^2$ of
the second BH-VCS interference term, which is, in the absence of
${\cal F}_L$, proportional to
\begin{eqnarray}
\label{Exp-cc210}
{\rm cc}_{10}^2 \propto \Re{\rm e}\left\{
\frac{\xi}{\eta} F_1 {\cal H}
-
\frac{\xi}{\eta} \frac{\Delta^2}{4M_N^2} F_2 {\cal E}
+
\eta\left(F_1  + F_2 \right) \widetilde {\cal H}
\right\}
\, ,
\end{eqnarray}
making use of Eq.\ (\ref{cc-INT2-10}). One realizes that ${\cal H}$
is now suppressed by a factor $\xi/\eta$ and, thus, with decreasing
$|\xi|$ the contribution of $\widetilde {\cal H}$ starts
to be important. However, for the sake of
simplicity, we will set $\widetilde {\cal H}$ for the illustration
purposes to zero. In Fig.\ \ref{Fig-CA} we display these asymmetries in
two different panels for the same kinematics and GPD models as employed
above for spin asymmetries. Clearly, the shape of these asymmetries is
in any case dictated by the GPD models. The left panel shows the charge
asymmetry. One realizes, that as in Fig.\ (\ref{Fig-CFFs}), the
$D$-term is responsible for the sign change of the asymmetries%
\footnote{We remind the known fact, that for $\xi = \eta$ there is a
competition in sign between the regular valence and sea quark GPDs. The
resulting sign at $\xi = \pm \eta$ is a consequence of chosen
parametrizations, and can not be taken as a clear-cut signature for the
$D$-term contribution.}, when $\xi$ approaches $\mp \eta$: compare solid
(dash-dotted) with dashed (dotted) line. The size of the asymmetries in
the range $ -|\eta| < \xi < |\eta|$ is driven by the parameterization of
the reduced GPD, see also Fig.\ \ref{Fig-CFFs}. We observe that in the
timelike region ($\xi < 0$) both asymmetries are rather similar, while
in the spacelike region the angular asymmetry, compared to the charge
asymmetry, is shifted downwards. This is caused by the interference of
both BH amplitudes, where the first one becomes small, compared to the
VCS amplitude, in the timelike region. Due to the additional power of
$\xi$, indicated in Eq.\ (\ref{Exp-cc210}), the charge asymmetry with
the DD-model goes to zero at $\xi\to 0$. This features will not be
changed, if one includes $\widetilde {\cal H}$, since its real part is
antisymmetric in $\xi$. For the FPD-model both asymmetries go to zero
which is caused by the fact that the normalization $\cal N$ as in the
case of beam-spin asymmetries strongly increase for $\xi\to 0$. We note
that the wriggles around the point $\xi \to 0$ of the solid and dashed
line in the left panel arise from the competition in the numerical
increase of the numerator and denominator.

We remark that as in the case of DVCS charge and angular asymmetries
might be contaminated stronger by twist-three effects than the beam-spin
asymmetries. They are mainly of kinematical origin, i.e., expressed by
twist-two GPDs and generate Fourier coefficients ${\rm cc}^{\rm INT}_{00},\
 {\rm ss}^{\rm INT}_{11} \simeq {\rm cc}^{\rm INT}_{11}$, and
${\rm ss}^{\rm INT}_{22} \simeq {\rm cc}^{\rm INT}_{22}$ that do not
necessarily vanish at the kinematical boundaries.

\subsection{Collider experiments}

Finally, we address the question of whether GPDs can also be measured in the
small-$\eta$ region with lepton-hadron collider experiments. Analogously to
the above Eqs.\ (\ref{ratioDVCS/BH}) and (\ref{ratioDVCS/BH-tim}) one can
estimate the ratio of the BH and VCS cross sections by
\begin{eqnarray}
\label{ratioVCS/BH}
\frac{\int\! d
\varphi_\ell \int\! d\phi\, |{\cal T}^{\rm VCS}|^2} {\int\!
d\varphi_\ell\! \int d\phi|{\cal T}^{\rm BH}|^2} \!\!\!&\sim&\!\!\!
\frac{2\eta^2 |\Delta^2|}{(\xi^2 + \eta^2)({\cal Q}^2 + {\cal
M}^2_{\ell\bar\ell})} \\ &\times&\!\!\! \frac{ 2\eta (1 - y)
\sin^2\theta_\ell }{ 4(\xi + \eta) (1 - y) - y^2 (\xi - \eta) \sin^2
\theta_\ell } \frac{ \eta^2\, |{\cal H} (\xi, \eta, \Delta^2)|^2 (1 +
\dots) }{ F_1(\Delta^2)^2 (1 + \dots) } \, . \nonumber
\end{eqnarray}
Here the ellipses stand for the ratio of terms that arise solely from
the scattering by longitudinally and transversely polarized photons.
They vanish in the DVCS limits and will be omitted since they do not
affect our qualitative considerations. The first kinematical factor,
essentially equal to $|\Delta^2|/({\cal Q}^2 + {\cal
M}^2_{\ell\bar\ell})$, should be much smaller than one to ensure the
applicability of perturbative QCD. The second term is, over a wide
kinematical range, (much) smaller than one, except when one approaches
the spacelike DVCS limit. In this case it goes like $1/y^2$ and gives
the desired kinematical enhancement, which allows to measure the DVCS
cross section at small value of $y$. Staying away from this limit we
already see that the VCS cross section is kinematically suppressed by at
least one order of magnitude, or so, compared to the BH one. Of course,
the behavior of $\eta {\cal H} (\xi, \eta, \Delta^2)$ is completely
unknown in the central region, since it has never been measured so far.
Theoretically, we are facing here new non-perturbative phenomena for
which the connection to forward parton densities is lost. So there is a
lot of space for speculations, which should be clarified by experimental
measurements. If we take the FPD-model, then ${\cal
H}(\xi,\eta,\Delta^2)$ grows like $\xi^{- 1 - \lambda}$ as $\xi$ goes to
zero. Certainly, the kinematical suppression will be overwhelmed by this
rise in the equal mass limit, ${\cal Q}^2 \simeq M_{\ell\bar\ell}^2$. We
have found no plausible arguments of why this should happen. We note,
that if the sea-quark GPD does not vanish at $\xi = 0$ as $x = 0$, the
real part of the Compton form factor will blow up. In case the GPD
is vanishing at this particular point, we still expect that the real
part, given by the integral (\ref{ReaParxi0}) over the whole $x$ range,
goes with $\# \eta^{- 1 - \lambda}$, which follows from rather general
considerations based on the DD-representation of GPDs and the fact that
they have to reduce to the parton densities in the kinematical forward
limit. The available DVCS measurements, analyzed with leading order formulas
of perturbative QCD, tell us that the sea-quark GPD is rather sizeable at
the point $\xi = - \eta$. Consequently, there is only a small amount of
the phase-space left, in which the GPD goes to zero at $x = 0$. The way
this happens will be essentially up to the prefactor $\#$, we do not know.
We conclude, that the only chance to have an experimental handle on the VCS
cross section is in the kinematical region in which $\Delta^2$ is small,
${\cal Q}^2 + {\cal M}^2_{\ell\bar\ell}$ should be several times larger
than the average $\langle\langle -\Delta^2 \rangle\rangle$ and at least
one or two ${\rm GeV}^2$ and $y$ should rather large to ensure that
$|\eta|$ is small. For instance, in the case of H1 and ZEUS experiments
we find for $y = 0.5,\ \eta= -2.2 \times 10^{-5},\ \langle\langle \Delta^2
\rangle\rangle = -0.1\ {\rm GeV}^2$ and $\theta_{\ell}=\pi/4$ within the
DD model that the ratio (\ref{ratioVCS/BH}) is of order one for a large
interval of $- |\eta|/2 < \xi \le |\eta|$. Similar estimates can be
done for the charge and angular asymmetries, where one should project
the Fourier coefficient $c^2_{01}$, which, compared to $c^2_{10}$, is
not plagued from an additional suppression factor $\xi/\eta$.

\section{Conclusions}

In the present paper we have studied the process $e N \to e' N' \ell
\bar\ell$. We have elaborated the structure of the cross section to
leading power in the hard momentum. The power suppressed twist-three
contributions, though have not been discussed presently, generate
further harmonics in the squared amplitude. For instance, they induce
the off-diagonal elements in the coefficient matrix of the squared VCS
amplitude, e.g., ${\rm cc}_{nm}$ with $n \neq m$ etc., however, they
will not contaminate already existing Fourier harmonics, e.g., ${\rm
cc}_{nn}$ etc. This process, we have discussed, is the most favorable
for experimental measurements of GPDs by a number of reasons:
\begin{itemize}
\item it is a clean electromagnetic process which does not involve other
unknown non-perturbative function and, thus, has no contamination from other
unknown sources;
\item the virtuality of the final state photon allows to disentangle the
dependence of GPDs on both scaling variables
and thus constrain the angular momentum sum rule;
\item studies of the angular dependence of the recoiled nucleon and of the
lepton pair are complimentary and lead to a rich angular structure of the cross
section that can be used for separation of diverse combinations of GPDs;
\item variation of the relative magnitude of space- and timelike photon
virtualities allows to access distributions of partons and anti-partons in the
``exclusive" domain $\xi > |\eta|$.
\item the higher one goes in skewness $\eta$, whose maximal value is limited
by the magnitude of the momentum transfer to be within the region of
applicability of QCD factorization $|\Delta^2| \ll p \cdot q$, the more surface
in the exclusive domain ($|\eta| > |\xi|$) one measures in experiment. This
diminishes the uncertainty coming from the inaccessible inclusive sector
($|\eta| < |\xi|$). The exclusive domain might saturate the spin sum rule
(\ref{SpinSumRule}) even for moderate $\eta$ since the second moment required
for it is not extremely sensitive to the large-$\xi$ behavior of GPDs where
the latter is known to decrease according to the quark-counting power law
of conventional parton distributions.
\end{itemize}
Another interesting feature of this process is that the zero value of
generalized Bjorken variable can be exactly attained when the incoming
and outgoing photons have about the same absolute values of virtualities
${\cal Q}^2 \simeq M_{\ell\bar\ell}^2$.

We have presently discussed the most favorable observables, namely,
diverse lepton-spin and azimuthal asymmetries that are sensitive to the
imaginary part of the Compton form factors and, thus, directly to
GPDs. We have not discussed in full, however, phenomenological consequences
of polarized targets, though, we have derived the complete set of formulas
with explicit angular dependence which can be used to extract complimentary
combinations of Compton form factors from experimental data. Longitudinal
and transverse nucleon-spin asymmetries combined with the Fourier analysis
will serve this purpose analogous to the lepton-spin asymmetries addressed
presently. For the complete analysis along this line of exclusive
electroproduction of a real photon see Ref.\ \cite{BelMulKir01}. The
process $e N \to e' N' \ell \bar\ell$ with both photons being virtual is
unique due to the independence of skewness $\eta$ from the generalized
Bjorken variable $\xi$. Unfortunately, it suffers from very low cross
sections, however, this drawback will be circumvented with future
high-luminosity machines. The current analysis of available events from
CLAS detector at Jefferson Laboratory is under way \cite{Gar03}.

\vspace{0.5cm}

We would like to thank A.V. Radyushkin for useful discussions on the
factorization issue and M.~Diehl, M. Guidal, and M. Vanderhaeghen for
discussions on phenomenological aspects of the paper. We thank the
Department of Energy's Institute for Nuclear Theory at the University
of Washington for its hospitality during the program ``Generalized
parton distributions and hard exclusive processes'' and the Department
of Energy for the partial support during the completion of this paper.
The present work was supported by the US Department of Energy under
contract DE-FG02-93ER40762.


\appendix

\setcounter{section}{0}
\setcounter{equation}{0}
\renewcommand{\theequation}{\Alph{section}.\arabic{equation}}

\section{Light-cone vectors}
\label{LCvectors}

The vectors defining the kinematics of the process can be used to construct
a pair of the light-cone vectors $n_\mu$ and $n^\star_\mu$, such that
$n^2 = n^{\star 2} = 0$ and $n \cdot n^\star = 1$, as follows
\begin{eqnarray}
n_\mu
\!\!\!&=&\!\!\!
\frac{
2 \xi
}{
Q^2 \sqrt{1 + 4 (\xi \delta)^2}
} \, q_\mu
-
\frac{
1 - \sqrt{1 + 4 (\xi \delta)^2}
}{
2 Q^2 \delta^2 \sqrt{1 + 4 (\xi \delta)^2}
} \, p_\mu
\, , \\
n^\star_\mu
\!\!\!&=&\!\!\!
- \frac{
\xi \delta^2
}{
\sqrt{1 + 4 (\xi \delta)^2}
} \, q_\mu
+
\frac{
1 + \sqrt{1 + 4 (\xi \delta)^2}
}{
4 \sqrt{1 + 4 (\xi \delta)^2}
} \, p_\mu \, ,
\nonumber
\end{eqnarray}
where $\delta^2 \equiv (M_N^2 - \Delta^2/4)/Q^2$. Throughout the paper we
use the following decomposition of a given four-vector $v_\mu$ in its
light-cone components
\begin{equation}
v_\mu = v_- n_\mu + v_+ n^\ast_\mu + v_{\perp \mu} \, ,
\end{equation}
so that a scalar product is written as
$v \cdot u = v_+ u_- + v_- u_+ - \bit{v}_{\perp} \cdot \bit{u}_{\perp}$.

The light-cone decomposition of the momenta reads
\begin{eqnarray}
p_\mu
\!\!\!&=&\!\!\!
2 n^\star_\mu + Q^2 \delta^2 n_\mu
\, , \\
q_\mu
\!\!\!&=&\!\!\!
-
\frac{
2 \xi
}{
1 + \sqrt{1 + 4 \left( \xi \delta \right)^2}
} n^\star_\mu
-
\frac{
\xi \, Q^2 \delta^2
}{
1 - \sqrt{1 + 4 \left( \xi \delta \right)^2}
} n_\mu
\, , \nonumber\\
\Delta_\mu
\!\!\!&=&\!\!\!
\frac{
2 \eta
}{
\sqrt{1 + 4 \left( \xi \delta \right)^2}
} n^\star_\mu
-
\frac{
\eta \, Q^2 \delta^2
}{
\sqrt{1 + 4 \left( \xi \delta \right)^2}
}
n_\mu
+
\Delta^\perp_\mu
\, . \nonumber
\end{eqnarray}
Setting $\Delta_\perp = 0$ we get from the last equation the minimal value of
$\Delta^2$, namely,
\begin{equation}
\Delta_{\rm min}^2
=
\frac{Q^2}{2 \xi^2}
\left\{
1 - \eta^2 + 4 M_N^2 \xi^2/Q^2
-
\sqrt{
\left( 1 - \eta^2 + 4 M_N^2 \xi^2/Q^2 \right)^2
+
16 M_N^2 \xi^2 \eta^2/Q^2
}
\right\} \, .
\end{equation}

\setcounter{equation}{0}
\renewcommand{\theequation}{\Alph{section}.\arabic{equation}}

\section{Results for polarized target}
\label{App-PolTar}

In the body of the paper we have given the results for the unpolarized
nucleon target only. In the subsequent two appendices we fill the gap
left and extend them by including polarization. We parametrize the
polarization vector of the nucleon by a polar and an azimuthal angles
${\mit\Theta}$ and ${\mit\Phi}$, respectively. In the target rest frame
TRF-I it has the form (see Fig.\ \ref{LeptonPairKinetic})
\begin{equation}
\label{Spin}
S
=
(0, \sin{\mit\Theta} \cos{\mit\Phi}, \sin{\mit\Theta} \sin{\mit\Phi},
\cos{\mit\Theta}) .
\end{equation}
Introduction of the transverse polarization results into addition of an
extra integration variable in the phase space given by Eqs.\ (\ref{LIPS4})
and (\ref{LIPS3-Phi}),
\begin{equation}
d {\rm LIPS}_4 \to  \frac{d {\mit\Phi}}{2 \pi} \times d {\rm LIPS}_4
\, .
\end{equation}
The Fourier coefficients
${\rm ab}_{nm}^i = ({\rm cc}_{nm}^i, {\rm cs}_{nm}^i, {\rm sc}_{nm}^i,
{\rm ss}_{nm}^i)$ depend on the nucleon polarization vector (\ref{Spin})
and admit the decomposition
\begin{equation}
\label{Def-DecFC}
{\rm ab}_{nm}^i
=
{\rm ab}_{nm, {\rm unp}}^i
+
\cos {\mit\Theta} \,
{\rm ab}_{nm, {\rm LP}}^i
+
\sin {\mit\Theta} \,
{\rm ab}_{nm, {\rm TP}}^i ({\mit\Phi})
\, .
\end{equation}

\subsection{Squared VCS amplitude}
\label{C-coef}

The general structure of products of Compton form factors reads
\begin{eqnarray}
\label{Def-VV-full}
\frac{1}{4} {\cal V} {\cal V}^\dagger
\!\!\!&\equiv&\!\!\!
{\cal C}^{\rm VCS}_{{\cal VV}, {\rm unp}}
+
{\cal C}^{\rm VCS}_{{\cal VV}, {\rm LP}}
\cos{\mit\Theta}
+
i
\sqrt{- \frac{\Delta^2}{4M^2}}
\sqrt{1 - \frac{\Delta^2_{\rm min}}{\Delta^2}}
\sqrt{ \frac{1 + \eta}{1 - \eta} } \,
{\cal C}^{\rm VCS}_{{\cal VV}, {\rm TP}}
\sin \left( {\mit\Phi} - \phi \right) \sin{\mit\Theta}
\, , \\
\nonumber\\
\label{Def-VA-full}
\frac{1}{4} {\cal V} {\cal A}^\dagger
\!\!\!&\equiv&\!\!\!
{\cal C}^{\rm VCS}_{{\cal VA}, {\rm unp}}
+
{\cal C}^{\rm VCS}_{{\cal VA}, {\rm LP}}
\cos{\mit\Theta}
+
\sqrt{- \frac{\Delta^2}{4M^2}}
\sqrt{1 - \frac{\Delta^2_{\rm min}}{\Delta^2}}
\sqrt{ \frac{1 + \eta}{1 - \eta} } \,
{\cal C}^{\rm VCS}_{{\cal VA}, {\rm TP}}
\cos \left( {\mit\Phi} - \phi \right) \sin{\mit\Theta}
\, ,
\end{eqnarray}
with ${\cal AA}^\dagger$ product having the decomposition similar to Eq.\
(\ref{Def-VV-full}). Here the polar ${\mit\Theta}$ and azimuthal ${\mit\Phi}$
angles parametrize the nucleon polarization vector $S$ according to Eq.\
(\ref{Spin}). The nonvanishing ${\cal C}$-coefficients in the squared VCS
amplitude, defined in Eq.\ (\ref{Def-VV-full}-\ref{Def-VA-full}) read for
the polarized nucleon target
\begin{eqnarray}
{\cal C}^{\rm VCS}_{{\cal VA}, {\rm LP}} ({\cal F}, {\cal F}^\ast)
\!\!\!&=&\!\!\!
(1 - \eta^2) {\cal H} \widetilde {\cal H}^\ast
-
\eta^2 ({\cal H} \widetilde {\cal E}^\ast
+
{\cal E} \widetilde {\cal H}^\ast)
+
\eta
\left(
\frac{\Delta^2}{4 M^2} + \frac{\eta^2}{1 - \eta}
\right)
{\cal E} \widetilde{\cal E}^\ast
\, , \\
{\cal C}^{\rm VCS}_{{\cal VV}, {\rm TP}} ({\cal F}, {\cal F}^\ast)
\!\!\!&=&\!\!\!
(1 - \eta)
\left(
{\cal H} {\cal E}^\ast
-
{\cal E} {\cal H}^\ast
\right)
\, , \\
{\cal C}^{\rm VCS}_{{\cal AA}, {\rm TP}} ({\cal F}, {\cal F}^\ast)
\!\!\!&=&\!\!\!
(1 - \eta) \eta
\left(
\widetilde {\cal H} \widetilde {\cal E}^\ast
-
\widetilde  {\cal E} \widetilde {\cal H}^\ast
\right)
\, , \\
{\cal C}^{\rm VCS}_{{\cal VA}, {\rm TP}} ({\cal F}, {\cal F}^\ast)
\!\!\!&=&\!\!\!
- (1 - \eta)
\left(
\eta\, {\cal H} \widetilde {\cal E}^\ast
+
{\cal E} \widetilde {\cal H}^\ast
\right)
+
\eta^2 {\cal E} \widetilde{\cal E}^\ast
\, .
\end{eqnarray}
We note that with this definitions, we have the following correspondence with those
given in Ref.\ \cite{BelMulKir01} for polarized target in the DVCS limit $\eta = -
\xi$
\begin{eqnarray}
\label{Cor-DVCS2}
{\cal C}^{\rm DVCS}_{{\rm LP}} ({\cal F}, {\cal F}^\ast)
\!\!\!&\stackrel{\mbox{\tiny DVCS}}{=}&\!\!\!
{\cal C}^{\rm VCS}_{{\cal VA}, {\rm LP}} ({\cal F}, {\cal F}^\ast)
+
{\cal C}^{\rm VCS}_{{\cal VA}, {\rm LP}} ({\cal F}^\ast, {\cal F})
\, , \nonumber\\
{\cal C}^{\rm DVCS}_{{\rm TP}+} ({\cal F}, {\cal F}^\ast)
\!\!\!&\stackrel{\mbox{\tiny DVCS}}{=}&\!\!\!
{\cal C}^{\rm VCS}_{{\cal VA}, {\rm TP}} ({\cal F}, {\cal F}^\ast)
+
{\cal C}^{\rm VCS}_{{\cal VA}, {\rm TP}} ({\cal F}^\ast, {\cal F})
\, , \\
{\cal C}^{\rm DVCS}_{{\rm TP}-} ({\cal F}, {\cal F}^\ast)
\!\!\!&\stackrel{\mbox{\tiny DVCS}}{=}&\!\!\!
{\cal C}^{\rm VCS}_{{\cal VV}, {\rm TP}} ({\cal F}, {\cal F}^\ast)
+
{\cal C}^{\rm VCS}_{{\cal AA}, {\rm TP}} ({\cal F}, {\cal F}^\ast)
\, . \nonumber
\end{eqnarray}
The remaining Fourier coefficients read for longitudinally polarized target
\begin{eqnarray}
{\rm cc}^{\rm VCS}_{00,{\rm LP}}
\!\!\!&=&\!\!\!
2 \lambda (2 - y) y (2 - 2 \widetilde y + \widetilde y^2)
\left\{
{\cal C}^{\rm VCS}_{{\cal VA}, {\rm LP}} ({\cal F}, {\cal F}^\ast)
+
{\cal C}^{\rm VCS}_{{\cal VA}, {\rm LP}} ({\cal F}^\ast, {\cal F})
\right\}
\, , \nonumber\\
{\rm cc}^{\rm VCS}_{11,{\rm LP}}
\!\!\!&=&\!\!\! \lambda
\frac{4  \sigma}{\xi}
y (2 - \widetilde y)
\sqrt{(1 - y)(1 - \widetilde y)(\xi^2 - \eta^2)}
\left\{
{\cal C}^{\rm VCS}_{{\cal VA}, {\rm LP}} ({\cal F}_L, {\cal F}^\ast)
+
{\cal C}^{\rm VCS}_{{\cal VA}, {\rm LP}} ({\cal F}_L^\ast, {\cal F})
\right\}
\, ,
\end{eqnarray}
and for transversely polarized target
\begin{eqnarray}
\label{Def-FC-TP}
{\rm cc}^{\rm VCS}_{00,{\rm TP}}
\!\!\!&=&\!\!\!
\sqrt{-\frac{\Delta^2}{M^2}}
\sqrt{ 1 - \frac{\Delta^2_{\rm min}}{\Delta^2} }
\sqrt{\frac{1 + \eta}{1 - \eta}}
\nonumber \\
&&\!\!\!
\times
\Bigg\{
\lambda
\cos \left( {\mit\Phi} - \phi \right)
(2 - y) y (2 - 2 \widetilde y + \widetilde y^2)
\left\{
{\cal C}^{\rm VCS}_{{\cal VA},{\rm TP}} ({\cal F},{\cal F}^\ast)
+
{\cal C}^{\rm VCS}_{{\cal VA},{\rm TP}} ({\cal F}^\ast,{\cal F})
\right\}
\nonumber\\
&&\quad + \
i \sin \left( {\mit\Phi} - \phi \right)
(2 - 2 y + y^2) (2 - 2 \widetilde y + \widetilde y^2)
\left\{
{\cal C}^{\rm VCS}_{{\cal VV},{\rm TP}} ({\cal F},{\cal F}^\ast)
+
{\cal C}^{\rm VCS}_{{\cal AA},{\rm TP}} ({\cal F},{\cal F}^\ast)
\right\}
\nonumber\\
&&\quad + \
8 i \sin \left( {\mit\Phi} - \phi \right)
(1 - y) (1 - \widetilde y) \frac{\xi^2 - \eta^2}{\xi^2}
{\cal C}^{\rm VCS}_{{\cal VV}, {\rm TP}} ({\cal F}_L, {\cal F}^\ast_L)
\Bigg\}
\, , \\
{\rm cc}^{\rm VCS}_{11,{\rm TP}}
\!\!\!&=&\!\!\!
\sqrt{-\frac{\Delta^2}{M^2}}
\sqrt{ 1 - \frac{\Delta^2_{\rm min}}{\Delta^2} }
\sqrt{\frac{1 + \eta}{1 - \eta}}\,
\frac{2 \sigma}{\xi}  \sqrt{(1 - y)(1 - \widetilde y)(\xi^2 - \eta^2)}
\nonumber\\
&&\!\!\!
\times
\Bigg\{
\lambda
\cos \left( {\mit\Phi} - \phi \right)
y (2 - \widetilde y)
\left\{
{\cal C}^{\rm VCS}_{{\cal VA}, {\rm TP}} ({\cal F}_L, {\cal F}^\ast)
+
{\cal C}^{\rm VCS}_{{\cal VA}, {\rm TP}} ({\cal F}_L^\ast, {\cal F})
\right\}
\nonumber\\
&&\;\;+
i
\sin \left( {\mit\Phi} - \phi \right)
(2 - y) (2 - \widetilde y)
\left\{
{\cal C}^{\rm VCS}_{{\cal VV}, {\rm TP}} ({\cal F}_L, {\cal F}^\ast)
+
{\cal C}^{\rm VCS}_{{\cal VV}, {\rm TP}} ({\cal F}_L^\ast, {\cal F})
\right\}
\Bigg\}
\, ,\\
{\rm cc}^{\rm VCS}_{22,{\rm TP}}
\!\!\!&=&\!\!\!
\sqrt{-\frac{\Delta^2}{M^2}}
\sqrt{ 1 - \frac{\Delta^2_{\rm min}}{\Delta^2} }
\sqrt{\frac{1 + \eta}{1 - \eta}}
\nonumber\\
&&\!\!\!\times 4 i \sin \left( {\mit\Phi} - \phi \right)
(1 - y) (1 - \widetilde y)
\left\{
{\cal C}^{\rm VCS}_{{\cal VV}, {\rm TP}} ({\cal F}, {\cal F}^\ast)
-
{\cal C}^{\rm VCS}_{{\cal AA}, {\rm TP}} ({\cal F}, {\cal F}^\ast)
\right\}
\, ,
\end{eqnarray}
where according to the general twist-two relation (\ref{Rel-FC-VCS2-Tw2})
\begin{equation}
{\rm ss}^{\rm VCS}_{11,{\rm LP}/{\rm TP}}
\simeq
{\rm cc}^{\rm VCS}_{11,{\rm LP}/{\rm TP}}
\, , \qquad
{\rm ss}^{\rm VCS}_{22,{\rm TP}}
\simeq
{\rm cc}^{\rm VCS}_{22,{\rm TP}}
\, .
\end{equation}

\subsection{Interference term}
\label{App-IntPolTar}

Similarly to the previous appendix, we present here the results for the
interference term on a polarized target.
\begin{eqnarray}
\label{Def-trace-SV}
\left\{
\!\!\!
\begin{array}{c}
{\cal S}_1
\\
{\cal S}_2
\end{array}
\!\!\!
\right\}\!
{\cal V}
\!\!\!&\equiv&\!\!\!
4 Q^2 \frac{(1 - \eta) \eta}{y \widetilde y \xi}
\left(
\left\{
\!\!\!
\begin{array}{c}
\widetilde y K
\\
y \widetilde K
\end{array}
\!\!\!
\right\} \left[
-{\cal C}_{{\cal V},{\rm unp}} ({\cal F})
\left\{
\!\!\!
\begin{array}{c}
\cos \phi
\\
\cos \varphi_\ell
\end{array}
\!\!\!
\right\}
-
\cos{\mit\Theta} \,
i {\cal C}_{{\cal V},{\rm LP}} ({\cal F})
\left\{
\!\!\!
\begin{array}{c}
\sin \phi
\\
\sin \varphi_\ell
\end{array}
\!\!\!
\right\}
\right] \right.
\\
&-&\!\!\!
\left. \sin{\mit\Theta}
\left\{
\!\!\!
\begin{array}{c}
\widetilde y L
\\
y \widetilde L
\end{array}
\!\!\!
\right\} \left[
i {\cal C}_{{\cal V},{\rm TP}+} ({\cal F})
\left\{
\!\!\!
\begin{array}{c}
\sin \phi
\\
\sin \varphi_\ell
\end{array}
\!\!\!
\right\}
\cos ({\mit\Phi} - \phi)
+
i {\cal C}_{{\cal V},{\rm TP}-} ({\cal F})
\left\{
\!\!\!
\begin{array}{c}
\cos \phi
\\
\cos \varphi_\ell
\end{array}
\!\!\!
\right\}
\sin ({\mit\Phi} - \phi)
\right] \right)
\, ,
\nonumber\\
\nonumber\\
\label{Def-trace-RV}
\left\{
\!\!\!
\begin{array}{c}
{\cal R}_1
\\
{\cal R}_2
\end{array}
\!\!\!
\right\}\!
{\cal V}
\!\!\!&\equiv&\!\!\!
4 Q^2 \frac{(1 - \eta) \eta}{y \widetilde y \xi}
\left(
\left\{
\!\!\!
\begin{array}{c}
\widetilde y K
\\
y \widetilde K
\end{array}
\!\!\!
\right\}
\left[
i {\cal C}_{{\cal V}, {\rm unp}} (\cal F)
\left\{
\!\!\!
\begin{array}{c}
\sin \phi
\\
\sin \varphi_\ell
\end{array}
\!\!\!
\right\}
+
\cos{\mit\Theta} \,
{\cal C}_{{\cal V}, {\rm LP}} (\cal F)
\left\{
\!\!\!
\begin{array}{c}
\cos \phi
\\
\cos \varphi_\ell
\end{array}
\!\!\!
\right\}\right]\right.
\\
&+&\!\!\!
\left. \sin{\mit\Theta}
\left\{
\!\!\!
\begin{array}{c}
\widetilde y L
\\
y \widetilde L
\end{array}
\!\!\!
\right\} \left[
{\cal C}_{{\cal V},{\rm TP}+} ({\cal F})
\left\{
\!\!\!
\begin{array}{c}
\cos \phi
\\
\cos \varphi_\ell
\end{array}
\!\!\!
\right\}
\cos ({\mit\Phi} - \phi)
-
{\cal C}_{{\cal V},{\rm TP}-} ({\cal F})
\left\{
\!\!\!
\begin{array}{c}
\sin \phi
\\
\sin \varphi_\ell
\end{array}
\!\!\!
\right\}
\sin({\mit\Phi} - \phi)
\right] \right)
\, ,
\nonumber\\
\nonumber\\
\label{Def-trace-SA}
\left\{
\!\!\!
\begin{array}{c}
{\cal S}_1
\\
{\cal S}_2
\end{array}
\!\!\!
\right\}\!
{\cal A}
\!\!\!&\equiv&\!\!\!
4  Q^2 \frac{(1 - \eta) \eta}{y \widetilde y \xi}
\left(\left\{
\!\!\!
\begin{array}{c}
\widetilde y K
\\
y \widetilde K
\end{array}
\!\!\!
\right\}
\left[
- i {\cal C}_{{\cal A},{\rm unp}} ({\cal F})
\left\{
\!\!\!
\begin{array}{c}
\sin \phi
\\
\sin \varphi_\ell
\end{array}
\!\!\!
\right\}
-
\cos{\mit\Theta} \,
{\cal C}_{{\cal A},{\rm LP}} ({\cal F})
\left\{
\!\!\!
\begin{array}{c}
\cos \phi
\\
\cos \varphi_\ell
\end{array}
\!\!\!
\right\}
\right]\right.
\\
&-&\!\!\!
\left. \sin{\mit\Theta}
\left\{
\!\!\!
\begin{array}{c}
\widetilde y L
\\
y \widetilde L
\end{array}
\!\!\!
\right\} \left[
{\cal C}_{{\cal A},{\rm TP}+} ({\cal F})
\left\{
\!\!\!
\begin{array}{c}
\cos \phi
\\
\cos \varphi_\ell
\end{array}
\!\!\!
\right\}
\cos ({\mit\Phi} - \phi)
-{\cal C}_{{\cal A},{\rm TP}-} ({\cal F})
\left\{
\!\!\!
\begin{array}{c}
\sin \phi
\\
\sin \varphi_\ell
\end{array}
\!\!\!
\right\}
\sin ({\mit\Phi} - \phi)
\right] \right)
\, ,
\nonumber\\
\nonumber\\
\label{Def-trace-RA}
\left\{
\!\!\!
\begin{array}{c}
{\cal R}_1
\\
{\cal R}_2
\end{array}
\!\!\!
\right\}\!
{\cal A}
\!\!\!&\equiv&\!\!\!
4 Q^2 \frac{(1 - \eta) \eta}{y \widetilde y \xi}
\left(
\left\{
\!\!\!
\begin{array}{c}
\widetilde y K
\\
y \widetilde K
\end{array}
\!\!\!
\right\} \left[
{\cal C}_{{\cal A},{\rm unp}} ({\cal F})
\left\{
\!\!\!
\begin{array}{c}
\cos \phi
\\
\cos \varphi_\ell
\end{array}
\!\!\!
\right\}
+
\cos {\mit\Theta} \,
i {\cal C}_{{\cal A},{\rm LP}} ({\cal F})
\left\{
\!\!\!
\begin{array}{c}
\sin \phi
\\
\sin \varphi_\ell
\end{array}
\!\!\!
\right\}
\right]\right.
\\
&+&\!\!\!
\left. \sin{\mit\Theta}
\left\{
\!\!\!
\begin{array}{c}
\widetilde y L
\\
y \widetilde L
\end{array}
\!\!\!
\right\} \left[
i {\cal C}_{{\cal A},{\rm TP}+} ({\cal F})
\left\{
\!\!\!
\begin{array}{c}
\sin \phi
\\
\sin \varphi_\ell
\end{array}
\!\!\!
\right\}
\cos ({\mit\Phi} - \phi)
+
i {\cal C}_{{\cal A},{\rm TP}-} ({\cal F})
\left\{
\!\!\!
\begin{array}{c}
\cos \phi
\\
\cos \varphi_\ell
\end{array}
\!\!\!
\right\}
\sin ({\mit\Phi} - \phi)
\right] \right)
\, ,
\nonumber
\end{eqnarray}
where we used in analogy to the definitions (\ref{Def-K})
the shorthand notation
\begin{equation}
\label{Def-L}
\left\{
\begin{array}{c}
L
\\
\widetilde L
\end{array}
\right\}
\approx
 - \frac{1}{2\eta}  \sqrt{\frac{\xi M^2}{Q^2}}
\left\{
\begin{array}{c}
\sqrt{(1-y)(\xi-\eta)}
\\
\sqrt{(1-\widetilde y)(\xi+\eta)}
\end{array}
\right\}
\end{equation}

For the longitudinally polarized nucleon, we get the following combinations of
the electromagnetic and Compton form factors:
\begin{equation}
\label{Def-C-int-LP}
{\cal C}_{{\cal V}, {\rm LP}}
=
-\eta (F_1 + F_2)
\left(
{\cal H} - \frac{\eta}{1 - \eta} {\cal E}
\right)
\, , \qquad
{\cal C}_{{\cal A}, {\rm LP}}
=
 F_1 \widetilde {\cal H}
-
\eta
\left(
\frac{\eta}{1 - \eta} F_1 - \frac{\Delta^2}{4 M_N^2} F_2
\right)
\widetilde {\cal E}
\, ,
\end{equation}
while for the transversal case we have four more combinations
\begin{eqnarray}
\label{Def-C-int-TP-V+}
{\cal C}_{{\cal V}, {\rm TP}+}
\!\!\!&=&\!\!\!
\frac{2 \eta}{1-\eta} (F_1 + F_2)
\left\{
\eta
\left(
{\cal H} - \frac{\eta}{1 - \eta} {\cal E}
\right)
- \frac{\Delta^2}{4 M_N^2} {\cal E}
\right\}
\, , \\
\label{Def-C-int-TP-A+}
{\cal C}_{{\cal A}, {\rm TP}+}
\!\!\!&=&\!\!\!
-
\frac{2}{1 - \eta}
\left\{
\eta^2 F_1
\left(
\widetilde {\cal H} -  \frac{\eta}{1 - \eta} \widetilde {\cal E}
\right)
- \frac{\Delta^2}{4 M_N^2}
\left(
(1 - \eta^2) F_2 \widetilde {\cal H}
+
\eta (F_1 - \eta F_2) \widetilde {\cal E}
\right)
\right\}
\, ,\\
\label{Def-C-int-TP-V-}
{\cal C}_{{\cal V}, {\rm TP}-}
\!\!\!&=&\!\!\!
\frac{2}{1 - \eta}
\left\{
\eta^2 F_1
\left(
{\cal H} + {\cal E}
\right)
-
\frac{\Delta^2}{4 M_N^2}
\left(
(1 - \eta^2) F_2 {\cal H} - (F_1 + \eta^2 F_2) {\cal E}
\right)
\right\}
\, ,\\
\label{Def-C-int-TP-A-}
{\cal C}_{{\cal A}, {\rm TP}-}
\!\!\!&=&\!\!\!
-
\frac{2\eta^2}{1 - \eta} (F_1 + F_2)
\left\{
\widetilde  {\cal H} + \frac{\Delta^2}{4 M_N^2} \widetilde {\cal E}
\right\}
\, .
\end{eqnarray}

First, we display the explicit form of the Fourier coefficient in the
interference term with the second BH process for an unpolarized nucleon
target, which we deduced by symmetry considerations in the main text. They are
\begin{eqnarray}
\label{cc-INT2-01}
{\rm cc}^2_{01,{\rm unp}}
\!\!\!&=&\!\!\!
8 K (2 - y) (1 - \widetilde y) (2 - \widetilde y)
\frac{\xi + \eta}{\eta} \Re{\rm e}\!
\left\{\!
{\cal C}_{{\cal V},{\rm unp}}({\cal F})
-
{\cal C}_{{\cal A},{\rm unp}}({\cal F})
-
\frac{\xi - \eta}{\xi} {\cal C}_{{\cal V},{\rm unp}} ({\cal F}_L)
\right\}\!
, \\
{\rm cs}^2_{01,{\rm unp}}
\!\!\!&=&\!\!\!
8 \lambda K  y (1 - \widetilde y) (2 - \widetilde y)
\frac{\xi + \eta}{\eta}
\; \Im{\rm m}\!
\left\{\!
-
{\cal C}_{{\cal V},{\rm unp}}({\cal F})
+
{\cal C}_{{\cal A},{\rm unp}}({\cal F})
-
\frac{\xi - \eta}{\xi} {\cal C}_{{\cal V},{\rm unp}}({\cal F}_L)
\right\}\!
, \\
\label{cc-INT2-10}
{\rm cc}^2_{10,{\rm unp}}
\!\!\!&=&\!\!\!
- 8 \widetilde K \; \Re{\rm e}
\left\{
(2 - 2 y + y^2) (2 - 2 \widetilde y + \widetilde y^2)
\left(
\frac{\xi}{\eta} {\cal C}_{{\cal V}, {\rm unp}} ({\cal F})
-
{\cal C}_{{\cal A},{\rm unp}}({\cal F})
\right)
\right.
\nonumber \\
&&\qquad\qquad
\left. - 8 (1 - y)(1 - \widetilde y)
\frac{\xi^2 - \eta^2}{\eta \xi} {\cal C}_{{\cal V},{\rm unp}} ({\cal F}_L)
\right\}
, \\
{\rm cc}^2_{12,{\rm unp}}
\!\!\!&=&\!\!\!
- 16 \widetilde K  (1 -  y)(1 - \widetilde y)
\frac{\xi - \eta}{\eta}
\Re{\rm e}
\left\{
{\cal C}_{{\cal V},{\rm unp}} ({\cal F})
-
{\cal C}_{{\cal A},{\rm unp}}({\cal F})
\right\}
, \\
{\rm sc}^2_{10,{\rm unp}}
\!\!\!&=&\!\!\!
8 \lambda \widetilde K y (2 - y) (2 - 2 \widetilde y + \widetilde y^2)
\Im{\rm m}
\left\{
{\cal C}_{{\cal V},{\rm unp}}({\cal F})
-
\frac{\xi}{\eta} {\cal C}_{{\cal A},{\rm unp}}({\cal F})
\right\}
, \\
{\rm cc}^2_{21,{\rm unp}}
\!\!\!&=&\!\!\!
8 K (2 - y) (1 - \widetilde y) (2 - \widetilde y)
\frac{\xi + \eta}{\eta} \Re{\rm e}\!
\left\{\!
{\cal C}_{{\cal V},{\rm unp}} ({\cal F})
+
{\cal C}_{{\cal A},{\rm unp}} ({\cal F})
-
\frac{\xi + \eta}{\xi} {\cal C}_{{\cal V},{\rm unp}} ({\cal F}_L)
\right\}\!
, \\
{\rm sc}^2_{21,{\rm unp}}
\!\!\!&=&\!\!\!-
8 \lambda K  y (1 - \widetilde y) (2 - \widetilde y)
\frac{\xi + \eta}{\eta}
\; \Im{\rm m}\!
\left\{\!
{\cal C}_{{\cal V},{\rm unp}}({\cal F})
+
{\cal C}_{{\cal A},{\rm unp}}({\cal F})
+
\frac{\xi + \eta}{\xi} {\cal C}_{{\cal V},{\rm unp}}({\cal F}_L)
\right\},
\\
\label{cc-INT2-32}
{\rm cc}^2_{32,{\rm unp}}
\!\!\!&=&\!\!\!
- 16 \widetilde K  (1 -  y)  (1 - \widetilde y)
\frac{\xi + \eta}{\eta}
\Re{\rm e}
\left\{
{\cal C}_{{\cal V}, {\rm unp}}({\cal F})
+
{\cal C}_{{\cal A},{\rm unp}} ({\cal F})
\right\}
.
\end{eqnarray}

Next we consider the Fourier coefficients for a polarized target. From the
results (\ref{Def-trace-SV})-(\ref{Def-trace-RA}) we immediately read off
several relations which allows us to obtain the expression for the Fourier
coefficients making use of simple substitution rules in the unpolarized case
(\ref{cc-INT1-01})-(\ref{cc-INT1-32}) and (\ref{cc-INT2-01})-(\ref{cc-INT2-32}):
\begin{eqnarray}
\label{Def-FC-INT-LP}
\left\{
{\rm cc}_{01},{\rm cc}_{10},{\rm cc}_{21}
\right\}_{\rm LP}^{\rm INT}
&\!\!\!=\!\!\!&
\left\{
{\rm cs}_{01}, {\rm sc}_{10},{\rm sc}_{21}
\right\}_{\rm unp}^{\rm INT}
\Big|_{
\Im{\rm m}{\cal C}_{\rm unp} \to \Re{\rm e}{\cal C}_{\rm LP}
}\, ,
\\
\left\{
{\rm cs}_{01},{\rm sc}_{10},{\rm cs}_{12},{\rm sc}_{21} ,{\rm sc}_{32}
\right\}_{\rm LP}^{\rm INT}
&\!\!\!=\!\!\!&
\left\{
{\rm cc}_{01}, {\rm cc}_{10},{\rm cc}_{12},{\rm cc}_{21},{\rm cc}_{32}
\right\}_{\rm unp}^{\rm INT}
\Big|_{
\Re{\rm e}{\cal C}_{\rm unp} \to \Im{\rm m}{\cal C}_{\rm LP}
}\, .
\nonumber
\end{eqnarray}
In the case of a transversely polarized target we have as before an additional
decomposition in $\cos({\mit\Phi} - \phi) $ and $\sin({\mit\Phi} - \phi)$ and
should replace $K$ ($\widetilde K$) by  $L$ ($\widetilde L$)
\begin{eqnarray}
\label{Def-FC-INT-TP}
\left\{
{\rm cc}_{01},{\rm cc}_{10},{\rm cc}_{21}
\right\}_{{\rm TP}+}^{\rm INT}
&\!\!\!=\!\!\!&
\cos({\mit\Phi} - \phi)
\left\{
{\rm cs}_{01},{\rm sc}_{10},{\rm sc}_{21}
\right\}_{\rm unp}^{\rm INT}
\Big|_{
\Im{\rm m}{\cal C}_{\rm unp} \to \Re{\rm e}{\cal C}_{{\rm TP}+}
\atop
K \to L\, , \widetilde K \to \widetilde L \hspace{0.7cm}
}\, ,
\\
\left\{
{\rm cs}_{01},{\rm sc}_{10},{\rm sc}_{21}
\right\}_{{\rm TP}-}^{\rm INT}
&\!\!\!=\!\!\!&
\sin({\mit\Phi} - \phi)
\left\{
{\rm cs}_{01}, {\rm sc}_{10},{\rm sc}_{21}
\right\}_{\rm unp}^{\rm INT} \Big|_
{ \Im{\rm m}{\cal C}_{\rm unp} \to -\Re{\rm e}{\cal C}_{{\rm TP}-}
\atop
K \to L\, , \widetilde K \to \widetilde L \hspace{0.7cm}
}\, ,
\nonumber\\
\left\{
{\rm cs}_{01},{\rm sc}_{10},{\rm cs}_{12},{\rm sc}_{21},{\rm sc}_{32}
\right\}_{{\rm TP}+}^{\rm INT}  &\!\!\! =\!\!\! &
\cos({\mit\Phi} - \phi)
\left\{
{\rm cc}_{01}, {\rm cc}_{10}, {\rm cc}_{12}, {\rm cc}_{21},{\rm cc}_{32}
\right\}_{\rm unp}^{\rm INT}
\Big|_{
\Re{\rm e}{\cal C}_{\rm unp} \to \Im{\rm m}{\cal C}_{{\rm TP}+}
\atop
K \to L\, , \widetilde K \to \widetilde L \hspace{0.7cm}
}\, ,
\nonumber\\
\left\{
{\rm cc}_{01},{\rm cc}_{10},{\rm cc}_{12},{\rm cc}_{23} ,{\rm cc}_{32}
\right\}_{{\rm TP}-}^{\rm INT}  &\!\!\! =\!\!\! &
\sin({\mit\Phi} - \phi)
\left\{
{\rm cc}_{01},{\rm cc}_{10}, {\rm cc}_{12}, {\rm cc}_{23} ,{\rm cc}_{32}
\right\}_{\rm unp}^{\rm INT} \Big|_
{ \Re{\rm e} {\cal C}_{\rm unp} \to \Im{\rm m}{\cal C}_{{\rm TP}-}
\atop
K \to L\, , \widetilde K \to \widetilde L \hspace{0.7cm}
}\, .
\nonumber
\end{eqnarray}
Applying the relation (\ref{Rel-FC-INT-Tw2}), the remaining
nonvanishing Fourier coefficients, i.e., ${\rm ss}_{21}^{\rm INT} $,
${\rm sc}_{12}^{\rm INT} $ ${\rm cs}_{21}^{\rm INT} $, ${\rm
cs}_{23}^{\rm INT} $ for LP and ${\rm TP}+$ as well as ${\rm
ss}_{12}^{\rm INT} $, ${\rm ss}_{21}^{\rm INT} $, ${\rm ss}_{23}^{\rm
INT} $ and ${\rm cs}_{21}^{\rm INT} $ for ${\rm TP}-$ are easily established.



\begin{thebibliography}{99}
\bibitem{MueRobGeyDitHor94}
D. M\"uller, D. Robaschik, B. Geyer, F.-M. Dittes, J. Horejsi,
Fortschr. Phys. 42 (1994) 101.
\bibitem{Ji97}
X. Ji,
Phys. Rev. D 55 (1997) 7114.
\bibitem{Rad97}
A.V. Radyushkin,
Phys. Rev. D 56 (1997) 5524.
\bibitem{Bel03}
A.V. Belitsky,
{\it Renormalons in exclusive meson electroproduction},
hep-ph/0307256.
\bibitem{Ji03}
X. Ji,
{\it Viewing the proton through ``color'' filters},
hep-ph/0304037.
\bibitem{Ji96}
X. Ji,
Phys. Rev. Lett. 78 (1997) 610.
\bibitem{GouDiePirRal97}
M. Diehl, T. Gousset, B. Pire, J.P. Ralston,
Phys. Lett. B 411 (1997) 193.
\bibitem{BelMulNieSch00}
A.V. Belitsky, D. M\"uller, L. Niedermeier, A. Sch\"afer,
Nucl. Phys. B 593 (2001) 289.
\bibitem{BelMulKir01}
A.V. Belitsky, D. M\"uller, A. Kirchner,
Nucl. Phys. B 629 (2002) 323.
\bibitem{BerDiePir01}
E.R. Berger, M. Diehl, B. Pire,
Eur. Phys. J. C 23 (2002) 675.
\bibitem{BelMul02}
A.V. Belitsky, D. M\"uller,
Phys. Rev. Lett. 90 (2003) 022001.
\bibitem{GuiVan02}
M. Guidal, M. Vanderhaeghen,
Phys. Rev. Lett. 90 (2003) 012001.
\bibitem{ColFre99}
J.C. Collins, A. Freund,
Phys. Rev. D 59 (1999) 074009.
\bibitem{ShiVanZak79}
M.A. Shifman, A.I. Vainshtein, V.I. Zakharov,
Nucl. Phys. B 147 (1979) 347.
\bibitem{Rad99}
A.V. Radyushkin,
Phys. Lett. B 449 (1999) 81.
\bibitem{EfrRad80}
A.V. Efremov, A.V. Radyushkin,
Theor. Math. Phys. 44 (1981) 774.
\bibitem{LabSte85}
J.M.F. Labastida, G. Sterman,
Nucl. Phys. B 254 (1985) 425.
\bibitem{ColSopSte88}
J.C. Collins, D.E. Soper, G. Sterman,
{\it Factorization of hard process in QCD},
in {\it Perturbative QCD}, ed.\ A.H. Mueller, World Scientific
(Singapore, 1989) p.\ 1.
\bibitem{BelJiYua02}
A.V. Belitsky, X. Ji, F. Yuan,
Nucl. Phys. B 656 (2003) 165.
\bibitem{BelMul00}
A.V. Belitsky, D. M\"uller,
Nucl. Phys. B 589 (2000) 611.
\bibitem{AniPirTer00}
I.V. Anikin, B. Pire, O.V. Teryaev,
Phys. Rev. D 62 (2000) 071501.
\bibitem{PenPolShuStr00}
M. Pentinen, M.V. Polyakov, A.G. Shuvaev, M. Strikman,
Phys. Lett. B 491 (2000) 96.
\bibitem{RadWei00}
A.V. Radyushkin, C. Weiss,
Phys. Rev. D 63 (2001) 114012.
\bibitem{JiOsb98}
X. Ji, J. Osborne,
Phys. Rev. D 58 (1998) 094018.
\bibitem{BelMul97}
A.V. Belitsky, D. M\"uller,
Phys. Lett. B 417 (1998) 129.
\bibitem{ManPilSteVanWei97}
L. Mankiewicz, G. Piller, E. Stein, M. V\"anttinen, T. Weigl,
Phys. Lett. B 425 (1998) 186.
\bibitem{JiHoo98}
P. Hoodbhoy, X. Ji,
Phys. Rev. D 58 (1998) 054006.
\bibitem{BelMul00d}
A.V. Belitsky, D. M\"uller,
Phys. Lett. B 486 (2000) 369.
\bibitem{PolWei98}
M.V. Polyakov, C. Weiss,
Phys. Rev. D 60 (1999) 114017.
\bibitem{SchBofRad02}
P. Schweitzer, S. Boffi, M. Radici,
Phys. Rev. D 66 (2002) 114004.
\bibitem{GeoPolVan01}
K. Goeke, M.V. Polyakov, M. Vanderhaeghen,
Prog. Part. Nucl. Phys. 47 (2001) 401.
\bibitem{Gar03}
M. Gar\c{c}on,
talk at {\it User group symposium and annual meeting: A selebration of JLab
physics},
{\tt http://www.jlab.org/intralab/calendar/archive03/ugm/talks/garcon.pdf}.
\end{thebibliography}
\end{document}